\documentclass[aps,pra,reprint,amsmath,amssymb,superscriptaddress,onecolumn,longbibliography, notitlepage]{revtex4-2}
\usepackage{mathtools}
\usepackage{siunitx}
\usepackage[hidelinks]{hyperref}
\hypersetup{colorlinks}
\usepackage{svg}
\usepackage{tikz}
\usepackage{caption}
\usepackage{float}
\usepackage{subcaption}
\usepackage[braket, qm]{qcircuit}
\usepackage{braket}
\usepackage{bbm}

\newcommand{\mI}{\mathcal{I}}
\newcommand{\qftd}{\left(\text{QFT}\right)^{\dagger}}
\usepackage{booktabs}
\usepackage{amsfonts}
\usepackage{multirow}
\usepackage{array}
\usepackage{amsmath}
\usepackage{amssymb}
\usepackage{soul}

\begin{document}
\title{Linear-depth quantum circuits for loading Fourier approximations of arbitrary functions}
\author{Mudassir~Moosa}
\thanks{These authors contributed equally to this work}
\email{mudassir@purdue.edu}
\affiliation{Department of Physics, Cornell University, Ithaca, NY 14853, USA}
\affiliation{Department of Physics and Astronomy, Purdue University, West Lafayette, IN 47907, USA}
\author{Thomas~W.~Watts}
\thanks{These authors contributed equally to this work}
\email{tww55@cornell.edu}
\affiliation{School of Applied and Engineering Physics, Cornell University, Ithaca, NY 14853, USA}
\author{Yiyou Chen}
\affiliation{School of Applied and Engineering Physics, Cornell University, Ithaca, NY 14853, USA}
\affiliation{Department of Electrical and Computer Engineering, Princeton University, Princeton, NJ 08544, USA}
\author{Abhijat Sarma}
\affiliation{Department of Physics, Cornell University, Ithaca, NY 14853, USA}
\author{Peter~L.~McMahon}
\email{pmcmahon@cornell.edu}
\affiliation{School of Applied and Engineering Physics, Cornell University, Ithaca, NY 14853, USA}

\begin{abstract}

The ability to efficiently load functions on quantum computers with high fidelity is essential for many quantum algorithms, including those for solving partial differential equations and Monte Carlo estimation. In this work, we introduce the Fourier Series Loader (FSL) method for preparing quantum states that exactly encode multi-dimensional Fourier series using linear-depth quantum circuits. Specifically, the FSL method prepares a ($Dn$)-qubit state encoding the $2^{Dn}$-point uniform discretization of a $D$-dimensional function specified by a $D$-dimensional Fourier series. A free parameter, $m$, which must be less than $n$, determines the number of Fourier coefficients, $2^{D(m+1)}$, used to represent the function. The FSL method uses a quantum circuit of depth at most $2(n-2)+\lceil \log_{2}(n-m) \rceil + 2^{D(m+1)+2} -2D(m+1)$, which is linear in the number of Fourier coefficients, and linear in the number of qubits ($Dn$) despite the fact that the loaded function's discretization is over exponentially many ($2^{Dn}$) points. The FSL circuit consists of at most $Dn+2^{D(m+1)+1}-1$ single-qubit and $Dn(n+1)/2 + 2^{D(m+1)+1} - 3D(m+1) - 2$ two-qubit gates; we present a classical compilation algorithm with runtime $O(2^{3D(m+1)})$ to determine the FSL circuit for a given Fourier series. The FSL method allows for the highly accurate loading of complex-valued functions that are well-approximated by a Fourier series with finitely many terms. We report results from noiseless quantum circuit simulations, illustrating the capability of the FSL method to load various continuous 1D functions, and a discontinuous 1D function, on 20 qubits with infidelities of less than $10^{-6}$ and $10^{-3}$, respectively. We also demonstrate the practicality of the FSL method for near-term quantum computers by presenting experiments performed on the Quantinuum H$1$-$1$ and H$1$-$2$ trapped-ion quantum computers: we loaded a complex-valued function on 3 qubits with a fidelity of over $95\%$, as well as various 1D real-valued functions on up to 6 qubits with classical fidelities $\approx 99\%$, and a 2D function on 10 qubits with a classical fidelity $\approx 94\%$.
\end{abstract}

\maketitle

\section{Introduction}
\label{sec:intro}

Efficiently loading classical data on quantum computers is an important ingredient in many quantum algorithms. For example, quantum-linear-solver algorithms require an efficient preparation of a state $\ket{b}$ in order to compute the solution $\ket{x} \propto A^{-1} \ket{b}$ to a large linear system $A$ \cite{hhl_original_paper,Childs_linear_systems, fast_lin_alg,preconditioned_hhl}. In general, exactly loading $2^{n}$ dimensional data into a state of $n$ qubits requires a quantum circuit of $O(2^{n})$ quantum operations
\cite{1998RSPSA.454..313Z,2002quant.ph..8112G,mottonen,arb_state_prep}. Exponential space and time complexities for the read-in component of a quantum algorithm will compromise any potential for an exponential speed-up over comparable classical algorithms \cite{2015NatPh..11..291A}. Hence, finding an efficient method for accurately loading classical input data into quantum computers is a problem with vast applications. 

In many practical applications, classical input data takes the form of values of a function or a distribution on a uniform discretized grid (or mesh) \cite{single_particle_tdse,Berry_2017_linear_de,lubasch, child_pdes_algorithm, nonlinear_pdes_algorithms,quantum_walk_MH, quantum_data_fitting, option_pricing_qc, risk_analysis, generalized_inner}. It is well-known that a large class of functions and distributions can be accurately approximated by a Fourier series of only a few terms. In this work, we exploit this fact in order to load a truncated Fourier-series approximation to a general multivariate target function $f : [0,1]^D \to \mathbb{C}$. More precisely, given a truncated Fourier series consisting of $2^{D(m+1)}$ Fourier modes, $f_{(m)}(x_{1},\cdots,x_{D})$, that approximates the target function of $D$ variables, $f(x_{1},\cdots,x_{D})$, we present a method to exactly prepare a quantum state of $Dn$ qubits of the form $\ket{f_{(m)}} \, = \, \sum_{k_{1}=0}^{2^{n}-1} \cdots \sum_{k_{D} = 0}^{2^{n}-1} f_{(m)}(k_{1}/2^{n},\cdots, k_{D}/2^{n}) \ket{k_{1}} \otimes \cdots \otimes \ket{k_{D}} \, $. For single variable functions (i.e., $D=1$), the target state becomes $\ket{f_{(m)}} \, = \, \sum_{k=0}^{2^{n}-1} f_{(m)}(k/2^{n}) \ket{k} \, $.  This allows us to load the target function with practically arbitrarily high fidelity so long as a sufficient number of Fourier modes are used. We refer to our method as the \textit{Fourier Series Loader} (FSL). 

The FSL method consists of three main steps: First, we find the dominant $2^{m+1}$ Fourier coefficients for a given function. This can be done classically with a time complexity of $O(m2^{m})$ using sparse Fourier transform algorithms \cite{6879613,Lawlor2012AdaptiveSF,Ghazi2013SampleoptimalAS,Pawar2013ComputingAK}. The second step is to load the $2^{m+1} \ll 2^{n}$ Fourier coefficients into a sparse quantum state of $n$ qubits. We achieve this using a quantum circuit of depth $O(\log(n) + 2^{m})$, which can be found in $O(8^m)$ classical runtime, and consists of $O(n+2^{m})$ and $O(2^{m})$ two-qubit and single-qubit gates, respectively. This is an improvement over a generic sparse state preparation algorithm that sports a classical runtime of $O(mn 4^{m})$ and produces a circuit of depth $O(n 2^{m})$ which consists of $O(n 2^{m})$ two-qubit gates and $O(m2^{m}+n)$ single-qubit gates \cite{efficient_sparse_state_prep}. Finally, the third step of the FSL method is to apply the inverse quantum Fourier transform to generate a `real-space' representation of the desired function from the loaded Fourier coefficients. The full quantum circuit implementing our method is shown in Fig.~(\hyperlink{fig:state_prep}{1a}) and has a depth of $O(n+2^{m})$ and consists of $O(n^{2} + 2^{m})$ two-qubit and $O(n+2^{m})$ single-qubit gates. The only error introduced by the FSL method is due to the approximation of the given function by a truncated Fourier series. This \textit{truncation error} is determined by the number of Fourier modes which dictates the requisite value of $m$. In particular, the infidelity between the target state and the state prepared by the FSL method decays exponentially with $m$.

It is worth mentioning that the FSL method is analogous to a state preparation method based on Fourier interpolation \cite{quantuminspired,GarcaMolina2022QuantumFA}. The interpolation-based method loads the values of the target function at $2^{m+1}$ discretized points and then interpolates to $2^{n}$ points using the quantum Fourier transform. The FSL method, as discussed above, instead loads $2^{m+1}$ Fourier coefficients and then produces a Fourier series representation of the target function using the quantum Fourier transform. Despite their similarities, the FSL method has various advantages over the interpolation-based method including, but not limited to, superior accuracy and reduced gate count.

\begin{figure}[!]
    \centering
    \hypertarget{fig:state_prep}{}
    \hypertarget{fig:zgr_circuit_diagram}{}
    \hypertarget{fig:svd_circuit}{}
    \hypertarget{fig:nonperiodic}{}
    \includegraphics[scale=0.90]{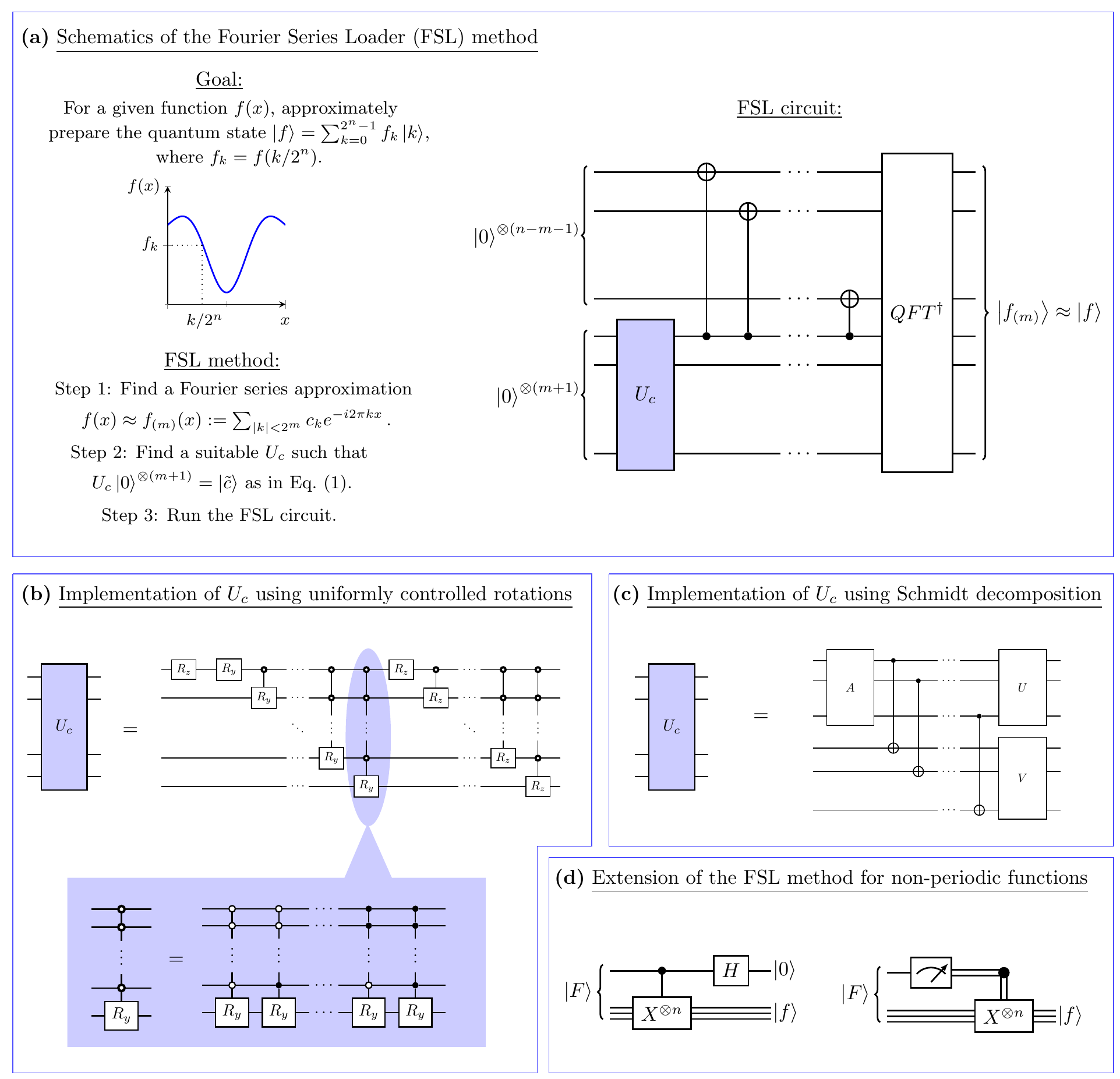} 
    \caption{\textbf{(a)} The Fourier Series Loader (FSL) method loads a Fourier series approximation of the target function into a quantum state of $n$ qubits. Implementing the FSL method requires approximating the target function with a Fourier series of $2^{m+1}$ terms and finding a unitary circuit $U_{c}$ to load the $2^{m+1}$ Fourier coefficients on $m+1$ qubits, both of which can be done on a classical computer. The integer $m$ should be chosen such that the truncation error of the Fourier series, and hence, the infidelity of the prepared state is within a desired threshold.  Once a suitable unitary circuit $U_{c}$ is found, the Fourier series approximation of the target function is loaded on $n$ qubits by running the FSL circuit on a quantum computer. The FSL circuit consists of three parts. First, the unitary $U_{c}$ loads $2^{m+1}$ Fourier coefficients in $m+1$ qubits and prepares the state given in Eq.~\eqref{eq-c-state}. Then, a cascade of CNOT gates entangles the remaining $n-m-1$ qubits with the $m+1$ qubits storing the Fourier coefficients which efficiently implements the technique of zero padding (see Eq.~(\ref{zero_padding})). Finally, the inverse QFT at the end of the circuit transforms the loaded Fourier coefficients into a state encoding the Fourier series approximation of the target function. \textbf{(b)} One possible implementation of the unitary $U_{c}$ is using a cascade of uniformly controlled rotations, which can be used to prepare an arbitrary quantum state. A uniformly $k$-controlled rotation operator, shown in the shaded region, decomposes as a series of controlled rotations conditioned on each computational basis state of $k$ control qubits. \textbf{(c)} Another possible implementation of the unitary $U_{c}$ is using a circuit that can prepare an arbitrary quantum state once the Schmidt decomposition of the target state is computed. The unitary $A$ in the circuit loads the Schmidt coefficients whereas unitaries $U$ and $V$ rotate the computational basis into the Schmidt basis. \textbf{(d)} The FSL method loads a non-periodic function $f$ by first loading a periodic extension $F$ on $n$ qubits and $1$ ancilla qubit. Then the desired state is prepared either through disentangling the ancilla qubit (left) or through measurement-based operations (right).} 
    \label{fig-fsl1}
\end{figure}

The FSL has several attractive features that make it appealing for loading classical functions on quantum computers. First, we have complete control over the accuracy of our results as it is determined by the number of Fourier coefficients used to represent the target function. Secondly, no time-consuming classical optimization of a parameterized quantum circuit or matrix product state is needed unlike other comparable methods for function loading in the literature \cite{qgan,mps,2020PhRvR...2d3442E,quantuminspired}. In fact, all the angles of rotation in our quantum circuits can be determined from the Fourier coefficients of the function of interest as we explain in Sec.~(\ref{sec:method}). Thirdly, the FSL works equally well for loading real-valued and complex-valued functions. More remarkably, the FSL can be easily generalized to load a $D$-dimensional Fourier series into a quantum state of $Dn$ qubits using a quantum circuit of $O(n)$ depth and $O(D n^{2})$ quantum gates. Finally, due to the low depth of the circuit in Fig.~(\hyperlink{fig:state_prep}{1a}), the FSL method is capable of loading functions on near-term noisy quantum computers as we demonstrate in Sec.~(\ref{sec:exp_results}).

The FSL method is particularly useful for quantum algorithms for solving differential equations that require the preparation of states encoding initial conditions \cite{child_pdes_algorithm, Berry_2017_linear_de, nonlinear_pdes_algorithms,single_particle_tdse, lubasch}. Other contexts in which the FSL method is useful include quantum image processing \cite{quantum_image_1}, Monte Carlo methods based on quantum amplitude estimation algorithms \cite{risk_analysis, option_pricing_qc}, and computing generalized inner products \cite{generalized_inner}. These last two contexts find applications in the domain of option pricing and risk analysis \cite{risk_analysis, option_pricing_qc,generalized_inner}.  Though this is not the goal of the present work, we elaborate in the supplementary Sec.~(\ref{app-image}) on how the FSL method can be utilized to load an image into a quantum state. 

The rest of this paper is organized as follows: In Sec.~(\ref{sec:method}), we present the details of the FSL method and the accompanying quantum circuit implementations. Starting with the simplest use cases, we demonstrate how our FSL method can be used to load periodic functions of a single variable. Then we discuss how the FSL method can be easily modified to load non-periodic, piece-wise discontinuous, and multivariate functions. For each case, we demonstrate high fidelity loading of various functions using the FSL method by quantum simulations performed using Qiskit \cite{Qiskit}.  Next, in Sec.~(\ref{sec:exp_results}), we discuss several experiments of loading functions of one and two variables on the Quantinuum H$1$-$1$ and H$1$-$2$ quantum computers \cite{quantinuum}. The results of these experiments demonstrate that the FSL method can load functions on up to $10$ qubits on a present-day noisy quantum computer. Finally, we conclude with a comparison of the FSL method with other state preparation methods for loading functions to quantum states and some possible future directions in Sec.~(\ref{sec:discussion}).

\section{Methods} \label{sec:method}

Our goal is to load the values of any arbitrary complex-valued function to the amplitudes of a quantum state of $n$ qubits. For example, given a one-dimensional function $f: [0,1] \to \mathbb{C}$, we want to prepare an $n$-qubit state $\ket{f} = \sum_{k=0}^{2^{n}-1} \, f_{k} \ket{k}$, where $\ket{k}$ is the $k^{\text{th}}$ computational basis state and $f_{k} = f(k/2^{n})$ is the value of the function at $k^{\text{th}}$ grid point. Without loss of generality, we assume that the function $f$ is such that the corresponding quantum state representation $\ket{f}$ is normalized. 

We begin by considering the simplest case of a function $f$ given by a finite Fourier series of the form $f(x) \, = \, \sum_{k=-M}^{M} c_{k} e^{-i 2\pi  k x} \, $, where $M := 2^{m} - 1 < 2^{n-1}$. We claim that the quantum state $\ket{f}$ corresponding to this Fourier series can be prepared exactly using the quantum circuit shown in Fig.~(\hyperlink{fig:state_prep}{1a}) in which $U_{c}$ denotes an arbitrary implementation of the unitary that maps $\ket{0}^{\otimes (m+1)}$ state to $\ket{\tilde{c}}$, where
\begin{align}
    \ket{\tilde{c}} \, := \, 2^{n/2} \sum_{k=0}^{M} c_{k} \ket{k} + 2^{n/2}\sum_{k=1}^{M} c_{-k} \ket{2^{m+1}-k} \, . \label{eq-c-state}
\end{align} 
The cascade of $n-m-1$ CNOT gates that follow $U_c$ entangle all $n$ qubits, mapping $\ket{0}^{\otimes (n-m-1)} \otimes \ket{\tilde{c}}$ to an $n$-qubit state $\ket{c}$, where
\begin{align}\label{zero_padding}
    \ket{c} \, := \, 2^{n/2} \sum_{k=0}^{M} c_{k} \ket{k} + 2^{n/2}\sum_{k=1}^{M} c_{-k} \ket{2^{n}-k} \, .
\end{align} 
Finally, the inverse Quantum Fourier Transform (QFT) maps $\ket{c}$ to the desired state $\ket{f}$. A more detailed derivation is provided in the Supplementary Sec.~(\ref{app-2d-fsl}).

The QFT and its inverse can be implemented with a quantum circuit of depth $O(n)$ and with $O(n^{2})$ quantum gates. The cascade of CNOT gates in Fig.~(\hyperlink{fig:state_prep}{1a}) can be implemented in $O(\log(n))$ depth. In the worst case, the implementation of $(m+1)$-qubit unitary $U_{c}$ will require $O(2^{m})$ quantum gates and a circuit of depth $O(2^{m})$. However, note that $m$ is entirely fixed by the number of Fourier modes in our function and does not scale with the total number of qubits, $n$. Hence, for a Fourier series with a fixed number of terms, the circuit in Fig.~(\hyperlink{fig:state_prep}{1a}) has $O(n)$ depth and has $O(n^{2})$ quantum gates.  

The problem of preparing the $n$-qubit state $\ket{f}$ has now reduced to the problem of preparing a $(m+1)$-qubit state $\ket{\tilde{c}}$. When the target Fourier series consists of only a few terms, the construction of an efficient circuit that prepares $\ket{\tilde{c}}$ is usually straightforward. However, when loading more complex functions that require many Fourier coefficients, we propose two general methods for constructing the state $\ket{\tilde{c}}$.

\subsection*{Implementations of $\boldsymbol{U_c}$}

Our first proposal is to prepare the state $\ket{\tilde{c}}$ using a cascade of uniformly controlled rotations as shown in Fig.~(\hyperlink{fig:zgr_circuit_diagram}{1b}), which can be used to prepare any arbitrary quantum state as was shown by M\"{o}tt\"{o}nen et al. \cite{mottonen}. M\"{o}tt\"{o}nen et al. proposed this circuit as an extension of a circuit of Zalka, Grover, and Rudolph \cite{1998RSPSA.454..313Z,2002quant.ph..8112G} which only involves a cascade of uniformly controlled $R_{y}$ gates, and hence, can only prepare states with positive, real-valued amplitudes. Moreover, M\"{o}tt\"{o}nen et al. made two interesting observations that render this circuit a practical choice for implementing $U_c$. First, this circuit can be constructed using $2^{m+2}$ single-qubit rotations and $2^{m+2}$ CNOT gates. Secondly, there exists a set of formulae for calculating the angles of rotation for all single-qubit gates in the circuit when provided access to the amplitudes of the target state $\ket{\tilde{c}}$ \cite{mottonen}. These formulae can be evaluated on a classical computer in $O(8^m)$ time. (See Supplementary Sec.~(\ref{app-ucr}) for more details.)

Our second proposal is based on the Schmidt decomposition. We can treat $\ket{\tilde{c}}$ as the overall quantum state of a bipartite system by partitioning $m+1$ qubits into $(m+1)/2$ and $(m+1)/2$ qubits if $m$ is odd or into $m/2$ and $m/2 + 1$ qubits if $m$ is even. Then, we can classically compute the Schmidt decomposition of $\ket{\tilde{c}}$ and express it as $\ket{\tilde{c}} = U \otimes V \, \sum_{k=0}^{r} \alpha_{k} \ket{k}\otimes\ket{k} \, $, where $\alpha_{k}$ are the Schmidt coefficients (singular values), $U$ and $V$ are unitary operators, and $r = 2^{(m+1)/2}$ ($r = 2^{m/2}$) if $m$ is odd (even). Given this decomposition, we can prepare the state $\ket{\tilde{c}}$ using the circuit shown in Fig.~(\hyperlink{fig:svd_circuit}{1c}) \cite{2018arXiv180403719A}. The gate $A$ in Fig.~(\hyperlink{fig:svd_circuit}{1c}) is taken to be the uniformly controlled rotations that encode the Schmidt coefficients in the state $\ket{\alpha} \, = \, \sum_{k=0}^{r} \alpha_{k}\ket{k} \, ,$ and the gates $U$ and $V$ are generated using methods for unitary synthesis methods such as those described in \cite{arb_unitaries}. The classical computational cost to implement this method grows exponentially in $m$, which is due to the cost of finding the Schmidt decomposition of the target state $\ket{\tilde{c}}$ and the cost of computing the gate decompositions of unitaries $A$, $U$, and $V$.

These two methods complement each other nicely. On the one hand, the first method has a lower classical pre-processing cost than the second method. On the other hand, the first method leads to a quantum circuit with a higher gate count than the second method for generic states. Since a lower gate count is more desirable when working with noisy quantum computers, we used the second method for most of the experiments we performed on real-world quantum computers as we discuss in Sec.~(\ref{sec:exp_results}). However, we used the first method when performing quantum simulations which we discuss in the next subsection.

Finally, even though the asymptotic scaling of the computational cost of finding $U_{c}$ using the aforementioned two methods is exponential in $m$, we find that the computational time to find implement $U_c$ is reasonable in practice. For example, for $m=10$ (i.e., around $2000$ Fourier coefficients), it takes less than $6$ and $3$ seconds to find $U_{c}$ using the first and the second method respectively. More details about the benchmarking of the CPU wall clock time can be found in Supplementary Sec.~(\ref{app-classical-time}).

\subsection*{Function Loading}

Generally, only a few terms in the truncated Fourier series are needed to well-approximate any periodic function $f$. Now, given a Fourier series approximation $f_{(m)} = \sum_{k=-M}^M c_k e^{-i 2\pi  k x}$ of a periodic function $f$, we can use the FSL method to prepare the state $\ket{f_{(m)}} \approx \ket{f}$. The only source of error between the prepared state $\ket{f_{(m)}}$ and the target state $\ket{f}$ is the truncation error from the Fourier series approximation. The error as measured by infidelity,  $\epsilon_{(m)} := 1 - \left| \braket{f | f_{(m)}} \right|^{2} \, $, decays exponentially with $m$ in the limit of large $n$ as we show in the supplementary Sec.~(\ref{app-fid-analysis}). Moreover, the rate of exponential decay of infidelity depends on the smoothness of the function $f$. The bounds on infidelity derived in the supplementary Sec.~(\ref{app-fid-analysis}) can be used to determine the number of Fourier modes needed to prepare the state $\ket{f}$ with a specified infidelity.

In order to approximate a function by a Fourier series, we first need to find its Fourier coefficients, which leads to additional classical pre-processing costs. The fast Fourier transform (FFT) has a computational cost of $O(n2^{n})$, which takes away any hope of exponential advantage. Fortunately, there exists various sparse Fourier transform algorithms that can be used to find $M \ll 2^{n}$ dominant Fourier coefficients efficiently \cite{6879613}. In particular,  algorithms developed in \cite{Lawlor2012AdaptiveSF,Ghazi2013SampleoptimalAS,Pawar2013ComputingAK} compute the $M$ non-zero coefficients in time $O(M\log M)$. However, for our implementations of the FSL method, we used the conventional FFT.

To demonstrate the usefulness of the FSL method, we performed the quantum simulation of loading various periodic functions into a $20$ qubit state using the `statevector\textunderscore simulator' provided by Qiskit \cite{Qiskit}. For concreteness, we chose $m = 6$ (i.e., $128$ Fourier modes) and loaded the Fourier coefficients using the uniformly controlled rotations shown in Fig.~(\hyperlink{fig:zgr_circuit_diagram}{1b}). We were able to load non-trivial functions such as $x^{x}$, a sinc function, the reflected put option function of Ref.~\cite{2021arXiv210104023G}, and the wavefunction of the first excited state of a quantum harmonic oscillator with an infidelity of less than $10^{-6}$, $10^{-7}$, $10^{-6}$, and $10^{-12}$, respectively. The results of the simulation are shown in Fig.~(\hyperlink{fig:dg}{2a})-(\hyperlink{fig:qho}{2d}).

\begin{figure}[t]
    \centering
    \hypertarget{fig:dg}{}
    \hypertarget{fig:put}{}
    \hypertarget{fig:sinc}{}
    \hypertarget{fig:qho}{}
    \hypertarget{fig:sigmoid}{}
    \hypertarget{fig:discontinuous}{}
    \includegraphics[scale=0.9]{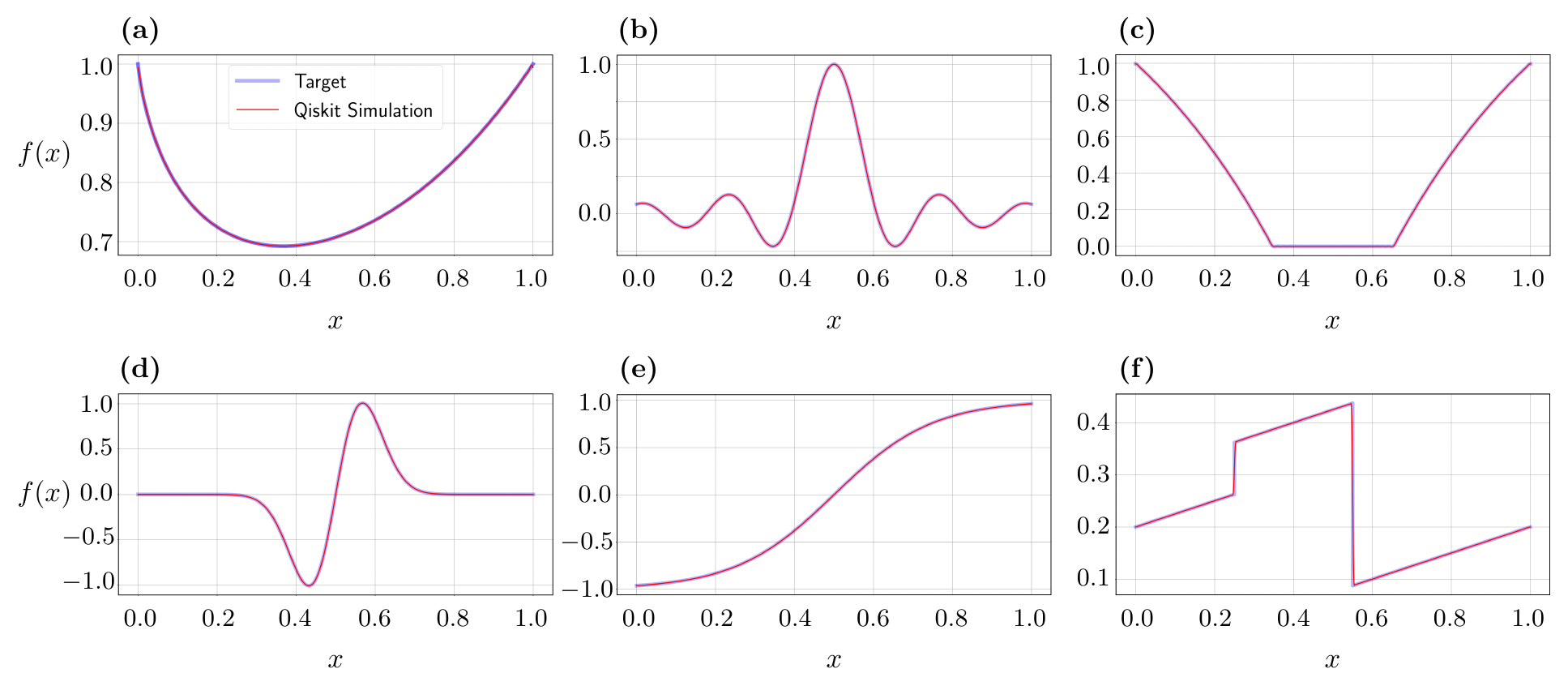} 
    \caption{Quantum circuit simulations of loading each of the following 1D functions into a quantum state of $n=20$ qubits using the FSL method: \textbf{(a)} $x^{x}$, \textbf{(b)} sinc, \textbf{(c)} reflected put option, \textbf{(d)} first excited state of a quantum oscillator, \textbf{(e)} hyperbolic tangent, and  \textbf{(f)} a piece-wise discontinuous function. Continuous functions in \textbf{(a)}-\textbf{(e)} were approximated with $128$ Fourier coefficients (i.e., $m=6$) whereas the piece-wise discontinuous function in \textbf{(f)} was approximated with $512$ Fourier coefficients (i.e., $m=8$). These functions were loaded with infidelity of less than \textbf{(a)} $10^{-6}$, \textbf{(b)} $10^{-7}$, \textbf{(c)} $10^{-6}$, \textbf{(d)} $10^{-12}$, \textbf{(e)} $10^{-8}$, and \textbf{(f)} $10^{-3}$. All of the simulations are performed using the `statevector\textunderscore simulator' provided by Qiskit \cite{Qiskit} and the Fourier coefficients were loaded using the uniformly controlled rotations shown in Fig.~(\protect\hyperlink{fig:zgr_circuit_diagram}{1b}). The results are presented up to normalization. }
    \label{fig:statevectors}
\end{figure}

The FSL method can be modified to load non-periodic functions as well. Given a non-periodic function $f$ on the domain $[0,1]$, we can define a function $F$ on the extended domain $[0,2]$ such that $F(x) = f(x)/\sqrt{2}$ for $0\le x \le 1$ and $F(x) = f(2-x)/\sqrt{2}$ for $1 \le x \le 2 $. By construction, $F$ is periodic on the domain $[0,2]$, and hence, can be loaded with high fidelity into a quantum state of $(n+1)$ qubits using the FSL method. We introduce an ancilla qubit so that there will be twice as many grid points on the extended domain which ensures that the grid spacing remains $1/2^{n}$. The state $\ket{F}$ of $n+1$ qubits is of the form
\begin{align}
    \ket{F} \, = \, \frac{1}{\sqrt{2}} \ket{0} \otimes \sum_{k=0}^{2^{n}-1} f_{k} \ket{k}  +  \frac{1}{\sqrt{2}} \ket{1} \otimes \sum_{k=0}^{2^{n}-1} f_{2^{n}-1-k} \ket{k} \, .
\end{align}
With the observation that $X^{\otimes n} \ket{k} \, = \, \ket{2^{n}-1-k}$, where $X$ is the Pauli-$X$ operator, we find that the state $\ket{F}$ can be written as
\begin{align}
    \ket{F} \, = \, \frac{1}{\sqrt{2}} \ket{0}\otimes\ket{f} + \frac{1}{\sqrt{2}} \ket{1} \otimes  X^{\otimes n} \ket{f} \, . 
\end{align}
Once we prepare $\ket{F}$ using the FSL, we measure the ancilla qubit in the computational basis. We do nothing to the $n$ qubits if the measurement output is $0$, but we apply $X$ to each of the $n$ qubits if the measurement output is $1$. In either case, the state of the $n$ qubits is $\ket{f}$ as desired. Alternatively, we can prepare the state $\ket{F}$ and apply CNOT gates on each of the $n$ qubits and a Hadamard gate on the ancilla qubit. This disentangles the ancilla qubit from the other $n$ qubits and maps $\ket{F}$ to $\ket{0}\otimes\ket{f} \, $. The circuits implementing both methods are provided in Fig.~(\hyperlink{fig:nonperiodic}{1d}). 

In order to demonstrate the FSL's ability to load non-periodic functions, we performed the simulation of loading a hyperbolic tangent function into a $20$ qubit state. Once again, we chose $m=6$ and used the uniformly controlled rotations to load the Fourier coefficients. We were able to load the hyperbolic tangent with infidelity of less than $10^{-8}$ and the result of this simulation is shown in Fig.~(\hyperlink{fig:sigmoid}{2e}).

Generalizing further, Fourier series approximations of piece-wise discontinuous functions typically exhibit large oscillations at the discontinuities resulting in overshooting; this is known as the Gibbs phenomenon. Depending on the application, these large oscillations may not be desirable. One way to suppress these oscillations is to apply Fourier filters, such as a Lanczos $\sigma$-factor, to the Fourier coefficients: $c_{k} \to \, \left(\text{sinc}(\pi k/M)\right)^{a} c_{k} \, $, where $M$ is the order of the partial Fourier series and $a \in \mathbb{R}^{+} \, $ \cite{lancoz}. The Lanczos filter is just one of many Fourier filters available and is by no means the best choice for every discontinuous function of interest \cite{expaccfilter, onesidedfilters, pdefilter}. The FSL method can easily incorporate these Fourier filters with a small additional classical pre-processing cost. Instead of directly loading in the Fourier coefficients using the unitary $U_{c}$ in Fig.~(\hyperlink{fig:state_prep}{1a}), one can load the filtered Fourier coefficients. This ability of the FSL method to suppress the Gibbs phenomenon gives it an edge over other QFT-based state preparation methods \cite{quantuminspired,2020PhRvR...2d3442E}. (See Sec.~(\ref{sec:discussion}) for more details.) As an example, we performed the quantum simulation of loading a piece-wise discontinuous shown in Fig.~(\hyperlink{fig:discontinuous}{2f}) into a $20$ qubit state. We found that with $m=8$, we were able to prepare the desired state with an infidelity of around $10^{-3}$.

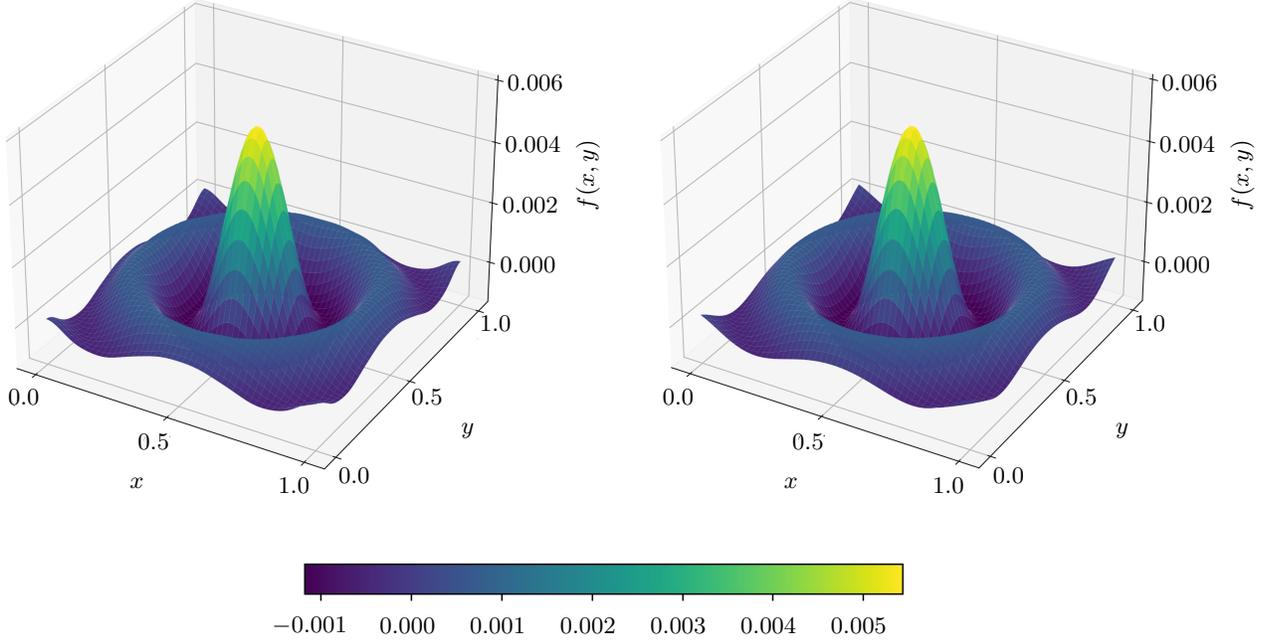
\begin{figure}
\begin{tikzpicture}
\node at (4.20,-4.85) {\includegraphics[scale=0.43]{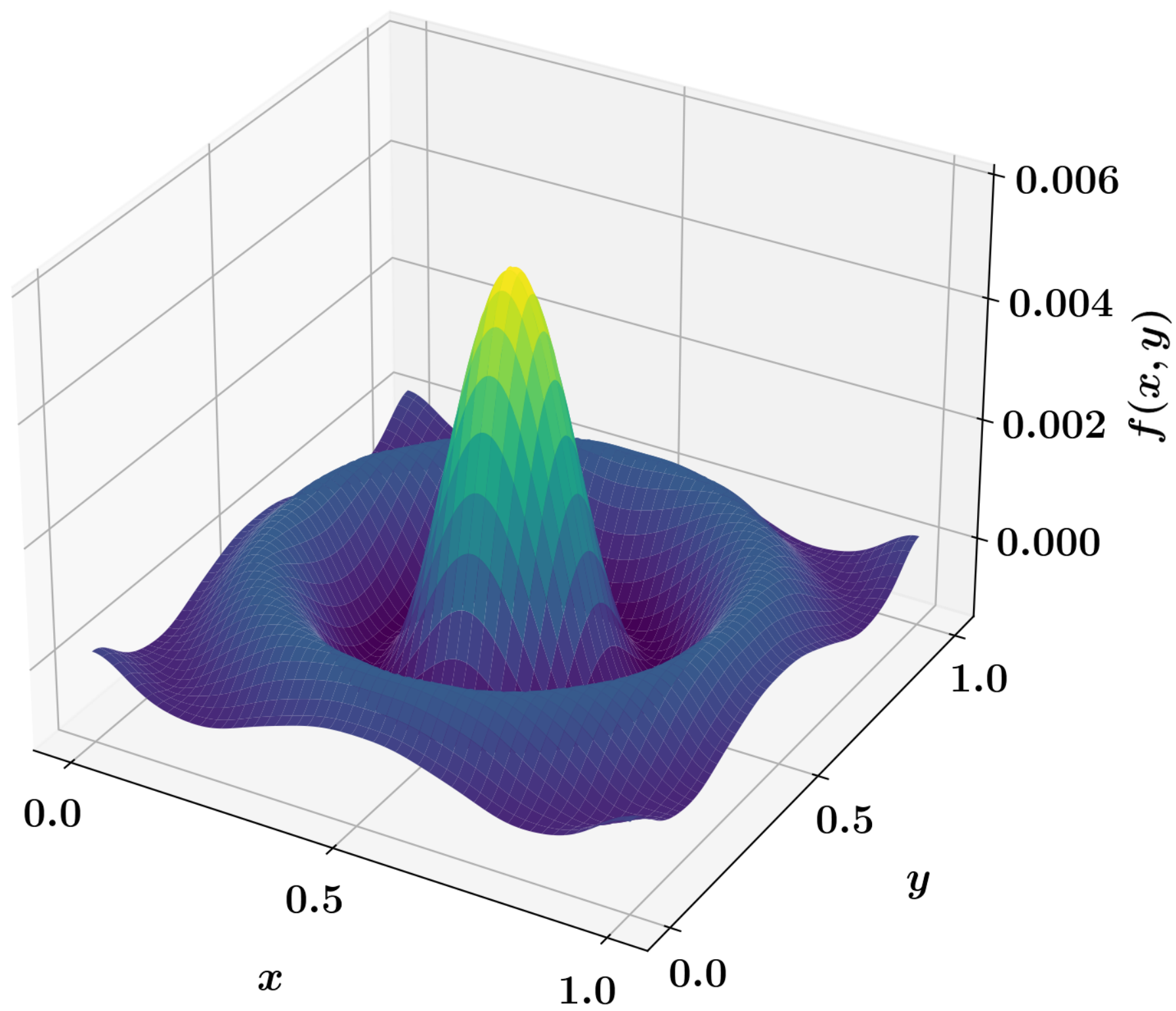}};
\node at (3.20,-0.95) {\textbf{(a)} \underline{Exact two-dimensional sinc function}};
\node at (12.90,-4.85) {\includegraphics[scale=0.43]{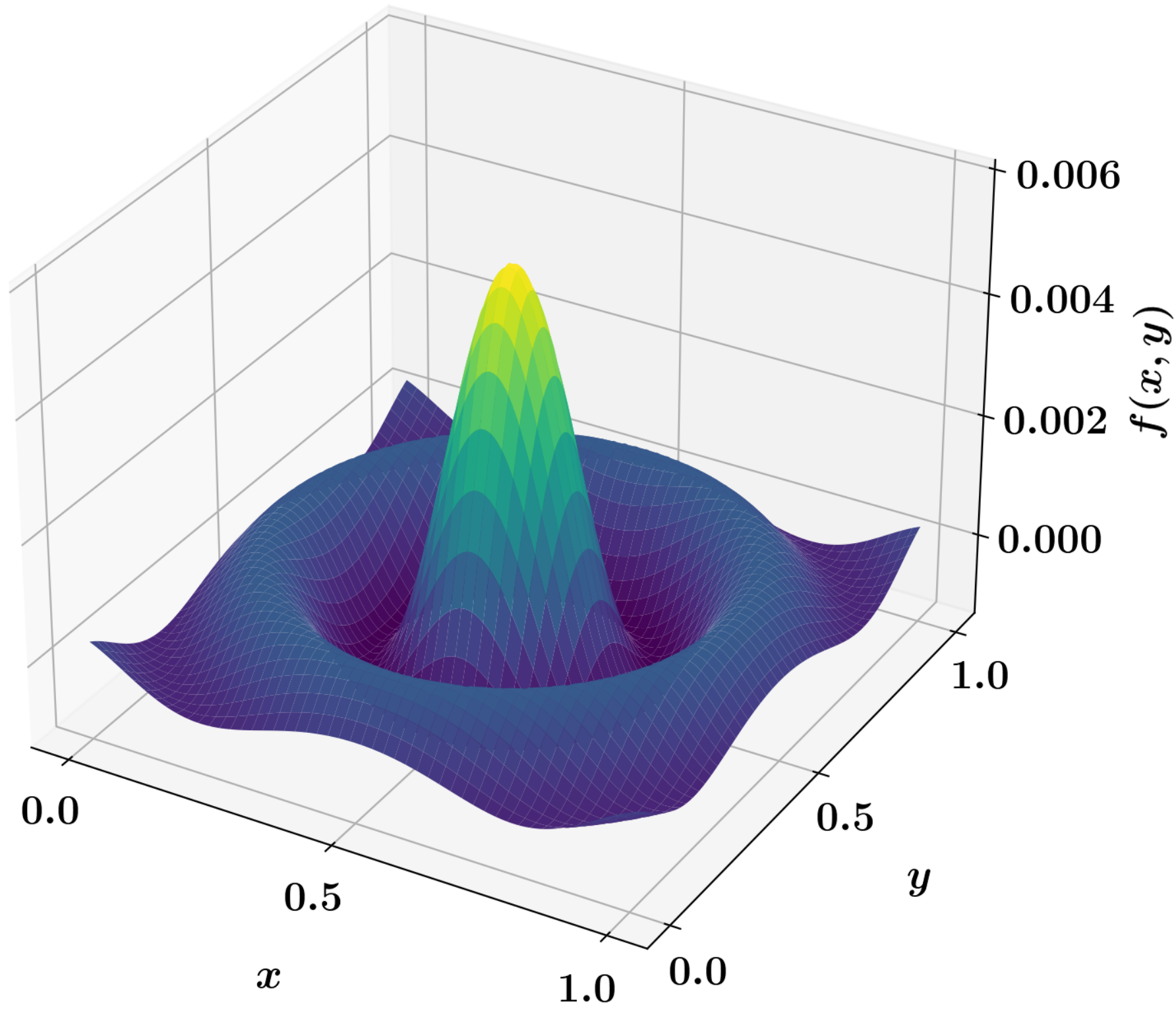}};
\node at (12.50,-0.95) {\textbf{(b)} \underline{Simulation of the two-dimensional FSL method}};
\node at (8.22,-9.5) {\includegraphics[scale=0.7]{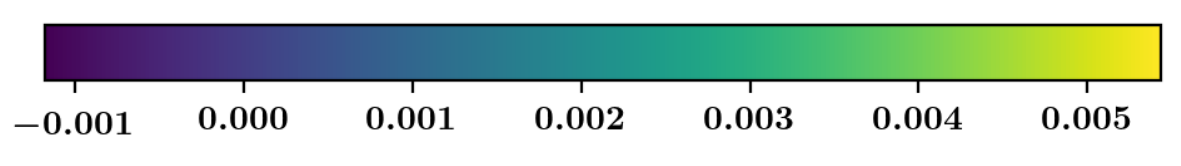}};
\draw[white,fill=white] (2.10,-8.0) circle (1 ex);
\node at (2.1,-8.0) {\small{$x$}};
\draw[white,fill=white] (6.40,-7.4) circle (1 ex);
\node at (6.5,-7.3) {\small{$y$}};
\draw[white,fill=white] (0.64,-6.85) circle (1.45 ex);
\node at (0.60,-6.85) {\small{$0.0$}};
\draw[white,fill=white] (2.38,-7.5) circle (1.4 ex);
\node at (2.325,-7.45) {\small{$0.5$}};
\draw[white,fill=white] (2.45+2.38-0.64,-8.1) circle (1.45 ex);
\node at (2.45+2.38-0.64,-8.03) {\small{$1.0$}};
\draw[white,fill=white] (3.19+2.38-0.64,-8.0) circle (1.55 ex);
\node at (3.25+2.38-0.64,-7.90) {\small{$0.0$}};
\draw[white,fill=white] (4.16+2.38-0.64,-6.95) circle (1.45 ex);
\node at (4.22+2.38-0.64,-6.85) {\small{$0.5$}};
\draw[white,fill=white] (5.07+2.38-0.64,-5.98) circle (1.45 ex);
\node at (5.13+2.38-0.64,-5.88) {\small{$1.0$}};
\draw [white,fill=white] (6.85,-5.5) -- (7.05,-2.2)--(8.18,-2.2)--(8.18,-5.5)--(6.85,-5.5);
\node at (7.3,-5.1) {\small{$0.000$}};
\node at (7.3333,-4.3) {\small{$0.002$}};
\node at (7.3666,-3.5) {\small{$0.004$}};
\node at (7.4,-2.67) {\small{$0.006$}};
\node at (8.1,-3.9) [rotate=90] {\small{$f(x,y)$}};
\draw[white,fill=white] (10.80,-8.0) circle (1 ex);
\node at (10.8,-8.0) {\small{$x$}};
\draw[white,fill=white] (8.7+6.40,-7.4) circle (1 ex);
\node at (8.7+6.5,-7.3) {\small{$y$}};
\draw[white,fill=white] (8.7+0.64,-6.85) circle (1.45 ex);
\node at (8.7+0.60,-6.85) {\small{$0.0$}};
\draw[white,fill=white] (8.7+2.38,-7.5) circle (1.4 ex);
\node at (8.7+2.325,-7.45) {\small{$0.5$}};
\draw[white,fill=white] (8.7+2.45+2.38-0.64,-8.1) circle (1.45 ex);
\node at (8.7+2.45+2.38-0.64,-8.03) {\small{$1.0$}};
\draw[white,fill=white] (8.7+3.19+2.38-0.64,-8.0) circle (1.55 ex);
\node at (8.7+3.25+2.38-0.64,-7.90) {\small{$0.0$}};
\draw[white,fill=white] (8.7+4.16+2.38-0.64,-6.95) circle (1.45 ex);
\node at (8.7+4.22+2.38-0.64,-6.85) {\small{$0.5$}};
\draw[white,fill=white] (8.7+5.07+2.38-0.64,-5.98) circle (1.45 ex);
\node at (8.7+5.13+2.38-0.64,-5.88) {\small{$1.0$}};
\draw [white,fill=white] (8.7+6.85,-5.5) -- (8.7+7.05,-2.2)--(8.7+8.18,-2.2)--(8.7+8.18,-5.5)--(8.7+6.85,-5.5);
\node at (8.7+7.3,-5.1) {\small{$0.000$}};
\node at (8.7+7.3333,-4.3) {\small{$0.002$}};
\node at (8.7+7.3666,-3.5) {\small{$0.004$}};
\node at (8.7+7.4,-2.67) {\small{$0.006$}};
\node at (8.7+8.1,-3.9) [rotate=90] {\small{$f(x,y)$}};
\draw [white,fill=white] (4.025,-9.6) -- (8.7+4.19,-9.6)--(8.7+4.19,-9.95)--(4.025,-9.95)--(4.025,-9.6);
\node at (4.4,-9.9)  {\small{$-0.001$}};
\node at (5.7,-9.9)  {\small{$0.000$}};
\node at (6.9,-9.9)  {\small{$0.001$}};
\node at (8.1,-9.9)  {\small{$0.002$}};
\node at (9.3,-9.9)  {\small{$0.003$}};
\node at (10.5,-9.9)  {\small{$0.004$}};
\node at (11.7,-9.9)  {\small{$0.005$}};
\end{tikzpicture}
    \caption{Quantum circuit simulation demonstrating the FSL method being used to load a 2D function. Specifically, a two-dimensional sinc function was loaded into a quantum state of $20$ qubits (i.e., $n=10$) with infidelity of $5.0\times 10^{-4}$. In this simulation, $256$ Fourier coefficients (i.e., $m=3$) were loaded using the uniformly controlled rotations shown in Fig.~(\protect\hyperlink{fig:zgr_circuit_diagram}{1b}). This simulation was performed using the `statevector\textunderscore simulator' provided by Qiskit \cite{Qiskit}.}
    \label{fig:2d-statevector}
\end{figure}

Remarkably, the FSL method can be generalized to functions of more than one variable. The approach is the same: we first find a multi-variable Fourier series approximation of the target function, prepare a sparse state of Fourier coefficients using either the uniformly controlled rotation or the Schmidt-decomposition circuit, and then apply inverse QFT operators. Further details, including the circuit diagrams for loading functions of two variables, are presented in the Supplementary Sec.~(\ref{app-2d-fsl}). Here, we simply demonstrate the practicality of the FSL method by presenting the simulation results of loading a two-dimensional sinc function into a $20$ qubits state, $10$ qubits for each dimension. We approximated this two-dimensional sinc function with $256$ Fourier coefficients, which we loaded to $8$ qubits using the cascade of uniformly controlled rotations method for implementing $U_c$. With only $256$ Fourier modes, we were able to load the sinc function with infidelity of $5.0\times 10^{-4}$. The results of the simulation performed using Qiskit's `statevector\textunderscore simulator' are presented in Fig.~(\ref{fig:2d-statevector}). 

We now conclude this introduction to the mechanisms of the FSL method. In the next section, we present the experimental results from running the FSL method on the Quantinuum H$1$-$1$ and H$1$-$2$ quantum computers. Our results indicate that the FSL method is able to load functions with high fidelity even in the presence of noise in near-term quantum computers.

\section{Experimental Results} \label{sec:exp_results}

In the previous section, we demonstrated that the FSL method is capable of preparing function-encoding states with high fidelity in an ideal noiseless setting. However, present-day quantum computers are very noisy and exhibit complex errors when running highly structured quantum programs \cite{mirror_circuits}. To test how the FSL method performs in the presence of noise, we performed experiments on the Quantinuum H$1$-$1$ and H$1$-$2$ quantum computers which we accessed remotely via an Open QASM-based API \cite{quantinuum}. System Model H$1$ quantum computers have, on average, single-qubit gate infidelities of $4\times 10^{-5}$, two-qubit gate infidelities of $3\times 10^{-3}$, state preparation and measurement (SPAM) error of $3\times 10^{-3}$, and a measurement cross-talk error of $2\times 10^{-5}$. More impressively, System Model H$1$ quantum computers support full connectivity between qubits and allows for mid-circuit measurements, the latter of which we used extensively as we discuss below. More details about the specifications for the System Model H$1$ quantum computers, H1-1 and H1-2, are provided in the supplementary Sec.~(\ref{app-experiments}). 

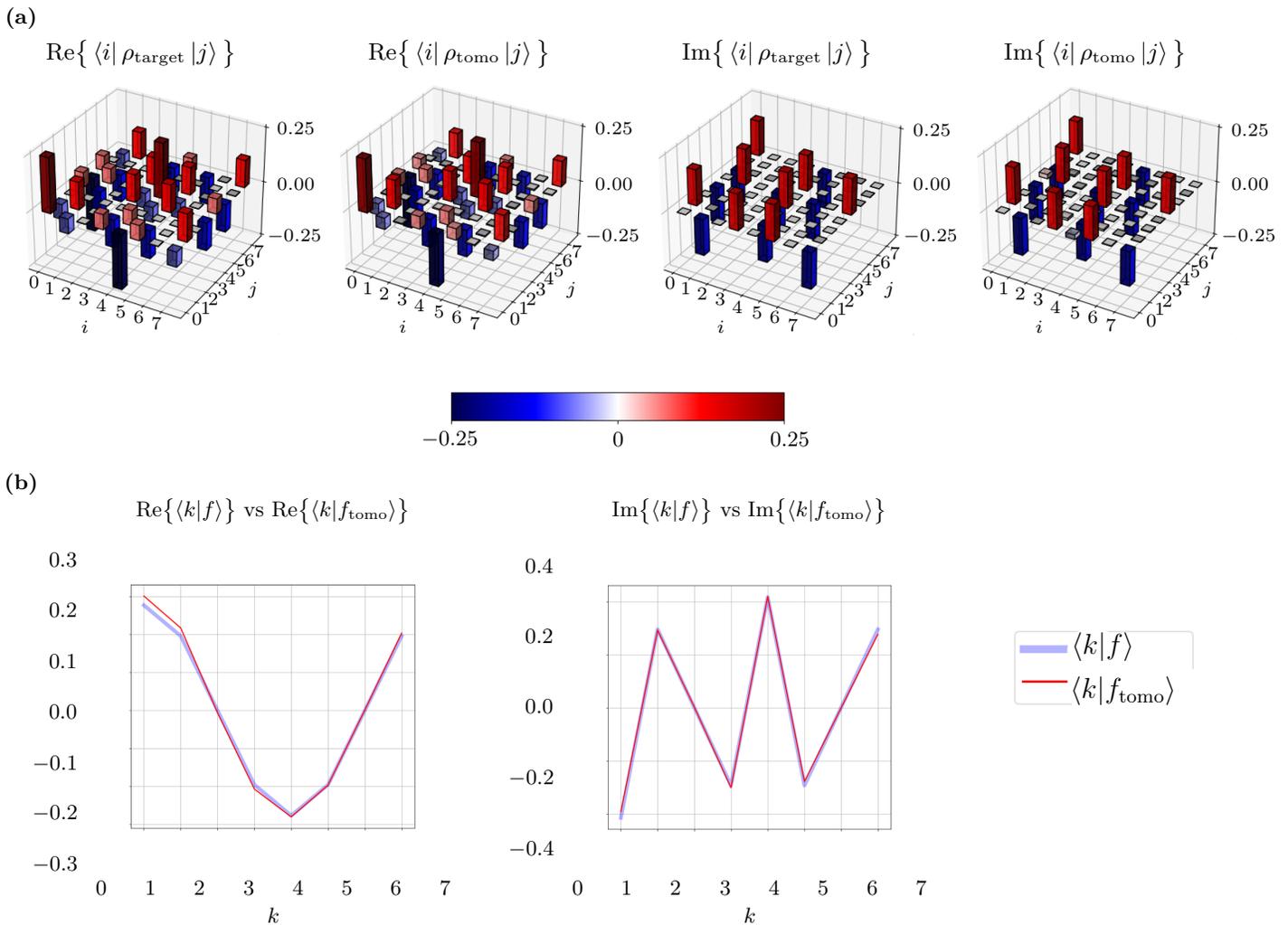
\begin{figure}
\hypertarget{fig:tomo-matrix}{}
    \hypertarget{fig:tomo-vector}{}
    \begin{tikzpicture}
    \node at (0,0) {\includegraphics[scale=1.1]{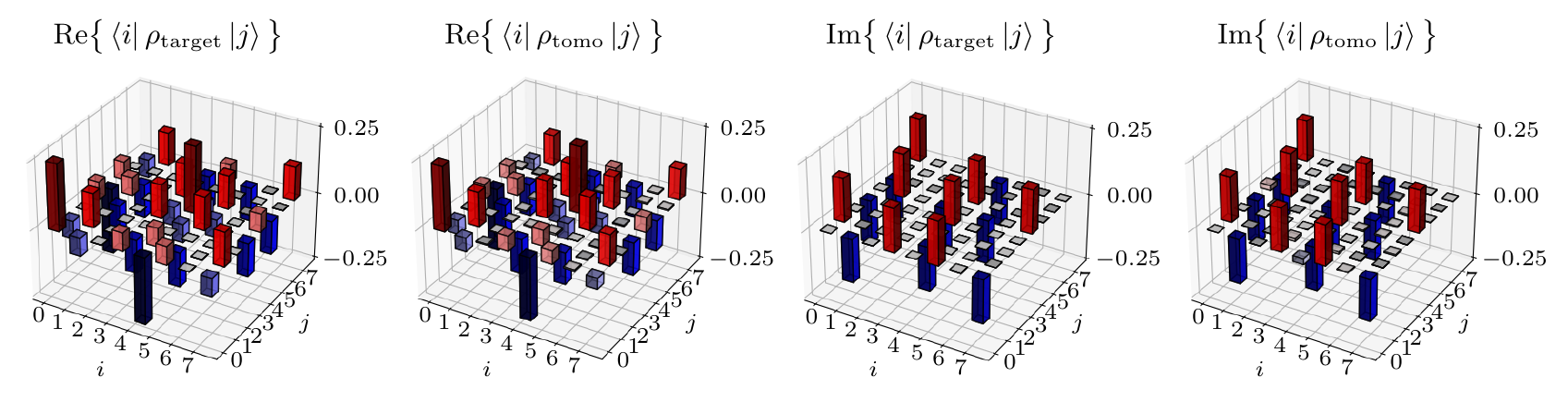}};
   \node at (-0.5,-3.35) {\includegraphics[scale=0.4]{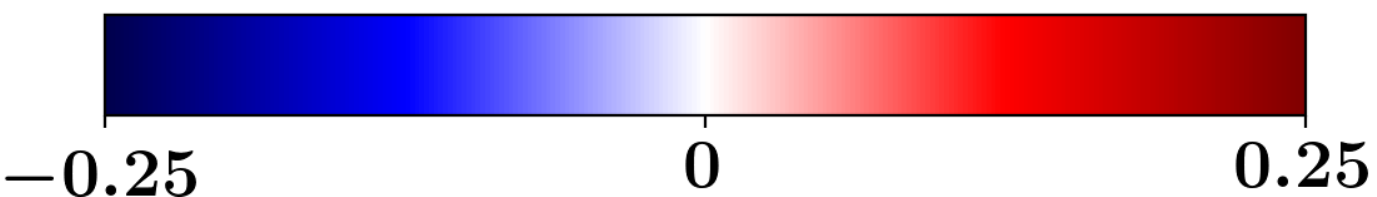}};
   \draw [white,fill=white] (-3.5,-3.5) -- (2.5,-3.5) -- (2.5,-4) -- (-3.5,-4) -- (-3.5,-3.5);
   \node at (-2.9,-3.7) {\small{$-0.25$}};
   \node at (-0.425,-3.7) {\small{$0$}};
   \node at (2.1,-3.7) {\small{$0.25$}};
   \node at (-9.2,2.5) {\textbf{(a)}};
   \node at (-9.2,-4.35) {\textbf{(b)}};
   \node at (-5.50,-4.75) {$\text{Re}\big\{ \langle k | f\rangle\big\}$ vs $\text{Re}\big\{ \langle k | f_{\text{tomo}}\rangle\big\}$ };
   \node at (7-5.50,-4.75) {$\text{Im}\big\{ \langle k | f\rangle\big\}$ vs $\text{Im}\big\{ \langle k | f_{\text{tomo}}\rangle\big\}$ };
    \node at (-5.5,-7.65) {\includegraphics[scale=0.3]{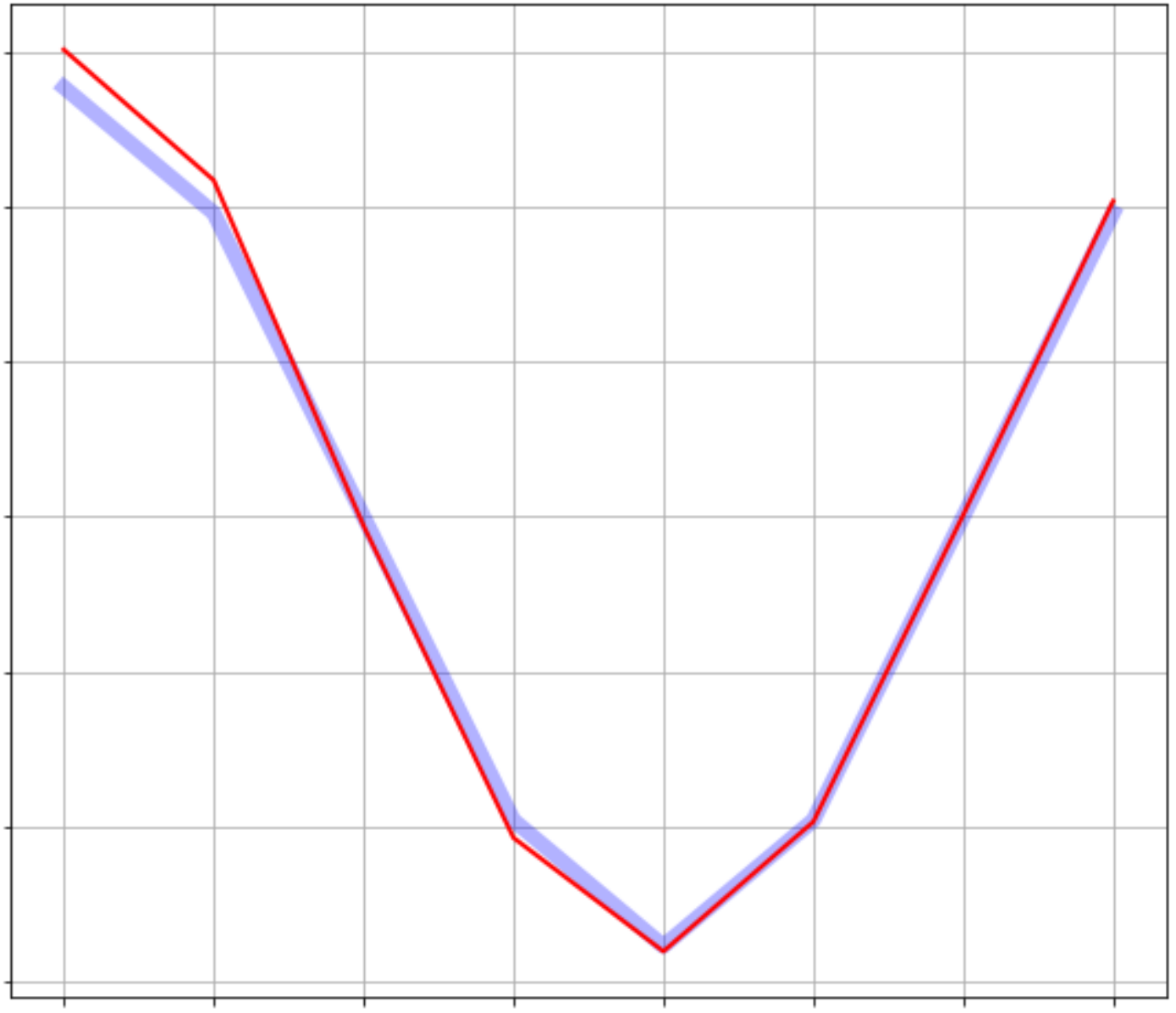}};
    \node at (1.5,-7.65) {\includegraphics[scale=0.3]{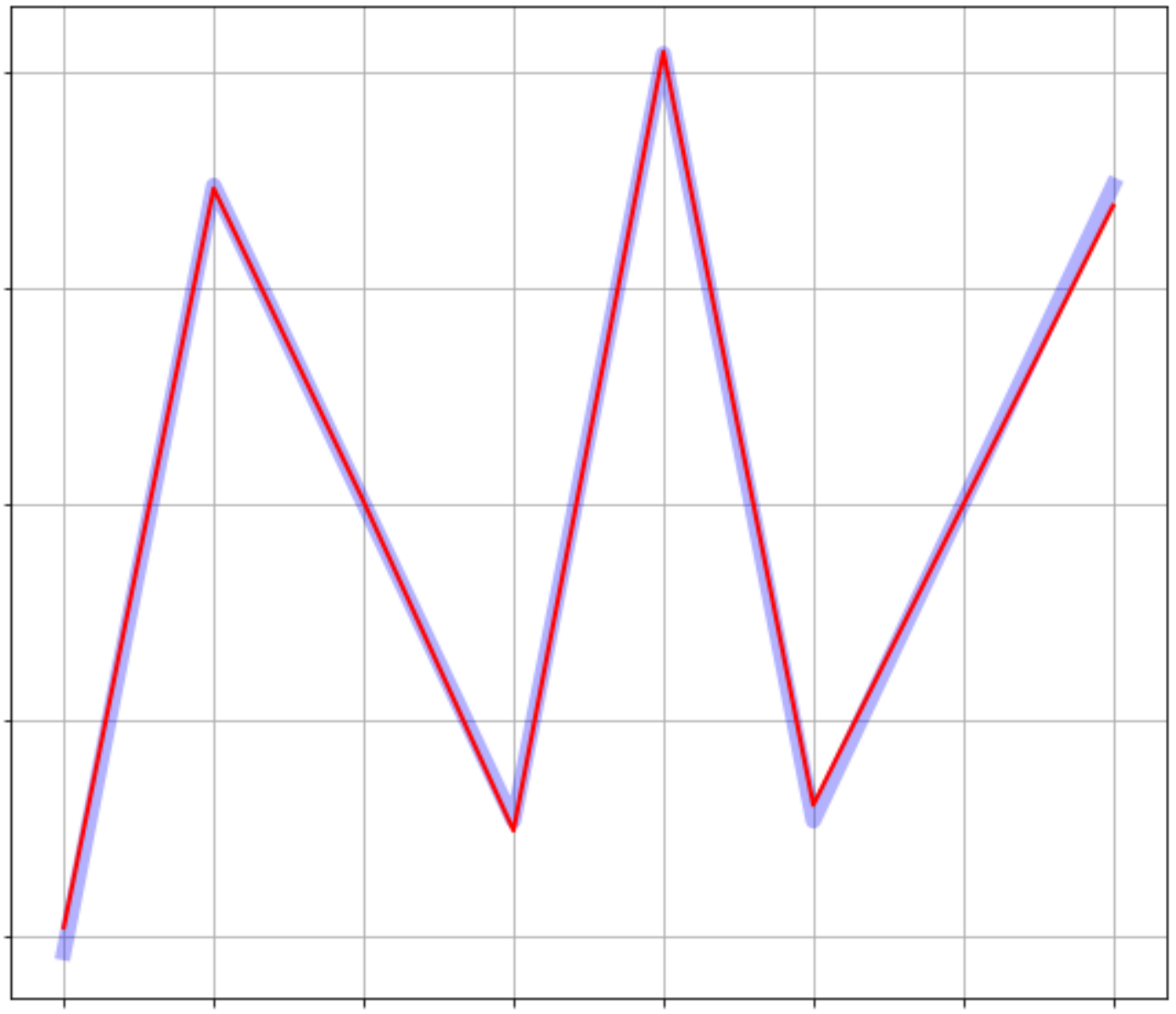}};
    \node at (0.1+6.6,-7.05) {\includegraphics[scale=0.6]{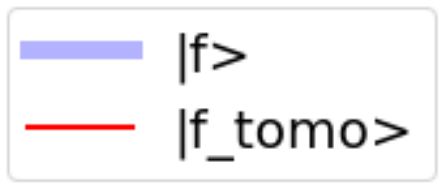}};
    \draw[white,fill=white] (0.1+6.5,-6.8) circle (2.33 ex);
    \draw[white,fill=white] (0.1+6.4,-6.8) circle (2.33 ex);
    \node at (0.2+6.50,-6.75) {\large{$\langle{k}|{f}\rangle$}};
    \draw[white,fill=white] (0.1+6.4,-7.4) circle (2.3 ex);
    \draw[white,fill=white] (0.1+6.5,-7.4) circle (2.3 ex);
    \draw[white,fill=white] (0.1+6.7,-7.4) circle (2.3 ex);
    \draw[white,fill=white] (0.1+6.9,-7.4) circle (2.3 ex);
    \draw[white,fill=white] (0.1+7.1,-7.4) circle (2.3 ex);
    \draw[white,fill=white] (0.1+7.3,-7.4) circle (2.3 ex);
    \draw[white,fill=white] (0.1+7.5,-7.4) circle (2.3 ex);
    \draw[white,fill=white] (0.1+7.7,-7.4) circle (2.3 ex);
    \draw[white,fill=white] (0.1+7.9,-7.4) circle (2.3 ex);
    \node at (0.2+6.80,-7.4) {\large{$\langle k|{f_{\text{tomo}}}\rangle$}};
    \node at (-8.02,-10.3) {\small{$0$}};
    \node at (-7.3,-10.3) {\small{$1$}};
    \node at (-6.58,-10.3) {\small{$2$}};
    \node at (-5.85,-10.3) {\small{$3$}};
    \node at (-5.49,-10.7) {\small{$k$}};
    \node at (-5.13,-10.3) {\small{$4$}};
    \node at (-4.41,-10.3) {\small{$5$}};
    \node at (-3.69,-10.3) {\small{$6$}};
    \node at (-2.97,-10.3) {\small{$7$}};
    \node at (7-8.02,-10.3) {\small{$0$}};
    \node at (7-7.3,-10.3) {\small{$1$}};
    \node at (7-6.58,-10.3) {\small{$2$}};
    \node at (7-5.85,-10.3) {\small{$3$}};
    \node at (7-5.49,-10.7) {\small{$k$}};
    \node at (7-5.13,-10.3) {\small{$4$}};
    \node at (7-4.41,-10.3) {\small{$5$}};
    \node at (7-3.69,-10.3) {\small{$6$}};
    \node at (7-2.97,-10.3) {\small{$7$}};
    \node at (-8.7,-9.95) {\small{$-0.3$}};
    \node at (-8.7,-9.20) {\small{$-0.2$}};
    \node at (-8.7,-9.22+0.75) {\small{$-0.1$}};
    \node at (-8.585,-9.20+1.5) {\small{$0.0$}};
    \node at (-8.585,-9.21+2.25) {\small{$0.1$}};
    \node at (-8.585,-9.21+3.0) {\small{$0.2$}};
    \node at (-8.585,-9.22+3.75) {\small{$0.3$}};
    \node at (7-8.7,-9.73) {\small{$-0.4$}};
    \node at (7-8.7,-8.68) {\small{$-0.2$}};
    \node at (7-8.585,-7.63) {\small{$0.0$}};
    \node at (7-8.585,-6.59) {\small{$0.2$}};
    \node at (7-8.585,-5.56) {\small{$0.4$}};
    \draw [white,fill=white] (-4.65,-2.1350) -- (0.5,-2.135) -- (0.5,-2.5) -- (-4.65,-2.5) -- (-4.65,-2.135);
    \draw [white,fill=white] (-0.9,-2.1350) -- (-0.7,-2.135) -- (-0.7,-0.750) -- (-0.9,-0.750) -- (-0.9,-2.135);
    \draw [white,fill=white] (-0.9,1.10) -- (-0.7,1.10) -- (-0.7,1.650) -- (-0.9,1.650) -- (-0.9,1.10);
    \draw [white,fill=white] (-0.9-4.7,-2.1350) -- (-0.7-4.7,-2.135) -- (-0.7-4.7,-0.750) -- (-0.9-4.7,-0.750) -- (-0.9-4.7,-2.135);
    \draw [white,fill=white] (-0.9-4.7,1.10) -- (-0.7-4.7,1.10) -- (-0.7-4.7,1.650) -- (-0.9-4.7,1.650) -- (-0.9-4.7,1.10);
    \draw [white,fill=white] (-0.9+4.7,-2.1350) -- (-0.7+4.7,-2.135) -- (-0.7+4.7,-0.750) -- (-0.9+4.7,-0.750) -- (-0.9+4.7,-2.135);
    \draw [white,fill=white] (-0.9+4.7,1.10) -- (-0.7+4.7,1.10) -- (-0.7+4.7,1.650) -- (-0.9+4.7,1.650) -- (-0.9+4.7,1.10);
    \draw [white,fill=white] (-0.9+4.7+4.7,-2.1350) -- (-0.7+4.7+4.7,-2.135) -- (-0.7+4.7+4.7,-0.750) -- (-0.9+4.7+4.7,-0.750) -- (-0.9+4.7+4.7,-2.135);
    \draw [white,fill=white] (-0.9+4.7+4.7,1.10) -- (-0.7+4.7+4.7,1.10) -- (-0.7+4.7+4.7,1.650) -- (-0.9+4.7+4.7,1.650) -- (-0.9+4.7+4.7,1.10);
    \end{tikzpicture}
    \caption{Experimental results of loading a complex-valued function, $f(x) \, = \, \left(\cos(2\pi x) - 1.5 i \cos(6\pi x)\right)/\sqrt{13} \, $, into a three qubit state on the Quantinuum H$1$-$1$ quantum computer using the quantum circuit shown in Fig.~(\ref{fig:tomo_circuit}). \textbf{(a)} The loaded quantum state was reconstructed through quantum state tomography based on maximum likelihood estimation, and has a fidelity of $95.5\%$ with the target state and a purity of $0.918$. \textbf{(b)} Comparison of the elements of the target function state with the eigenvector of $\rho_{\text{tomo}}$ corresponding to the dominant eigenvalue. As evident from these plots, the reconstructed state, $\rho_{\text{tomo}}$, is in good agreement with the target state, $\rho_{\text{target}} \, = \, |f\rangle\langle f|$. Hence, the FSL method can accurately load both the phases and the amplitudes of a complex-valued target state.}
    \label{fig:3qubit_tomography}
\end{figure}

For our first experiment, we considered  a complex-valued function $f(x) \, = \, \left(\cos(2\pi x) - 1.5 i \cos(6\pi x)\right)/\sqrt{13}$. We loaded this function into a quantum state of three qubits using the quantum circuit shown in Fig.~(\ref{fig:tomo_circuit}) in the supplementary Sec.~(\ref{app-experiments}). To check how close the prepared state is to the target state, we performed the state tomography where we measured each qubit in the eigenbasis of Pauli-$Z$, Pauli-$X$, and Pauli-$Y$ operators by applying $\{I, H, R_{x}(\pi/2)\}$ on each qubit before measuring it. From the results of $800$ measurements for each of the $27$ experiments, we estimated the probabilities of projection, $P_{\mu_{1},\mu_{2},\mu_{3}}$, onto states $\ket{\psi_{\mu_{1},\mu_{2},\mu_{3}}} \, \equiv \, \ket{\psi_{\mu_{1}}}\otimes \ket{\psi_{\mu_{2}}}\otimes \ket{\psi_{\mu_{3}}}\, $, where $\mu_i \in \{0,1,2,3\}$ and $\ket{\psi_{0}} = \ket{0} \, $, $\ket{\psi_{1}} = \ket{1} \, $, $\ket{\psi_{2}} = \left(\ket{0}+\ket{1}\right)/\sqrt{2} \, $, and $\ket{\psi_{3}} = \left(\ket{0}+i\ket{1}\right)/\sqrt{2} \, $. We assumed the density matrix to be of the form $\rho(T) \, = \, \left(T^{\dagger}T\right)/\text{tr}\left(T^{\dagger}T\right) \, $, where $T$ is a $8\times 8$ lower triangular matrix with real-valued diagonal elements. The matrix $\rho(T)$ which best fits the measured probabilities of projection onto states $\ket{\psi_{\mu_{1},\mu_{2},\mu_{3}}}$ was determined by minimizing the cost function, \cite{James2001MeasurementOQ, 2021NatNa..16..965T}
\begin{align}
    C(T) \, = \, \sum_{\mu_{1}=0}^{3}\sum_{\mu_{2}=0}^{3}\sum_{\mu_{3}=0}^{3} \, \frac{ \Big( \bra{\psi_{\mu_{1},\mu_{2},\mu_{3}}} \rho(T) \ket{\psi_{\mu_{1},\mu_{2},\mu_{3}}} \, - \, P_{\mu_{1},\mu_{2},\mu_{3}} \Big)^{2} }{\bra{\psi_{\mu_{1},\mu_{2},\mu_{3}}} \rho(T) \ket{\psi_{\mu_{1},\mu_{2},\mu_{3}}}} ,
\end{align}
with respect to $T$. The reconstructed state, $\rho_{\text{tomo}}$ through this tomography method has a fidelity of $95.5\%$ with the target state, $\rho_{\text{target}} = |f\rangle \langle f|$, and a purity of $\text{tr}(\rho_{\text{tomo}}^2) = 0.918$. The matrix elements of the target density matrix, $\rho_{\text{target}}$, and the measured density matrix, $\rho_{\text{tomo}}$, are shown in Fig.~(\hyperlink{fig:tomo-matrix}{4a}). As evident from this figure, the matrix elements of $\rho_{\text{target}}$ are in good agreement with those of $\rho_{\text{tomo}}$. Moreover, the comparison of the target function with the `measured' function is shown in Fig.~(\hyperlink{fig:tomo-vector}{4b}). Since the measured density matrix $\rho_{\text{tomo}}$ is not a pure state, the notion of the measured function is not uniquely defined. For concreteness, we defined the measured function such that its values at $8$ discretized points are the elements of the state $\ket{f_{\text{tomo}}}$ in the computational basis, where $\ket{f_{\text{tomo}}}$ is the eigenvector of $\rho_{\text{tomo}}$ corresponding to the dominant eigenvalue \footnote{Of course, $\ket{f_{\text{tomo}}}$ is only defined up to a global phase. We fix this phase by demanding that $\langle f|f_{\text{tomo}}\rangle$ is real and positive.}.   

The above approach, based on the maximum likelihood estimation, is especially useful as it restricts the resulting density matrix, $\rho(T)$, to being Hermitian, positive, and normalized. In the supplementary Sec.~(\ref{app-experiments}), we discuss an alternate `direct reconstruction' approach where we measured the density matrix without imposing any constraints. When using the `direct reconstruction' of the density matrix, we calculated a fidelity of  around $94.0\%$ between the measured and the target state. However, as is usually the case with such unconstrained approaches \cite{James2001MeasurementOQ}, the reconstructed matrix has negative eigenvalues, and hence, is not a valid density matrix. 

We now discuss experiments we performed to test the FSL method's potential for loading functions into a quantum state of five or more qubits. Note that due to the high cost of running a quantum computer, performing similar quantum state tomography experiments to verify the accuracy of the FSL method is not a viable option for more than three qubits. Even though there are economical algorithms to  perform state tomography using randomized measurements \cite{Huang2020PredictingMP,Elben2022TheRM} or machine learning techniques \cite{2018NatPh..14..447T,Neugebauer2020NeuralnetworkQS,Cha2021AttentionbasedQT,Zuo2021AllopticalNN}, they are beyond the scope of this work. 

To circumvent the issue of the verification of the prepared quantum state, we simplified our task in the rest of the experiments and only verified the amplitudes of the prepared quantum state by performing measurements in the computational basis. In particular, we first loaded $\sqrt{f(x)}$ for a given function $f(x)$ into a quantum state $\ket{\sqrt{f}} = \sum_{k=0}^{2^{n}-1} \sqrt{f_{k}} \ket{k} $ using the FSL method and then estimated the measurement probabilities, $|\langle k | \sqrt{f}\rangle|^{2}$, by measuring all the qubits in the computational basis sufficiently many times. This allowed us to compare the measurement probabilities with the value of the target function $f(x)$ at $x = k/2^{n}$. It is worth mentioning that even though we are unable to verify the accuracy of the local phases in the quantum state prepared using the FSL method, we still expect the experiments described above to test the performance of the FSL method on a present-day noisy quantum computer. Note that we already have enough evidence that the FSL method is, in principle, capable of loading correct amplitudes and local phases based on the simulation results in Fig.~(\ref{fig:statevectors}) and Fig.~(\ref{fig:2d-statevector}) and the proof of the FSL method provided in the supplementary Sec.~(\ref{app-2d-fsl}). Therefore, the only thing that needs to be checked is if the quantum state prepared using the FSL circuit can survive the hardware noise present in noisy quantum computers. For example, if the depth of the FSL quantum circuit, despite being $O(n)$, is not small enough, the resultant noisy state will be far from the target state, in which case even the measured amplitudes will be far from the actual amplitudes. Thus, measuring amplitudes of the prepared state and comparing them with those of the target state provides a nice alternate to more expensive state tomography to check the practicality of the FSL method for present-day noisy quantum computers.

Before we proceed, let us also point out that since we are only comparing the amplitudes of the prepared state to the target function, it seems more reasonable to quantify the accuracy of the FSL method in terms of the classical fidelity between the measurement probabilities (in the computational basis) and the target function. Recall that the classical fidelity between probability distributions $p$ and $q$ is defined as $F(p,q) \, = \, \left(\sum_{i} \sqrt{p_{i}}\sqrt{q_{i}} \, \right)^{2} \, $.

For our first experiment involving five or more qubits, we considered a bi-modal Gaussian function given by a superposition of two Gaussian functions; see Eq.~\eqref{eq-1d-bimodal} in the supplementary Sec.~(\ref{app-experiments}). We loaded this function into a $5$-qubit state and ran $5000$ shots of measurements. The measurement probabilities are in good agreement with the target function as shown in Fig.~(\hyperlink{fig:bimodal_exp}{5a}).  In fact, we obtained a classical fidelity of $99.8\%$ between the measured probabilities and the target function. In this experiment, we used the Schmidt-decomposition circuit shown in Fig.~(\hyperlink{fig:svd_circuit}{1c}) to load the Fourier coefficients. Further details about this circuit can be found in the supplementary Sec.~(\ref{app-experiments}).

\begin{figure}
\begin{tikzpicture}
\hypertarget{fig:bimodal_exp}{}
\node at (7.25,-1.85) {\includegraphics[scale=0.25]
{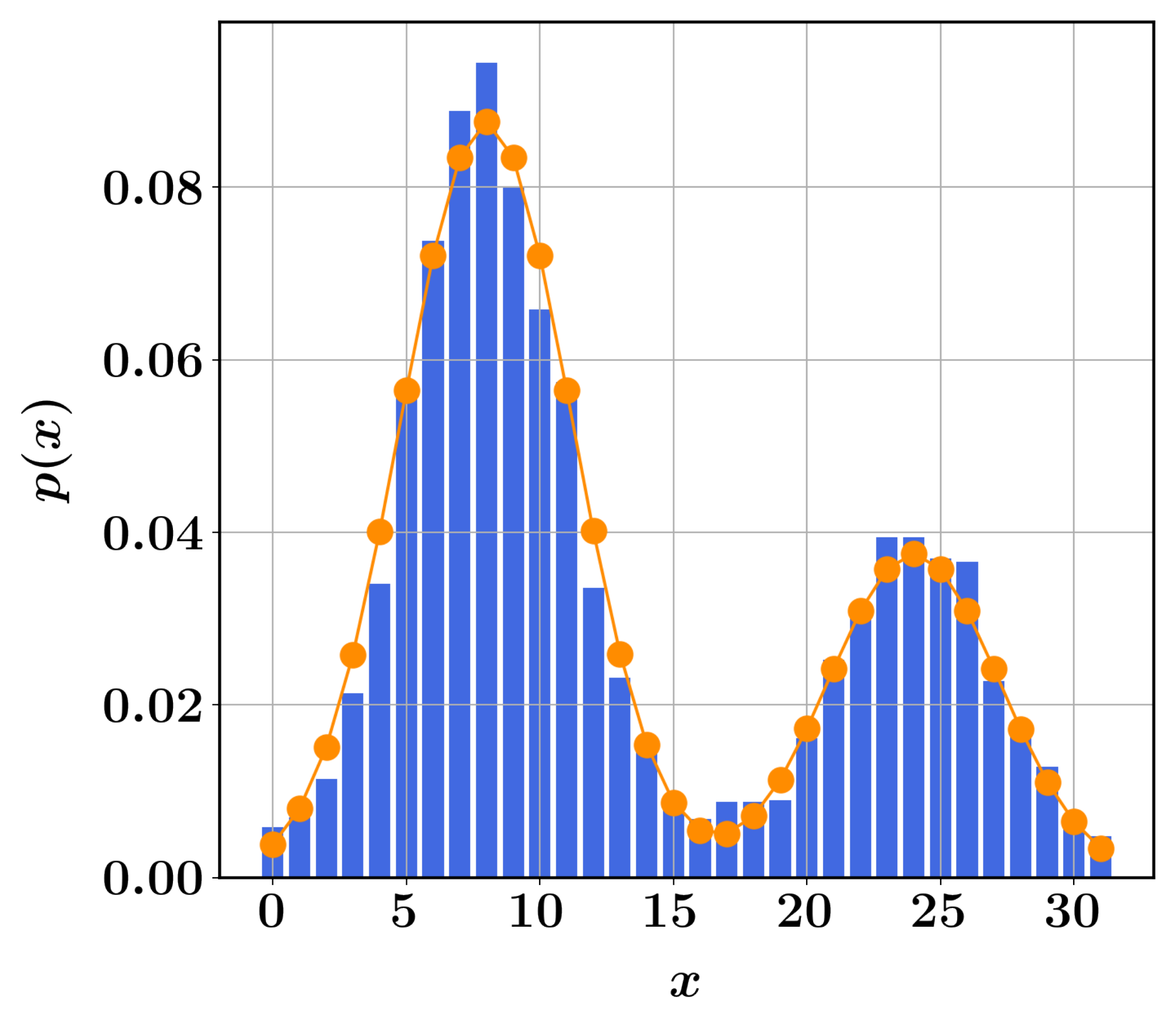}};
\node at (1.30,-1.85) {\includegraphics[scale=0.25]{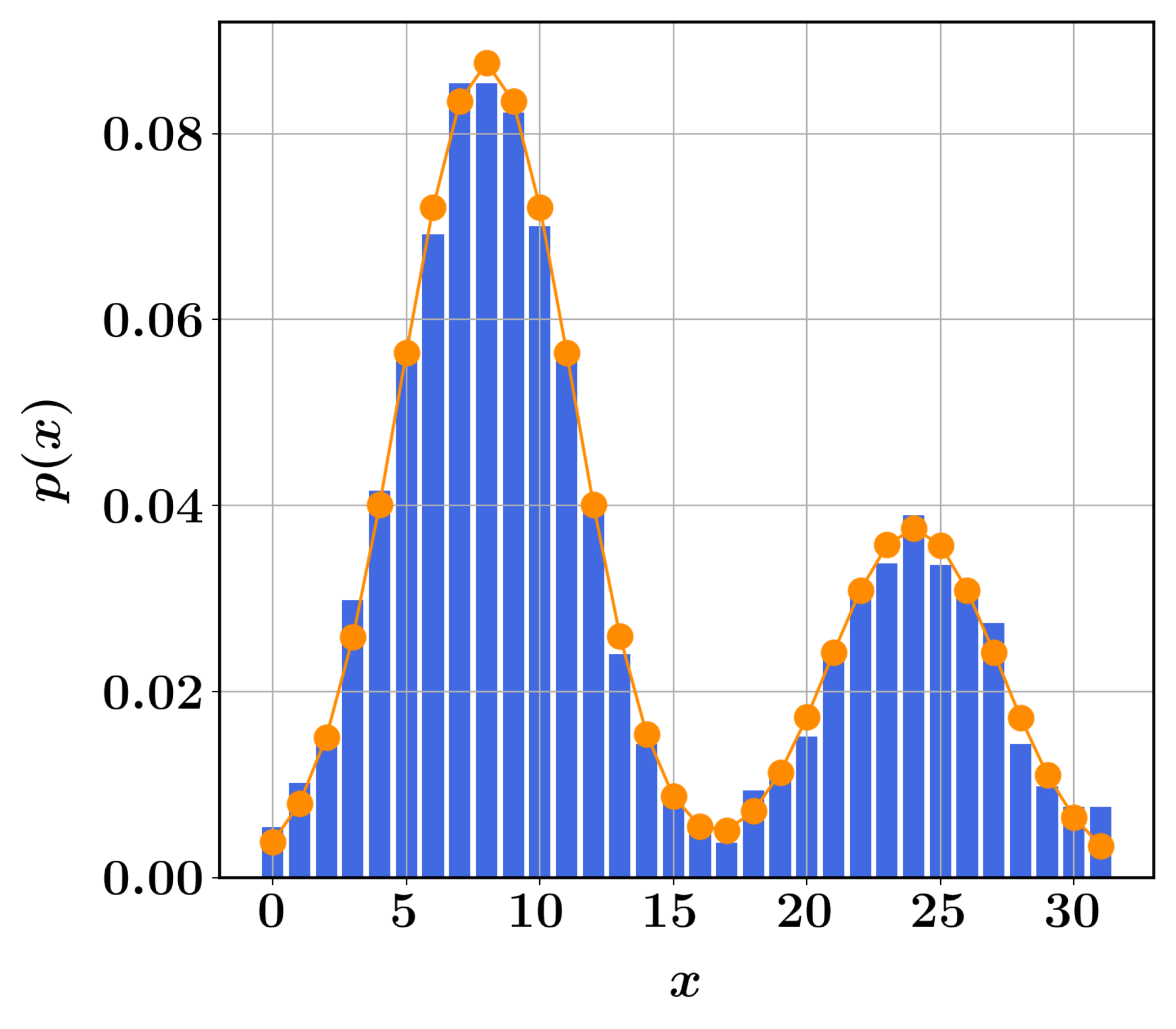}};
\node at (12.25,-0.1) {\includegraphics[scale=0.35]{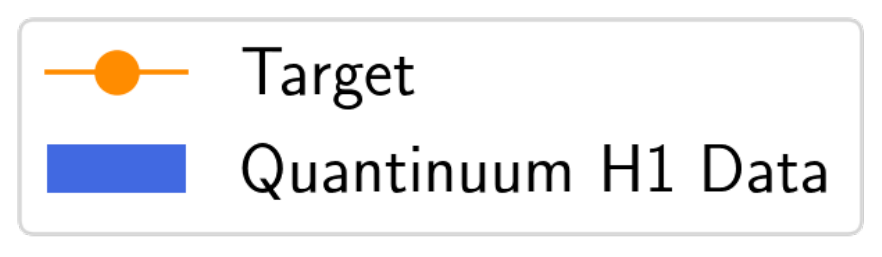}};
\node at (0,-7.1) {\includegraphics[scale=0.25]{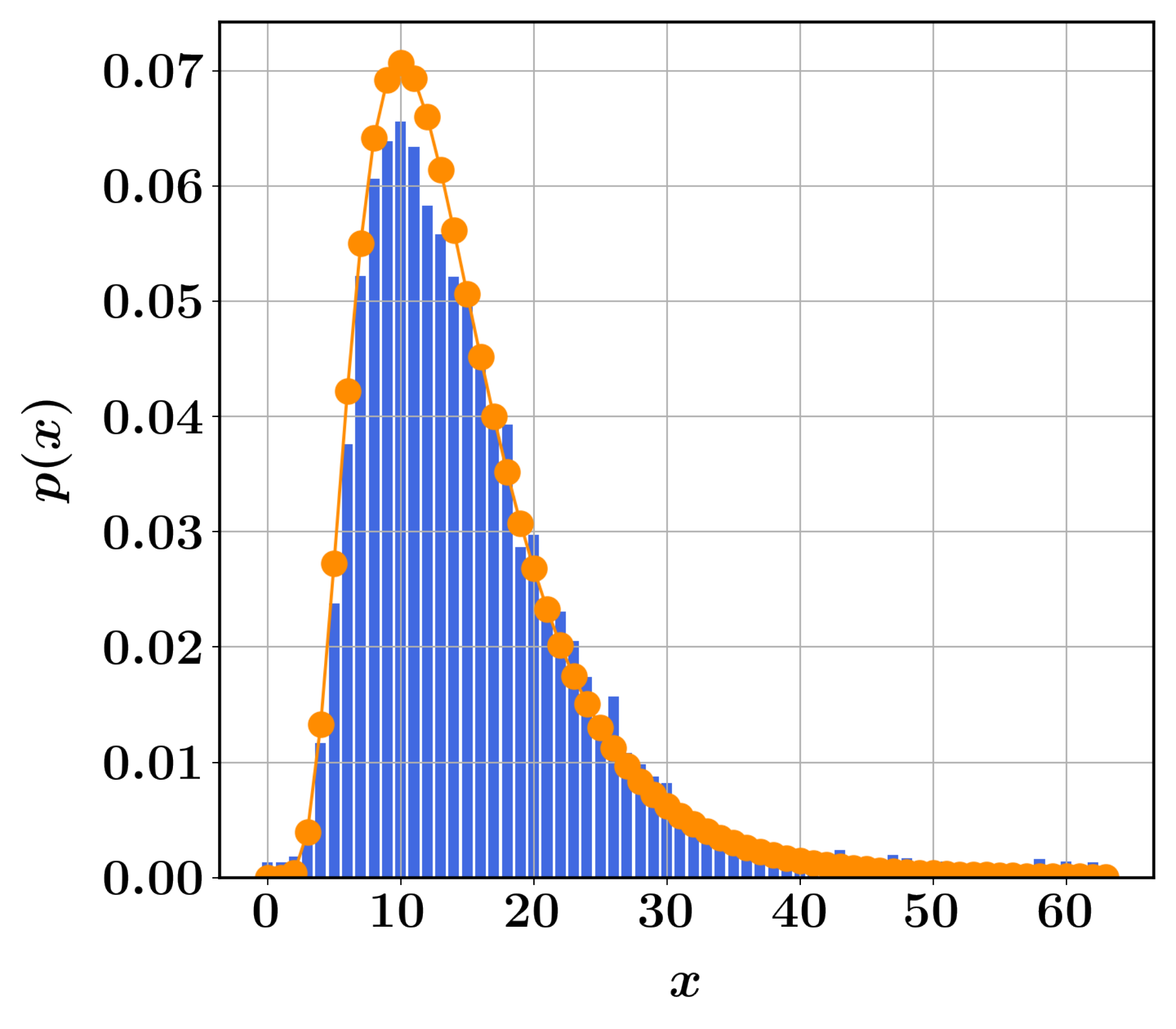}};
\node at (5.75,-7.10) {\includegraphics[scale=0.25]{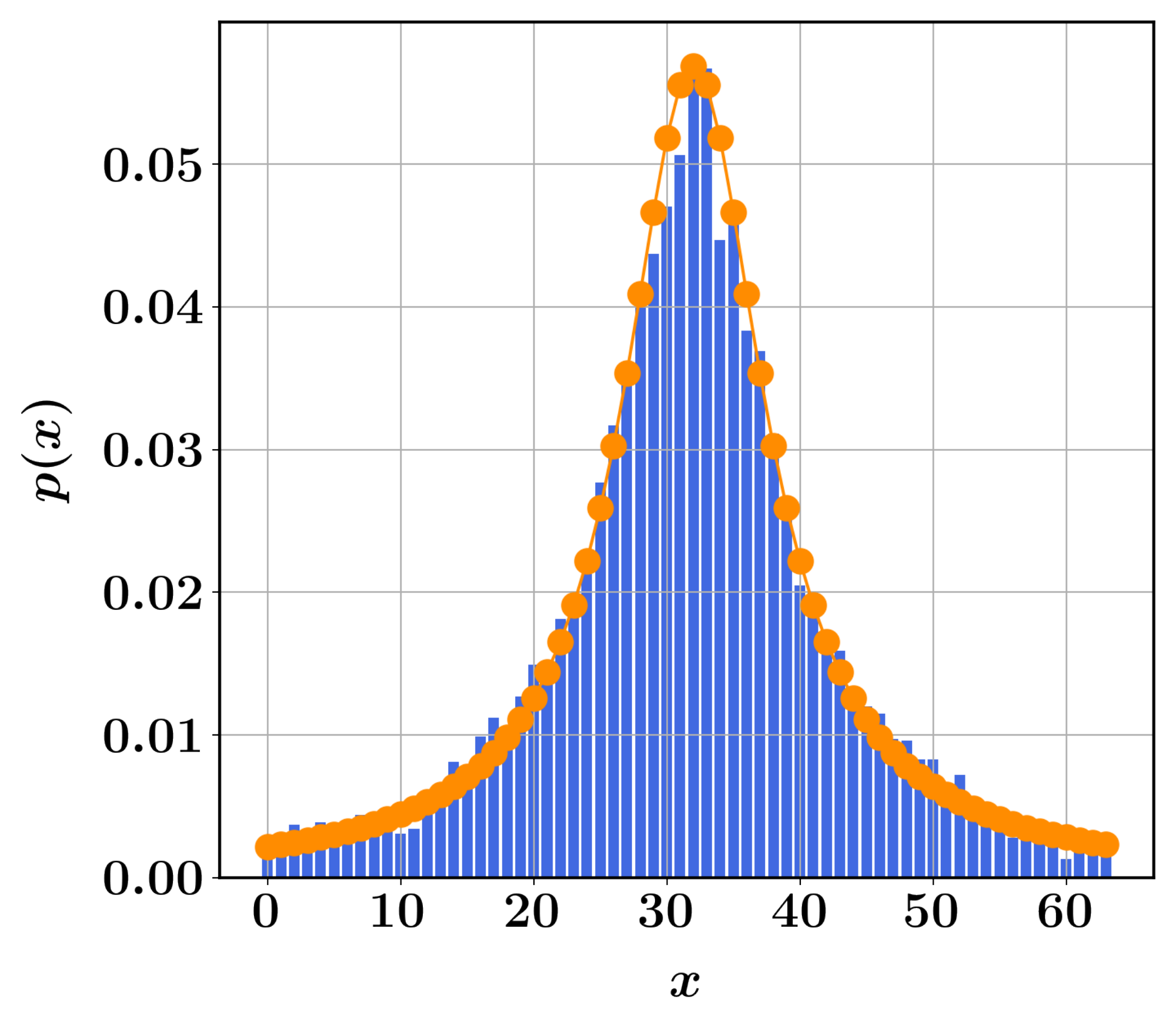}};
\node at (11.35,-7.1) {\includegraphics[scale=0.25]{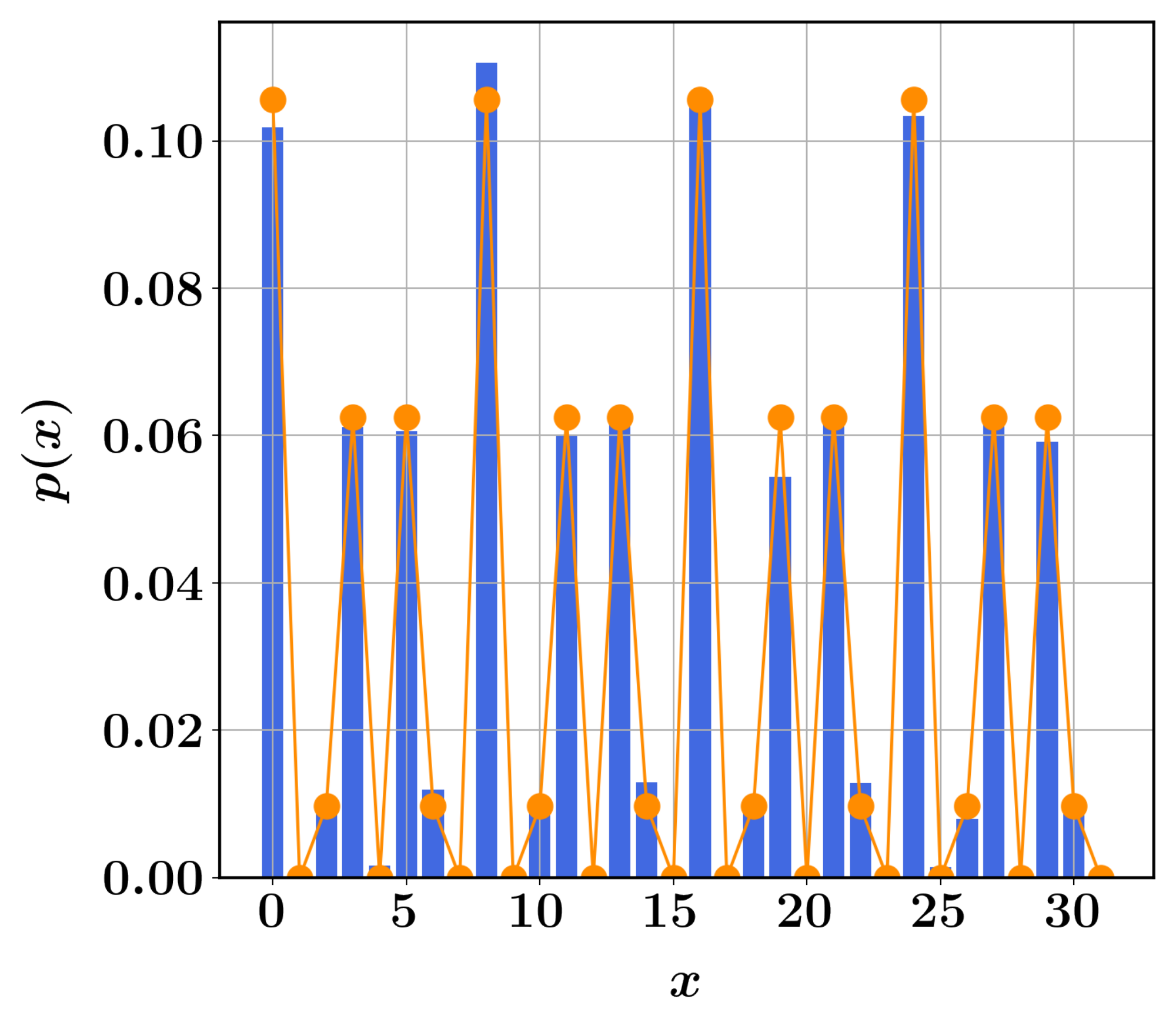}};
\node at (0.5,1.15) {\textbf{(a)} \underline{Bi-modal Gaussian function on $5$ qubits}};
\node at (7.75,0.55) {\small{Classically-controlled QFT}};
\node at (1.75,0.55) {\small{Textbook QFT}};
\draw[white,fill=white] (1.7,-3.9) circle (1 ex);
\node at (1.7,-3.95) {\small{$x$}};
\draw[white,fill=white] (-0.11,-3.60) circle (1 ex);
\node at (-0.06,-3.60) {\small{$0$}};
\draw[white,fill=white] (0.48,-3.60) circle (1 ex);
\node at (0.52,-3.60) {\small{$5$}};
\draw[white,fill=white] (1.07,-3.60) circle (1 ex);
\node at (1.07,-3.60) {\small{$10$}};
\draw[white,fill=white] (1.66,-3.60) circle (1 ex);
\node at (1.66,-3.60) {\small{$15$}};
\draw[white,fill=white] (2.25,-3.60) circle (1 ex);
\node at (2.25,-3.60) {\small{$20$}};
\draw[white,fill=white] (2.84,-3.60) circle (1 ex);
\node at (2.84,-3.60) {\small{$25$}};
\draw[white,fill=white] (3.43,-3.60) circle (1 ex);
\node at (3.43,-3.60) {\small{$30$}};
\draw[white,fill=white] (-0.95,-1.55) circle (2 ex);
\node at (-1.25,-1.45) [rotate=90] {\small  {$f(x)$}};
\draw[white,fill=white] (-0.6,-3.40) circle (2 ex);
\node at (-0.645,-3.40) {\small  {$0.00$}};
\draw[white,fill=white] (-0.6,-2.70) circle (2 ex);
\node at (-0.645,-2.60) {\small  {$0.02$}};
\draw[white,fill=white] (-0.6,-1.9) circle (2 ex);
\node at (-0.645,-1.80) {\small  {$0.04$}};
\draw[white,fill=white] (-0.6,-1.10) circle (2 ex);
\node at (-0.645,-1.00) {\small  {$0.06$}};
\draw[white,fill=white] (-0.6,-0.20) circle (2 ex);
\node at (-0.645,-0.20) {\small  {$0.08$}};
\draw[white,fill=white] (7.65,-3.9) circle (1 ex);
\node at (7.65,-3.95) {\small{$x$}};
\draw[white,fill=white] (5.84,-3.60) circle (1 ex);
\node at (5.89,-3.60) {\small{$0$}};
\draw[white,fill=white] (6.43,-3.60) circle (1 ex);
\node at (6.47,-3.60) {\small{$5$}};
\draw[white,fill=white] (7.02,-3.60) circle (1 ex);
\node at (7.02,-3.60) {\small{$10$}};
\draw[white,fill=white] (7.61,-3.60) circle (1 ex);
\node at (7.61,-3.60) {\small{$15$}};
\draw[white,fill=white] (8.2,-3.60) circle (1 ex);
\node at (8.2,-3.60) {\small{$20$}};
\draw[white,fill=white] (8.79,-3.60) circle (1 ex);
\node at (8.79,-3.60) {\small{$25$}};
\draw[white,fill=white] (9.38,-3.60) circle (1 ex);
\node at (9.38,-3.60) {\small{$30$}};
\draw[white,fill=white] (4.95,-1.55) circle (2 ex);
\node at (4.65,-1.5) [rotate=90] {\small  {$f(x)$}};
\draw[white,fill=white] (5.35,-3.40) circle (2 ex);
\node at (5.3,-3.40) {\small  {$0.00$}};
\draw[white,fill=white] (5.35,-2.70) circle (2 ex);
\node at (5.3,-2.65) {\small  {$0.02$}};
\draw[white,fill=white] (5.35,-1.9) circle (2 ex);
\node at (5.3,-1.925) {\small  {$0.04$}};
\draw[white,fill=white] (5.35,-1.10) circle (2 ex);
\node at (5.3,-1.2) {\small  {$0.06$}};
\draw[white,fill=white] (5.35,-0.35) circle (2 ex);
\node at (5.3,-0.45) {\small  {$0.08$}};
\draw[white,fill=white] (0.4,-9.15) circle (1 ex);
\node at (0.4,-9.2) {\small{$x$}};
\draw[white,fill=white] (-1.37,-8.85) circle (1.1 ex);
\node at (-1.38,-8.85) {\small{$0$}};
\draw[white,fill=white] (-0.8,-8.85) circle (1.1 ex);
\node at (-0.8,-8.85) {\small{$10$}};
\draw[white,fill=white] (-0.225,-8.85) circle (1.1 ex);
\node at (-0.225,-8.85) {\small{$20$}};
\draw[white,fill=white] (0.36,-8.85) circle (1 ex);
\node at (0.355,-8.85) {\small{$30$}};
\draw[white,fill=white] (0.95,-8.85) circle (1.1 ex);
\node at (0.925,-8.85) {\small{$40$}};
\draw[white,fill=white] (1.475,-8.85) circle (1.1 ex);
\node at (1.5,-8.85) {\small{$50$}};
\draw[white,fill=white] (2.08,-8.85) circle (1.1 ex);
\node at (2.09,-8.85) {\small{$60$}};
\draw[white,fill=white] (-2.25,-6.80) circle (2 ex);
\node at (-2.55,-6.95) [rotate=90] {\small  {$f(x)$}};
\draw[white,fill=white] (-2.15,-9.0) -- (-2.15,-5.0)-- (-1.65,-5.0) -- (-1.65,-9.0)--(-2.15,-9.0);
\node at (-1.95,-8.65) {\small  {$0.00$}};
\node at (-1.95,-8.175) {\small  {$0.01$}};
\node at (-1.95,-7.675) {\small  {$0.02$}};
\node at (-1.95,-7.175) {\small  {$0.03$}};
\node at (-1.95,-6.675) {\small  {$0.04$}};
\node at (-1.95,-6.175) {\small  {$0.05$}};
\node at (-1.95,-5.675) {\small  {$0.06$}};
\node at (-1.95,-5.175) {\small  {$0.07$}};
\draw[white,fill=white] (11.75,-9.15) circle (1 ex);
\node at (11.75,-9.2) {\small{$x$}};
\draw[white,fill=white] (9.98,-8.85) circle (1.1 ex);
\node at (9.99,-8.85) {\small{$0$}};
\draw[white,fill=white] (10.55,-8.85) circle (1.1 ex);
\node at (10.57,-8.85) {\small{$5$}};
\draw[white,fill=white] (11.125,-8.85) circle (1.1 ex);
\node at (11.135,-8.85) {\small{$10$}};
\draw[white,fill=white] (11.71,-8.85) circle (1 ex);
\node at (11.75,-8.85) {\small{$15$}};
\draw[white,fill=white] (12.3,-8.85) circle (1.1 ex);
\node at (12.2825,-8.85) {\small{$20$}};
\draw[white,fill=white] (12.825,-8.85) circle (1.1 ex);
\node at (12.9,-8.85) {\small{$25$}};
\draw[white,fill=white] (13.43,-8.85) circle (1.1 ex);
\node at (13.478,-8.85) {\small{$30$}};
\draw[white,fill=white] (9.05,-6.80) circle (2 ex);
\node at (8.75,-6.95) [rotate=90] {\small  {$f(x)$}};
\draw[white,fill=white] (9.15,-9.0) -- (9.15,-5.0)-- (9.7,-5.0) -- (9.7,-9.0)--(9.15,-9.0);
\node at (9.375,-8.65) {\small  {$0.00$}};
\node at (9.375,-8.025) {\small  {$0.02$}};
\node at (9.375,-7.4) {\small  {$0.04$}};
\node at (9.375,-6.775) {\small  {$0.06$}};
\node at (9.375,-6.1) {\small  {$0.08$}};
\node at (9.375,-5.475) {\small  {$0.10$}};
\draw[white,fill=white] (6.15,-9.15) circle (1 ex);
\node at (6.15,-9.2) {\small{$x$}};
\draw[white,fill=white] (4.38,-8.85) circle (1.1 ex);
\node at (4.37,-8.85) {\small{$0$}};
\draw[white,fill=white] (4.95,-8.85) circle (1.1 ex);
\node at (4.9,-8.85) {\small{$10$}};
\draw[white,fill=white] (5.525,-8.85) circle (1.1 ex);
\node at (5.525,-8.85) {\small{$20$}};
\draw[white,fill=white] (6.11,-8.85) circle (1 ex);
\node at (6.10,-8.85) {\small{$30$}};
\draw[white,fill=white] (6.7,-8.85) circle (1.1 ex);
\node at (6.675,-8.85) {\small{$40$}};
\draw[white,fill=white] (7.225,-8.85) circle (1.1 ex);
\node at (7.25,-8.85) {\small{$50$}};
\draw[white,fill=white] (7.83,-8.85) circle (1.1 ex);
\node at (7.84,-8.85) {\small{$60$}};
\draw[white,fill=white] (3.5,-6.80) circle (2 ex);
\node at (3.2,-6.95) [rotate=90] {\small  {$f(x)$}};
\draw[white,fill=white] (3.6,-9.0) -- (3.6,-5.0)-- (4.1,-5.0) -- (4.1,-9.0)--(3.6,-9.0);
\node at (3.8,-8.65) {\small  {$0.00$}};
\node at (3.8,-8.05) {\small  {$0.01$}};
\node at (3.8,-7.45) {\small  {$0.02$}};
\node at (3.8,-6.825) {\small  {$0.03$}};
\node at (3.8,-6.2) {\small  {$0.04$}};
\node at (3.8,-5.575) {\small  {$0.05$}};
\node at (5.95,-4.55) {\textbf{(c)} \underline{Lorentzian function on $6$ qubits}};
\hypertarget{fig:lognormal_exp}{}
\hypertarget{fig:cauchy_exp}{}
\hypertarget{fig:spiky_exp}{}
\node at (0.0,-4.55) {\textbf{(b)} \underline{Log-normal distribution on $6$ qubits}};
\node at (11.35,-4.55) {\textbf{(d)} \underline{`Spiky' function on $5$ qubits}};
\end{tikzpicture}
    \caption{Experimental results of loading functions on the Quantinuum H$1$-$1$ and H$1$-$2$ quantum computers using the FSL method. The histograms (blue bars) represent the measurement probabilities (in the computational basis) determined from $5,000$ measurements for \textbf{(a)} and \textbf{(d)} and $10,000$ measurements for \textbf{(b)} and \textbf{(c)}, whereas the orange dots represent the values of target functions. The classical fidelities between the empirical measurement probabilities and the target function are \textbf{(a)} more than $99.7\%$, \textbf{(b)} $98.8\%$, \textbf{(c)} $99.7\%$, and \textbf{(d)} $98.9\%$. The Schmidt-decomposition circuit shown in Fig.~(\protect\hyperlink{fig:svd_circuit}{1c}) was used to load the Fourier coefficients in experiments \textbf{(a)}, \textbf{(b)}, and \textbf{(c)}, whereas the quantum circuit in Fig.~(\ref{fig:spiky-circuit}) was used to load the `spiky' function in experiment \textbf{(d)}.}
   \label{fig:1d_function_loading}
\end{figure}
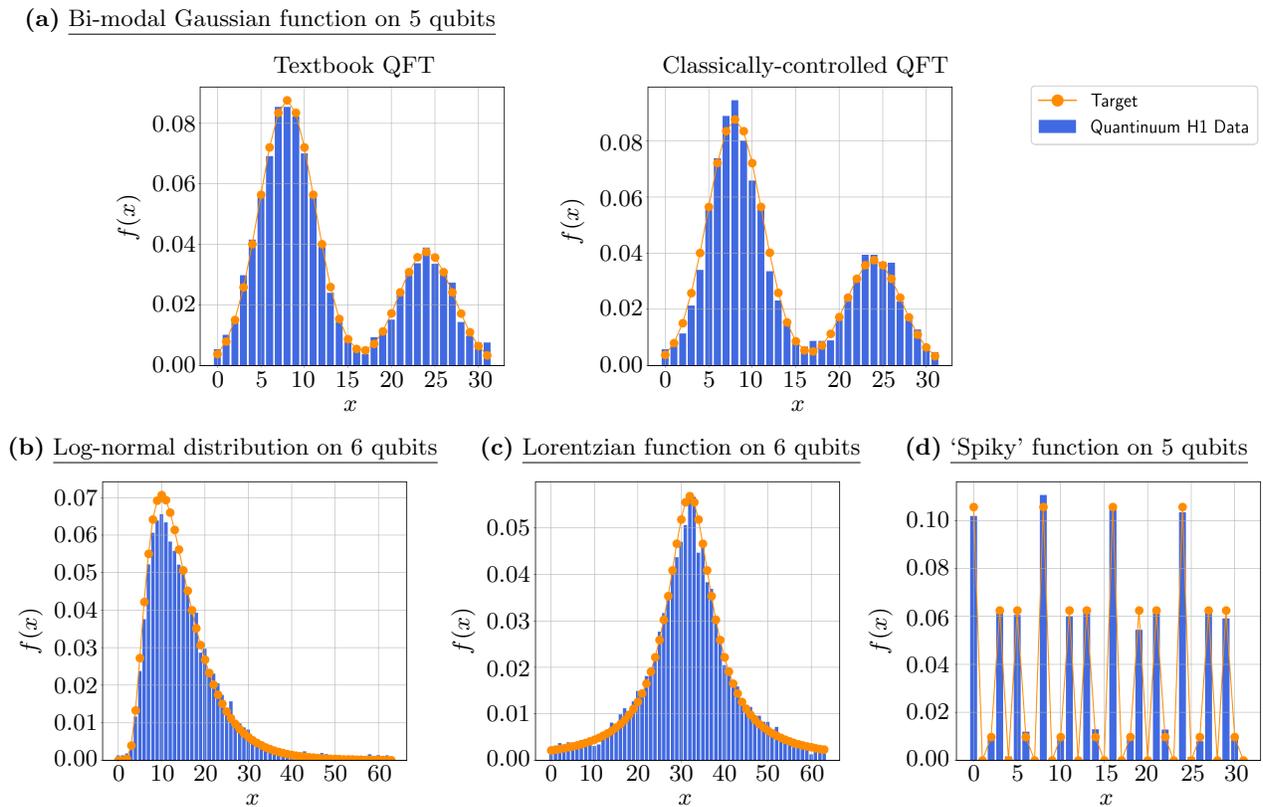

When all the qubits are measured right after the application of the inverse QFT, as is the case in the above experiment, it is possible to replace all the controlled phase gates in the inverse QFT with classically-controlled single-qubit rotation gates \cite{PhysRevLett.76.3228}. Utilizing this classically-controlled implementation of the inverse QFT, we repeated the above bi-modal Gaussian function experiment and obtained a fidelity of $99.7\%$ between the measured probabilities and the target function. The measurement probabilities for this experiment are also shown in Fig.~(\hyperlink{fig:bimodal_exp}{5a}).

As evident from the measured fidelities and the experimental results in Fig.~(\hyperlink{fig:bimodal_exp}{5a}), the FSL method works equally well with the `textbook' implementation and the classically-controlled implementation of the inverse QFT. In other words, the effect of hardware noise is similar for both implementations of the inverse QFT. However, the implementation of the classically-controlled QFT was more economical as the experiment using the classically-controlled QFT required almost three times fewer compute credits than the experiment with the textbook QFT. For this reason, we used the classically-controlled QFT to perform the rest of the function-loading experiments that we discuss.

Next, we loaded a log-normal distribution and a Lorentzian function given in Eq.~\eqref{eq-lognormal} and Eq.~\eqref{eq-cauchy}, respectively, to six qubits. From the results of $10,000$ measurements, we obtained a classical fidelity of $98.8\%$ for the log-normal distribution and a classical fidelity of $99.7\%$ for the Lorentzian function. The measured probabilities for these experiments are shown in Fig.~(\hyperlink{fig:lognormal_exp}{5b}) and Fig.~(\hyperlink{fig:cauchy_exp}{5c}) respectively. For these experiments, we used the Schmidt-decomposition circuit shown in Fig.~(\hyperlink{fig:svd_circuit}{1c}) to load the Fourier coefficients. Further details about these circuits are provided in the supplementary Sec.~(\ref{app-experiments}).

After testing the FSL method's ability to load smooth functions in the presence of noise, we next considered a `spiky' function given by a superposition of low-frequency ($\omega = 4\pi$) and high-frequency ($\omega=20\pi$) cosine waves; see Eq.~\eqref{eq-spiky-app}. We loaded this function into a $5$-qubit state using the quantum circuit shown in Fig.~(\ref{fig:spiky-circuit}) and performed $5000$ measurements. From these measurements, we determined the measurement probabilities which are shown in Fig.~(\hyperlink{fig:spiky_exp}{5d}) and obtained a classical fidelity of $98.9\%$ between the measurement probabilities and the target function.

Encouraged by the success of the FSL method to load functions of a single variable on five and six qubits states, we wanted to verify how the FSL method performs when loading a function of two variables on $10$ qubits of a noisy quantum computer. Specifically, we considered a function $f(x,y) \, = \, \exp\left(-(x-\mu_{1})^{2}/\sigma_{11}^{2} \, - \, (y-\mu_{1})^{2}/\sigma_{12}^{2} \right) \, + \, \lambda \, \exp\left(-(x-\mu_{2})^{2}/\sigma_{21}^{2} \, - \, (y-\mu_{2})^{2}/\sigma_{22}^{2} \right) \, $ with $\mu_{1} = 0.65$, $\mu_{2} = 0.35$, $\sigma_{11} = \sigma_{22} = \sqrt{1/50}$, $\sigma_{12} = \sqrt{1/40}$, $\sigma_{21} = \sqrt{1/30}$, and $\lambda = 0.5 \, $. We then loaded $\sqrt{f(x,y)}$ into a $10$-qubit state (five qubits for each variable). A graph of the function $f(x,y)$ is shown in Fig.~(\hyperlink{fig:exact_2d}{6a}). For concreteness, we used the Schmidt-decomposition circuit shown in Fig.~(\hyperlink{fig:svd_circuit}{1c}) to load the Fourier coefficients of this function. Performing this experiment is expensive due to the large number of measurements required to convincingly compare the amplitudes of the prepared state with the target function, $f(x,y)$.

\begin{figure}
    \includegraphics[scale=0.8]{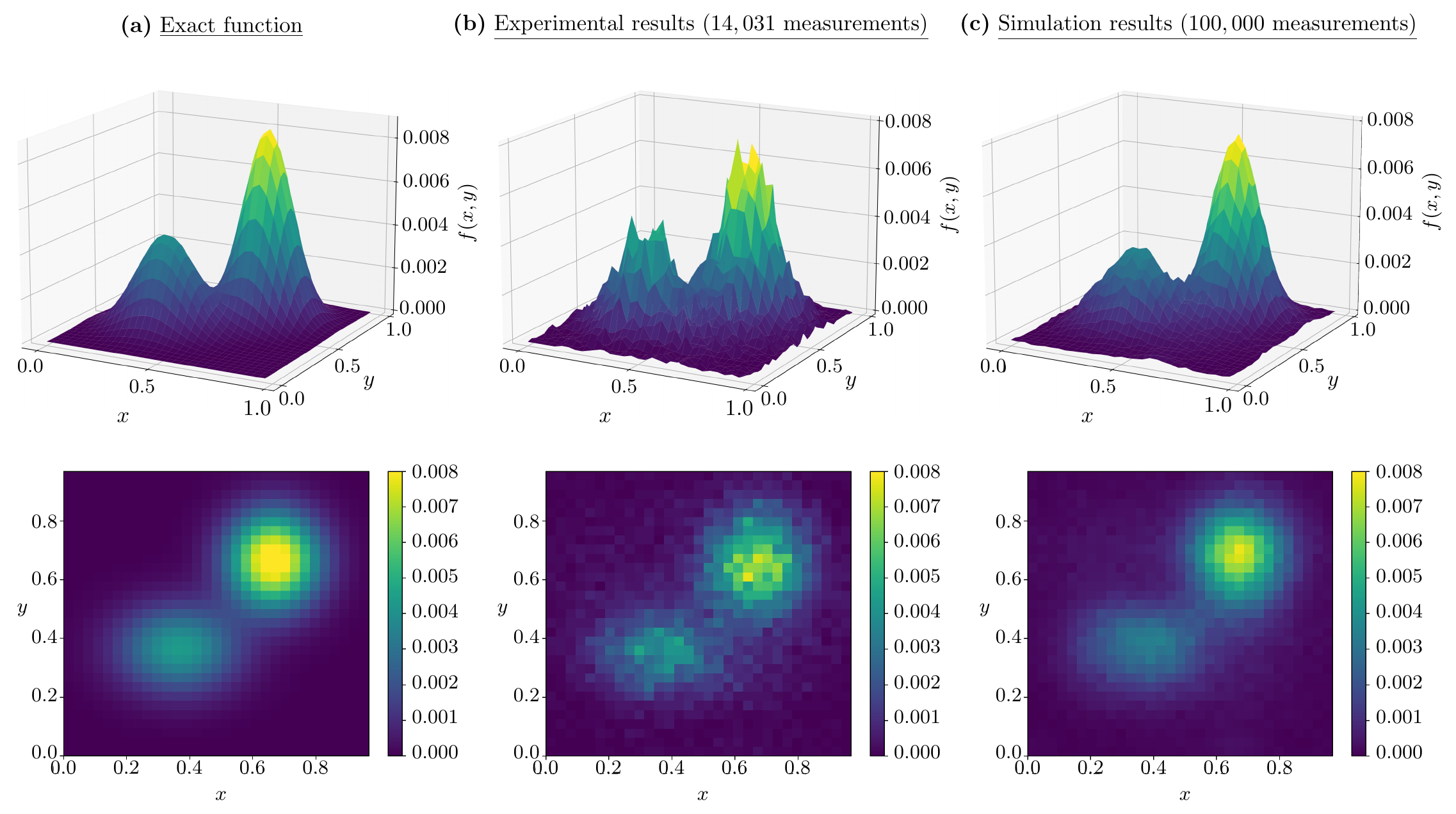}
    \hypertarget{fig:exact_2d}{}
    \hypertarget{fig:sim_2d}{}
    \hypertarget{fig:exp_2d}{}
    \caption{\textbf{(a)} The function $f(x,y) = \exp\left(-(x-\mu_{1})^{2}/\sigma_{11}^{2} - (y-\mu_{1})^{2}/\sigma_{12}^{2} \right) + \lambda \exp\left(-(x-\mu_{2})^{2}/\sigma_{21}^{2} - (y-\mu_{2})^{2}/\sigma_{22}^{2} \right) $ with $\mu_{1} = 0.65$, $\mu_{2} = 0.35$, $\sigma_{11} = \sigma_{22} = \sqrt{1/50}$, $\sigma_{12} = \sqrt{1/40}$, $\sigma_{21} = \sqrt{1/30}$, and $\lambda = 0.5 \, $ we use to test the performance of the FSL method to load functions of two variables on a present-day noisy quantum computer. \textbf{(b)} Measurement probabilities are determined by first loading $\sqrt{f(x,y)}$ into a quantum state of $10$ qubits of the Quantinuum H$1$-$1$ and H$1$-$2$ quantum computers and then measuring all the qubits in the computational basis. The Schmidt-decomposition circuit shown in Fig.~(\protect\hyperlink{fig:svd_circuit}{1c}) was used to load the Fourier coefficients of this function. Even with $14,031$ measurements, we observe that the general shape of the target function has emerged and we obtain a fidelity of $94.0\%$ between the measured probabilities and the target function. \textbf{(c)} To see how the experimental results in \textbf{(b)} would improve with the number of measurements, we perform the simulation of $100,000$ measurements on the Quantinuum H$1$-$1$ and H$1$-$2$ emulators whose noise models are calibrated to simulate the noise present in the Quantinuum H$1$-$1$ and H$1$-$2$ quantum computers. These results demonstrate that the FSL method can load multi-variable functions on up to $10$ qubits on a near-term noisy quantum computer.}
    \label{fig:2d_state_expr}
\end{figure}

Due to limited compute resources, we only managed to perform $14,031$ measurements on the Quantinuum H$1$-$1$ and H$1$-$2$ quantum computers. Even with this small number of measurements, we observe the correct overall shape of the target function as shown in Fig.~(\hyperlink{fig:exp_2d}{6b}). Furthermore, we obtained a fidelity of $94.0\%$ between the measured probabilities and the target function. We expect these results to improve further with the number of measurements. To justify this expectation, we performed a simulation of the experiment on the Quantinuum H$1$-$1$ and H$1$-$2$ emulators whose noise models and parameters match those of the Quantinuum H$1$-$1$ and H$1$-$2$ quantum computers \cite{h1-emulator}. The resulting probabilities determined from $100,000$ simulated measurements are shown in Fig.~(\hyperlink{fig:exp_2d}{6c}) and have a fidelity of $94.1\%$ with the target function. 

The results presented in this section clearly demonstrate that the FSL method can load functions into a quantum state of up to $10$ qubits on present-day noisy quantum computers. We now conclude with a comparison of the FSL method with previously known state preparation methods and some possible future directions.

\section{Discussion}
\label{sec:discussion}

\subsection{Comparison with Previous Work}

Many function-loading algorithms have been proposed in the literature. In this section, we discuss a variety of algorithms that are comparable with the FSL method. A comparative summary of the algorithms we consider is provided in Table~(\ref{table:my_label}).

We start by discussing other function-loading methods that also use the quantum Fourier transform (QFT) and have a close resemblance with the FSL method. One such method is based on Fourier interpolation or trigonometric interpolation, which uses a Fourier transform to approximate the values of a function everywhere given the values of the function on a subset of points in its domain. A quantum circuit implementing this interpolation technique was proposed in Refs.~\cite{quantuminspired,GarcaMolina2022QuantumFA}, and was generalized for loading images on a quantum computer in Ref.~\cite{2022arXiv220306196R}. This circuit is similar to the circuit that we have proposed in Fig.~(\hyperlink{fig:state_prep}{1a}) but with $U_{c}$ replaced with QFT$\mbox{}_{m'} U_{f}$, where $U_{f}$ is any unitary that loads the values of the function $f$ sampled at $2^{m'}$ equidistant grid points to an $m'$-qubit state, and QFT$\mbox{}_{m'}$ is a QFT operator acting on $m'$ qubits. Hence, the circuit implementing Fourier interpolation also has $O(n^2)$ gate count and $O(n)$ depth. However, despite their similarity, the FSL method has several advantages over the interpolation method. Firstly, depending on how the unitary $U_{f}$ is implemented, the circuit implementing the Fourier interpolation may have a higher gate count than the FSL circuit in Fig.~(\hyperlink{fig:state_prep}{1a}) due to an additional QFT operator. Secondly, it is known that the error associated with Fourier interpolation is always greater than the truncation error due to aliasing \cite{Kompenhans2016AdaptationSF}. Therefore, the FSL method is capable of higher accuracy than an interpolation method. Thirdly, as we demonstrated in Sec.~(\ref{sec:method}), it is easy to incorporate Fourier filters when using the FSL method, which can suppress the Gibbs phenomenon present in the Fourier-series approximation of piece-wise discontinuous functions. It is not clear how to suppress the Gibbs phenomenon using the interpolation method. Finally, we do not expect the interpolation method to perform well with functions having sharp peaks as these peaks may not be captured by the values of the functions at some subset of points. Consider the $5$-qubit spiky function state in Fig.~(\hyperlink{fig:spiky_exp}{4d}) as an example. Since this spiky function has a Fourier mode with frequency $\omega = 10 \pi$, the Nyquist theorem dictates that we need the value of the function at $2^{m'} \ge 20$ points to get an accurate result using the interpolation method. This implies that the interpolation method with $m'<n=5$ cannot accurately load this spiky function into a $5$-qubit state. The FSL method, on the other hand, can accurately load this function into a $5$-qubit state as we have demonstrated. 

\begin{table}
    \centering
    \begin{tabular}{|m{8.05em}||m{5.75em} m{6.5em} m{6.0em} m{6.0em} m{5.5em} m{5.5em} m{4.25em} m{4.25em}|} 
    \hline
     &  {Type} & {Requires} & Classical & {Classical} & {Gate} & {Circuit} & {Ancilla} & {}  \\  
        {Methods} & of  & Classical & Time  & Space &  Count  & Depth & Qubits & NISQ-y?  \\
         & State  & Optimization & Complexity & Complexity &  &  & & \\ \hline\hline
        \footnotesize{Fourier $\,\,\,\,\,\,\,\,\,\,\,\,\,\,\,\,\,\,\,\,\,\,\,\,\,\,\,\,\,\,\,\,$ Interpolation} \cite{quantuminspired} & \footnotesize{Smooth functions} & \footnotesize{No} & $O(\log\omega_{\text{max}})$ & {$O(\log\omega_{\text{max}})$} & $O(n^{2})$  & $O(n)$  & $0$ to $1$  & Yes \\
        \hline
        \footnotesize{{QFT Sampler}} \cite{2020PhRvR...2d3442E} & \footnotesize{Periodic distributions} & \footnotesize{Yes} & $\Omega(p)$  & $O(p)$  & $O(n^{2})$  & $O(n)$  & $0$ & Yes \\  
        \hline 
        {\footnotesize{Quantum GAN} } \cite{qgan} & \footnotesize{Arbitrary distributions} & \footnotesize{Yes} & $\Omega(p)$  & $O(p)$  & $O(n L)$  & $O(n L)$  & $0$ & Yes \\
        \hline 
        {\footnotesize{Matrix $\,\,\,\,\,\,\,\,\,\,\,\,\,\,\,\,\,$ Product State}} \cite{mps} & \footnotesize{Smooth functions} & \footnotesize{Yes}  & $O(n)$ & $O(n)$  & $O(n)$  & $O(n)$ & 0 & Yes \\ 
        \hline 
        {\footnotesize{Variational ZGR}} \cite{2021arXiv211107933M} & \footnotesize{Smooth functions} & \footnotesize{Yes} & $O(2^{n})$ & $O(2^{n})$  & $\Omega(n^{2})$  & $\Omega(n)$ & $0$ & Yes \\
        \hline 
        {\footnotesize{Hamiltonian $\,\,\,\,\,\,\,\,\,\,\,\,$ Simulation} } \cite{Hsim_state_prep} & \footnotesize{Arbitrary $\,\,\,\,\,\,\,\,\,\,\,\,$
        functions} & \footnotesize{No} & $O(1)$ & $O(1)$ & $O(T_{\text{oracle}})$ & $O(T_{\text{oracle}})$ & $O(n+n')$ & No \\
        \hline
        {\footnotesize{QSVT} } \cite{AWS} & \footnotesize{Arbitrary $\,\,\,\,\,\,\,\,\,\,\,\,$
        functions} & \footnotesize{No} & $O(d_\text{poly}^3)$ & $O(d_\text{poly})$ & $O(nd_\text{poly}R)$ & $O(nd_\text{poly}R)$ & $3$ to $4$ & No \\
        \hline 
        {\footnotesize{Kitaev-Webb} \cite{2008arXiv0801.0342K}} & \footnotesize{Gaussian distribution} & \footnotesize{No}  &  $O(1)$ & $O(1)$ & $O(\text{poly}(n))$  & $O(\text{poly}(n))$ & $\Omega(n')$ & No \\ 
        \hline 
        {\footnotesize{Discrete Random $\,\,\,\,\,\,\,\,\,\,\,\,$  Walks}} \cite{Rattew_Gaussian_prep} & \footnotesize{Gaussian distribution} & \footnotesize{No}  & $O(1)$  & $O(1)$  & $O(n \log n)$ & $O(n \log n)$ & $O(n)$ & Yes \\ 
        \hline
        {\footnotesize{Powers of Cosine $\,\,\,\,\,\,\,\,\,\,\,\,$ Approximation}} \cite{markov2022generalized} & \footnotesize{Gaussian distribution} & \footnotesize{No}  & $O(1)$  & $O(1)$  & $O(n^{2})$ & $O(n)$  & $0$ & Yes \\ 
        \hline
        {\footnotesize{FSL (This work) }} & \footnotesize{Arbitrary functions} & \footnotesize{No} & $O(8^m)$  & $O(2^{m})$ & $O(n^{2})$  & $O(n)$ & $0$ to $1$ & Yes\\ 
        \hline 
    \end{tabular}
    \caption{A comparison between various function-loading algorithms. The classical cost of the Fourier interpolation method is determined by the maximum frequency, denoted by $\omega_{\max}$, in the spectrum of the target function as per the Shannon-Nyquist theorem. In variational algorithms, such as Variational QFT and Quantum GAN, the number of parameters and the number of layers in a variational circuit are denoted by $p$ and $L$ respectively. The degree of the polynomial used to approximate the target function and the number of rounds of amplitude amplification needed in the QSVT method are denoted by $d_{\text{poly}}$ and $R$ respectively. In fully quantum algorithms, such as Hamiltonian simulation and the Kitaev-Webb method, the function/Hamiltonian/angles of rotation are stored using quantum registers. The number of qubits used to store this information is denoted by $n'$. Furthermore, $T_{\text{oracle}}$ denotes the circuit depth of the oracle used in the implementation of the corresponding method. Finally, the `NISQ-y' category classifies those methods whose circuits are relatively shallow [e.g., circuit depth of less than $1000$ for states of less than $100$ qubits] thereby allowing for the preparation of states with high fidelity [e.g., $>90\%$] on noisy, intermediate-scale quantum (NISQ) devices.}
    \label{table:my_label}
\end{table}

The Fourier transform has also been used in Ref.~\cite{2020PhRvR...2d3442E} to load probability distributions as the amplitudes of a quantum state. In fact, the circuit proposed in Ref.~\cite{2020PhRvR...2d3442E} to load a distribution consists of a unitary operator acting on a small subset of qubits followed by a QFT operator acting on all of the qubits, and hence, is similar to the circuit that we have proposed in Fig.~(\hyperlink{fig:state_prep}{1a}). However, there are several key differences. First, the unitary operator that precedes the QFT operator is taken to be a function of variational parameters which needs to be determined through classical optimization. Secondly, and more importantly, the subset of qubits on which the variational unitary acted on were not entangled with the rest of the qubits before the application of the QFT operator. In other words, there are no CNOT gates similar to those in Fig.~(\hyperlink{fig:state_prep}{1a}). This has interesting consequences as it can be shown that this circuit can only load functions of the form $f(x) = \sum_{k=0}^{2^{m}-1} c_{k} e^{-i2\pi k x} $, where $c_{k}$ depend on the variational parameters. This is only half of the Fourier series as it is missing the complex conjugate terms $c_{-k} = c^*_k$. Therefore, even though this circuit can be trained to sample a target probability distribution, it can not be used to load distributions as a subroutine of another quantum algorithm.

We now discuss other state preparation methods that are not based on Fourier transforms and compare them with the FSL method. Consider the variational algorithm based on the Generative Adversarial Network (GAN) which involves a variational quantum circuit (generator) and a classical neural network (discriminator) \cite{qgan}. The generator and the discriminator are simultaneously optimized until the discriminator cannot distinguish the samples from the generator with the given samples from the target distribution. The variational circuit that was used in Ref.~\cite{qgan} consists of $L$ consecutive layers, each of depth $O(n)$ and containing $O(n)$ quantum gates. However, the computational cost of optimizing a variational quantum circuit scales at least linearly in the number of parameters $p$, which in this case is $O(nL)$. Despite the usefulness of a quantum GAN for state preparations, there are limitations. For example, by construction, a quantum GAN can only learn to load distributions and not arbitrary complex-valued functions into a quantum state. Moreover, the performance of a quantum GAN-based method highly depends on the choice of the initial state on which the variational circuit acts \cite{qgan}. In general, it can take a significant amount of time to train a quantum GAN, further hindering the overall time complexity of this state preparation method. Therefore, we expect that the FSL method, which does not suffer from these limitations, offers a good alternative for loading functions and distributions into a quantum state.  

Another interesting state preparation method is based on matrix product states (MPS) \cite{mps,lubasch}. The main insight that was used in Ref.~\cite{mps} was that {smooth, differentiable, real-valued} (SDR) functions can be well-approximated by piece-wise polynomials. This led to an approximation of SDR function states using MPS of bond dimension $K(p+1)$, where $K$ is the number of subdomains and $p$ is the order of the polynomials \cite{Oseledets2013ConstructiveRO}. This MPS was further approximated by a MPS of bond dimension $2$ using the MPS compression algorithm which has a time complexity of $O(n)$ \cite{Schollwoeck2011TheDR}. The resulting compressed MPS was converted into a quantum circuit of $O(n)$ depth and gate count \cite{Ran2020EncodingOM}. The linear depth circuit and gate count make this approach exceedingly efficient for loading SDR functions into a quantum state. However, the approximation of the SDR functions by piece-wise polynomials and the compression of the MPS lead to a build-up of error. Indeed, the fidelities reported in \cite{mps} are lower than the fidelities that we obtained in Sec.~(\ref{sec:method}) from the FSL method for which the only source of error is the truncation error in the Fourier series. Furthermore, the FSL method can load a larger class of functions than the MPS method. Specifically, the FSL method can load complex-valued and non-smooth functions.

In in Sec.~(\ref{sec:method}), we proposed a method for encoding the target Fourier coefficients via a unitary $U_c$ where $U_c$ is taken to be the circuit of Zalka, Grover, and Rudolph (ZGR) which consists of a cascade of uniformly controlled rotations. Recently, another state preparation algorithm also based on the ZGR circuit was proposed in Ref.~\cite{2021arXiv211107933M}. In fact, it was observed in Ref.~\cite{2021arXiv211107933M} that for a class of positive integrable functions for which $|\partial_{x}^{2}\log f^{2}(x)| \le 8\pi$ on the domain $[0,1]$, most of the angles appearing in the ZGR circuits are close to each other. Hence, it is possible to replace uniformly $k$-controlled rotation gates with single rotation gates for all $k>k_{0}$ for some $k_0 > 0$. This approximation results in a reduction of the gate count and the depth of the circuit from $O(2^{n})$ to $O(2^{k_{0}})$. However, for functions that are non-positive or piece-wise defined, not all of the $k$-controlled rotation gates for $k>k_{0}$ can be replaced with single qubit gates. Furthermore, the angles of these gates have to be determined through classical optimization. Hence, the total gate count for this variational algorithm is $\Omega(n^2)$ whereas the classical time and space complexity is exponential in $n$ \cite{2021arXiv211107933M}.

The methods that we have discussed until now have all required some sort of classical pre-processing. A state preparation method that leverages adiabatic time evolution was recently proposed in Ref.~\cite{Hsim_state_prep} which does not require classical pre-processing at the cost of significantly deeper circuits. The idea behind this method is to encode the target state as a rank-$1$ projector Hamiltonian $\hat{H} = |f \rangle \langle f |$, then perform the adiabatic evolution using low-rank, $1$-sparse Hamiltonian simulation techniques \cite{PhysRevA.97.012327,2003quant.ph..1023A,2002quant.ph..9131C,2007CMaPh.270..359B}. The time evolution is simulated using Trotterization-like time steps that suppress adiabatic and simulation errors. Each time step requires $O(1)$ call to an oracle which implements the Hamiltonian using the arithmetic operations. Interestingly, it was proven in Ref.~\cite{Hsim_state_prep} that the query complexity of this algorithm approaches a constant asymptotically in the number of qubits. However, note that the simulation results reported in Ref.~\cite{Hsim_state_prep} show that on the order of $10^{5}$ Trotterization steps are required to achieve an error of $10^{-5}$ for $n>5$ qubits. Given that the implementation of the oracle requires quadratic gate count and circuit depth \cite{2018arXiv180512445H}, the total resources needed to run this fully-quantum state preparation algorithm are orders of magnitude higher than the present-day technology. This is not surprising as the authors intended the method would be executed on a fault-tolerant quantum computer. It is worth appreciating that this method is also capable of preparing arbitrary quantum states (i.e., loading arbitrary discrete functions), however, the associated complexity can differ greatly from the function loading use case.

Recently, a function-loading algorithm based on the quantum singular value transformation (QSVT), was proposed \cite{AWS}. The basic idea of the QSVT is to find a block encoding of $h(A)$ given a block-encoding $U_{A}$ of a Hermitian operator $A = \sum_{k} a_{k} |k \rangle\langle k|$, where $h(x)$ is a polynomial of degree $d_{poly}$. By cleverly choosing $A = \sum_{k} \sin(k/N) |k \rangle\langle k|$ and choosing $h(x)$ to be a polynomial approximation of $f(\arcsin(x))$, the authors were able to prepare a block-encoding of $\sum_{k} f(k/N) |k \rangle\langle k|$. The desired state could then be prepared with high probability by applying this block-encoding to $\ket{+}^{\otimes n} \otimes \ket{0}_{\text{ancilla}}$. In many cases, several rounds of amplitude amplification are required to boost the probability of success closer to 1. The number of rounds of amplitude estimation, $R$, needed is inversely related to the ``$L_{2}$-norm filling ratio'' of the target function which is approximately the ratio $\|f\|_2/\|f\|_\infty$. Given that quantum circuit that implements the QSVT requires $d_{poly}/2$ applications of $U_{A}$ and $U_{A}^{\dagger}$ \cite{AWS}, and $U_{A}$ for $A = \sum_{k} \sin(k/N) |k \rangle\langle k|$ can be implemented with a circuit of depth $n$, the full quantum circuit needed to prepare the desired state has a depth of $O(n d_{poly}  R)$. Moreover, as discussed in Ref. \cite{AWS}, the full quantum circuit consists of $O(n d_{poly} R)$ quantum gates and requires at most $4$ ancilla qubits. Just like the Hamiltonian simulation method discussed above, this method is only suitable for fault-tolerant quantum computers. However, this method replaces the requirement for an oracle needed in the Hamiltonian simulation method with a classical pre-processing step in which all the angles of rotations in the quantum circuit can be determined using algorithms such as those described in Refs.~\cite{2019Quant...3..190H,2021PhRvA.103d2419D,2020arXiv200302831C}. 

There are many other state preparation algorithms that are tailored to a specific distribution. The most notable example is the algorithm of Kitaev and Webb that loads a Gaussian distribution using a recursive algorithm \cite{2008arXiv0801.0342K}. At each recursive step, the algorithm calculates the angles of rotations directly on an ancilla quantum register and rotates the non-ancillary qubits by that angle. Even though this algorithm asymptotically only requires polynomial resources, a recent study argued that this algorithm is more costly than a generic exponentially scaling algorithm for $n \le 15$ \cite{Deliyannis2021che}. This is due to the large complexity of the quantum arithmetic circuits \cite{2018arXiv180512445H}. The method of Kitaev and Webb can also be generalized to load $D$-dimensional correlated Gaussian distributions. The algorithm first loads $D$ uncorrelated one-dimensional Gaussian distributions and then applies a `shearing' transformation which implements the change of basis required to introduce the correlations between different dimensions. The implementation of the shearing transformation requires a circuit involving $(2n+1) + 2\log_{2}(D+1)$ ancillae qubits and $O(n^{2}D^{2})$ CNOT gates \cite{Deliyannis2021che}. In contrast, the FSL method does not require the shearing transformation as it can load the correlated distribution directly and has a gate count of $O(n^2D)$. 

There are a few other proposals to load a Gaussian distribution that are, unlike the Kitaev-Webb algorithm, suitable for near-term quantum devices For example, Rattew et al. have proposed a Gaussian distribution loading method that is provably robust to bit-flip and phase-flip errors \cite{Rattew_Gaussian_prep}. This method is based on a discrete random walk inspired by Galton machines whose walk operator maps a computational basis state $\ket{j}$ to a superposition state $\ket{j} + \ket{j+1}$ up to a normalization factor. Rattew et al. propose first loading a low-resolution Gaussian distribution on a small number of qubits using a method such as the cascade of controlled rotations depicted in Fig.~(\hyperlink{fig:state_prep}{1b}), then iteratively qubits are added in the $\ket{+} = \left(\ket{0}+\ket{1}\right)/\sqrt{2}$  state followed by a small number of applications of a quantum walk operator until the target number of qubit is reached. The computational complexity of this approach depends upon the specific implementation of the walk operator which requires a choice of quantum adder circuit. Hence, if the logarithmic depth adder presented by Draper et al. \cite{Draper2006ALQ} is used, then the overall quantum gate complexity of this Gaussian loading method is $O(n \log n)$ and $O(n)$ ancilla qubits are required. Alternatively, a QFT-based adder can be used which leads to an overall quantum gate complexity of $O(n^2)$ and only a single ancilla is required.

Another method to approximately load Gaussian distributions into a quantum state was introduced by Markov et al. in Ref.~\cite{markov2022generalized}. This method is based on approximations of Gaussian distributions by powers of trigonometric functions. In particular, these approximations can be derived from the limit $\lim_{r \to \infty} \cos^r (x/\sqrt{r}) = e^{-x^2/2}$. To illustrate their method, consider the task of loading the 2nd power approximation $p_2(x) = \sigma^{-1} \cos^{2}(\pi(x-\mu)/2\sigma)$ for $\mu - \sigma \le x \le \mu+\sigma$. The main insight of Ref.~\cite{markov2022generalized} is that a quantum state $\ket{\sqrt{p_2}}$ can be prepared by applying the inverse QFT operator applied to the state $(\ket{0}-\ket{2^{n}-1})/\sqrt{2}$. This method is nearly a special case of the FSL method introduced in this work with the exception that their method neglects to control for complex phases introduced by the QFT operator.

\subsection{Future Directions}

The FSL method inherits the majority of its circuit complexity from the quantum Fourier transform. Therefore, a natural question to ask is how well the FSL method performs when using an approximate QFT sporting which would lead to reduced circuit complexity. Specifically, it has been shown that an approximate QFT (AQFT) with error bounded by $\varepsilon$ can be implemented by a circuit with depth bounded above by $O(\log n + \log \log (1/\varepsilon))$ and bounded below by $\Omega(\log n)$ \cite{log_depth_qft}. Therefore, using an AQFT to implement an approximate FSL would yield a logarithmic-depth function loading circuit. In order to justify the usage of an AQFT for practical applications, a rigorous study of the performance of an approximate FSL is needed.

In this work, we have only focused on amplitude embedding where the values of the functions are loaded as the amplitudes of the quantum state. There exists another type of embedding known as the key-value embedding, in which the function  $f : \{0, \ldots, 2^{n}-1\} \to [0,2^{m}-1]$ is encoded as an $(n+m)$-qubit state, $\frac{1}{\sqrt{2^{n}}} \sum_{k} \ket{k}_{n} \otimes \ket{f_{k}}_{m}$, where $\ket{k}_{n}$ is the $n$-qubit `key' register and $\ket{f_{k}}$ is the $m$-qubit `value' register. The difficulty in preparing such a key-value pair state is that the circuit preparing the state on the value register has to be controlled by the key register, which leads to a large gate count. Various efficient algorithms have been proposed for the key-value embedding of the functions/distributions \cite{Araujo2021ADA,Zhang2022QuantumSP,2022arXiv220308758S}. An interesting future direction is to see if the FSL method introduced in this work can be generalized to efficiently generate key-value embeddings.

It is also worth emphasizing that the Fourier series is just one possible orthonormal series representation of functions of interest. There are many other basis functions -- such as wavelets, Chebyshev polynomials, etc. -- that could be more suitable than a Fourier series expansion for some classes of functions. In this work, we only considered the Fourier-series representation given that the implementation of the Fourier transform on a quantum computer requires fewer gates than the implementation of the wavelet transform \cite{1998quant.ph..9004F,Li2018TheMA} and discrete cosine transform \cite{2001quant.ph.11038K}. Nevertheless, it could be worthwhile to investigate the performance of the truncated wavelet series, and other possible representations, for function loading on quantum computers.

\vspace{8mm}

\section*{Data and code availability}

A user-friendly implementation of the FSL method, along with various examples, is available at \url{https://github.com/mcmahon-lab/Fourier-Series-Loader}. All the experimental data gathered from the experiments performed on Quantinuum System Model H$1$ and the code to analyze the experimental results and to produce the figures presented in this work are available at \url{https://doi.org/10.5281/zenodo.8331675}. 

\section*{Acknowledgements}

This research used resources of the Oak Ridge Leadership Computing Facility, which is a DOE Office of Science User Facility supported under Contract DE-AC$05$-$00$OR$22725$. We thank Travis Humble (Oak Ridge National Laboratory) for his support of our research. We also thank Ryan Landfield (ORNL/OLCF User Support) and Brian Neyenhuis (Quantinuum) for technical assistance and for providing the hardware specifications included in this work. It is also a pleasure to thank Javier Gonzalez-Conde, Keisuke Fujii, Constantin Gonciulea, Vanio Markov, Juan Jos\'{e} Garcia-R\'{i}poll, Arthur Rattew, Mikel Sanz, Charlee Stefanski, Naoki Yamamoto, and Christa Zoufal for useful feedback on a draft of this manuscript. The work of M.M. was supported by the US Department of Energy under grant numbers DE-SC$0014123$ and DE-SC$0007884$, and the QuantiSED Fermilab consortium. P.L.M. gratefully acknowledges support from a Sloan Foundation Fellowship and a David and Lucile Packard Foundation Fellowship, and acknowledges membership in the CIFAR Quantum Information Science Program as an Azrieli Global Scholar.

{\textbf{Software used:}} The circuit diagrams in this paper and the supplementary sections were prepared using quantikz package \cite{Kay2018TutorialOT}. All the plots were prepared using Matplotlib \cite{Hunter:2007}. To perform the maximum likelihood estimation for the tomography experiment, the cost function was minimized using the generic minimize function in SciPy \cite{2020SciPy-NMeth}. 

\bibliography{references}

\appendix

\section{Derivation of the Fourier Series Loader method} \label{app-2d-fsl}

In Sec.~(\ref{sec:method}), we claimed that the FSL method can load a Fourier series of one or more variables into a quantum state. We demonstrated this capability by loading a function of one and two variables on the Quantinuum H$1$-$1$ and H$1$-$2$ quantum computers. In this appendix, we provide a detailed derivation of the FSL method. 

\subsection{FSL method for a Fourier series of single variable}

Consider a truncated Fourier series of the form
\begin{align}
    f(x) \, = \, \sum_{p = -M}^{M} \, c_{p} \, e^{-i2\pi px} \, ,
\end{align}
where $M = 2^{m}-1$ denotes the number of Fourier coefficients. Our goal is then to prepare the following state of $n$ qubits:
\begin{align}
    \ket{f}_{n} \, = \, \sum_{k=0}^{2^{n}-1} \, f_{k} \, \ket{k}_{n} \, , \label{revised_f1-der-1}
\end{align}
where $f_{k} \, = \, f\left(x=k/2^{n}\right) \, = \, \sum_{p = -M}^{M} \, c_{p} \, e^{-i2\pi pk/2^{n}} \, $. Here, the subscript $n$ in $\ket{ \hspace{1mm} \cdot \hspace{1mm}}_{n}$ specifies the number of qubits.

Let us assume the unitary $U_{c}$ in the quantum circuit in Fig.~(\hyperlink{fig:state_prep}{1a}) acts on $(m+1)$ qubits and maps $\ket{0}^{\otimes (m+1)}$ state to $\ket{\tilde{c}}$, where
\begin{align}
    \ket{\tilde{c}}_{m+1} \, := \, 2^{n/2} \sum_{p=0}^{M} c_{p} \ket{p}_{m+1} + 2^{n/2}\sum_{p=1}^{M} c_{-p} \ket{2^{m+1}-p}_{m+1} \, . \label{new-app-eq-c-state}
\end{align} 
Then we claim that the circuit in Fig.~(\hyperlink{fig:state_prep}{1a}) maps initial $\ket{0}^{\otimes n}$ state to the target $\ket{f}_{n}$ state in Eq.~\eqref{revised_f1-der-1}. To verify this claim, we traverse through the circuit in Fig.~(\hyperlink{fig:state_prep}{1a}) and analyze the state after each step. After the action on $U_{c}$ on the lower $(m+1)$ qubits, the initial $\ket{0}^{\otimes n}$ state becomes
\begin{align}
    2^{n/2} \sum_{p=0}^{M} c_{p} \ket{0}^{\otimes (n-m-1)}\otimes\ket{p}_{m+1} + 2^{n/2}\sum_{p=1}^{M} c_{-p} \ket{0}^{\otimes (n-m-1)}\otimes\ket{2^{m+1}-p}_{m+1} \, ,
\end{align}
which can equivalently be written as
\begin{align}
    2^{n/2} \sum_{p=0}^{M} c_{p} \ket{0}^{\otimes (n-m-1)}\otimes\ket{0}\otimes\ket{p}_{m} + 2^{n/2}\sum_{p=1}^{M} c_{-p} \ket{0}^{\otimes (n-m-1)}\otimes\ket{1}\otimes\ket{2^{m}-p}_{m} \, .
\end{align}
The next step following the action of $U_{c}$ is the action of the cascade of $n-m-1$ CNOT gates as shown in Fig.~(\hyperlink{fig:state_prep}{1a}). After the action of these CNOT gates, the state of $n$ qubits becomes
\begin{align}
    2^{n/2} \sum_{p=0}^{M} c_{p} \ket{0}^{\otimes (n-m)}\otimes\ket{p}_{m} + 2^{n/2}\sum_{p=1}^{M} c_{-p} \ket{1}^{\otimes (n-m)}\otimes\ket{2^{m}-p}_{m} \, ,
\end{align}
which can equivalently be written as
\begin{align}
    2^{n/2} \sum_{p=0}^{M} c_{p} \ket{p}_{n} + 2^{n/2}\sum_{p=1}^{M} c_{-p} \ket{2^{n}-p}_{n} \, .
\end{align}
The last step in the FSL circuit shown in Fig.~(\hyperlink{fig:state_prep}{1a}) is the action of the inverse Quantum Fourier Transform (QFT). Recall that the inverse QFT acts on a computational basis state as
\begin{align}
    \qftd \ket{p}_{n} \, = \, \frac{1}{2^{n/2}} \, \sum_{k=0}^{2^{n}-1} \, e^{-i2\pi p k/2^{n}} \, \ket{k}_{n} \, .
\end{align}
This implies that the state of $n$ qubits after the action of the inverse QFT becomes
\begin{align}
    2^{n/2} &\sum_{p=0}^{M} c_{p} \qftd \ket{p}_{n} + 2^{n/2}\sum_{p=1}^{M} c_{-p} \qftd \ket{2^{n}-p}_{n} \, , \nonumber\\
    = &\sum_{p=0}^{M} c_{p} \sum_{k=0}^{2^{n}} e^{-i2\pi kp/2^{n}} \ket{k}_{n} + \sum_{p=1}^{M} c_{-p} \sum_{k=0}^{2^{n}} e^{-i2\pi k (2^{n}-p)/2^{n}} \ket{k}_{n} \, , \nonumber\\
    = & \sum_{k=0}^{2^{n}} \, \sum_{p=-M}^{M} c_{p}e^{-i2\pi k p/2^{n}} \, \ket{k}_{n} \, , 
\end{align}
which is precisely the target state in Eq.~\eqref{revised_f1-der-1}. This finishes our derivation of the FSL method for single-variable functions.

\subsection{FSL method for a multi-variate Fourier series}


Here, we provide the details for extending the FSL method to loading higher-dimensional functions and provide the corresponding quantum circuit. For simplicity, we consider the case of a Fourier series of two variables. Although, our construction can easily be generalized to Fourier series of more than two variables.  

Any arbitrary truncated Fourier Series of two variables can be written as
\begin{align}
    f(x,y) \, = \, \sum_{p = -M}^{M} \sum_{q=-M}^{M} \, c_{p,q} \, e^{-i2\pi px} \, e^{-i2\pi qy} \, ,
\end{align}
where $M = 2^{m}-1$. Our goal is then to prepare the following state of $2n$ qubits ($n$ qubits for each variable):
\begin{align}
    \ket{f}_{2n} \, = \, \sum_{k=0}^{2^{n}-1} \sum_{j=0}^{2^{n}-1} \, f_{k,j} \, \ket{k}_{n}\otimes\ket{j}_{n} \, , \label{app-f2-state}
\end{align}
where $f_{k,j} \, = \, f\left(k/2^{n},j/2^{n}\right) \, = \, \sum_{p = -M}^{M} \sum_{q=-M}^{M} \, c_{p,q} \, e^{-i2\pi pk/2^{n}} \, e^{-i2\pi q j/2^{n}} \, $. Here, the subscript $n$ in $\ket{ \hspace{1mm} \cdot \hspace{1mm}}_{n}$ specifies the number of qubits.

\begin{figure}
    \centering
    \includegraphics{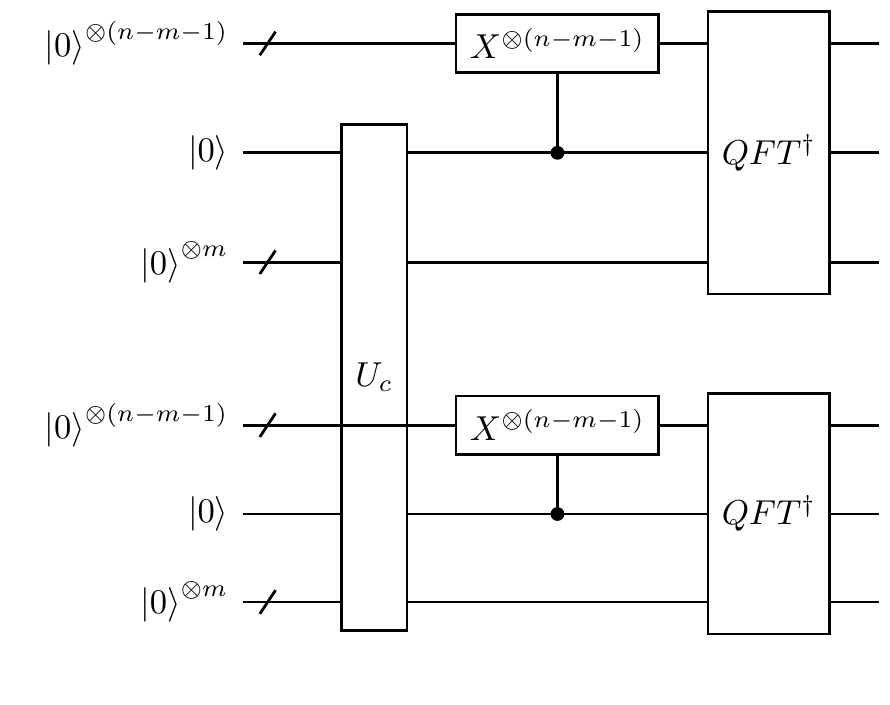}
    \caption{The quantum circuit implementing the two-dimensional FSL method. Similarly to the one-dimensional FSL circuit in Fig.~(\ref{fig-fsl1}), this circuit consists of an unitary $U_{c}$ which loads the Fourier coefficients, a cascade of CNOT gates, and inverse QFTs.}
  \label{fig:circuit-2d}
\end{figure}

We claim that the state in Eq.~\eqref{app-f2-state} can be prepared using the circuit shown in Fig.~(\ref{fig:circuit-2d}). In this circuit, the unitary operator $U_{c}$ is any unitary operator that maps $\ket{0}^{\otimes 2(m+1)}$ to state $\ket{\tilde{c}}_{2(m+1)}$, where
\begin{align}
    \ket{\tilde{c}}_{2(m+1)} \, =&\, \, 2^{n} \, \sum_{p=0}^{M}\sum_{q=0}^{M} \, c_{p,q} \, \ket{p}_{m+1}\otimes\ket{q}_{m+1} \, + \, 2^{n} \, \sum_{p=0}^{M}\sum_{q=1}^{M} \, c_{p,-q} \, \ket{p}_{m+1}\otimes\ket{2^{m+1}-q}_{m+1} \label{eq-2d-fsl-tc}\\
    +&\, \, \, 2^{n} \, \sum_{p=1}^{M}\sum_{q=0}^{M} \, c_{-p,q} \, \ket{2^{m+1}-p}_{m+1}\otimes\ket{q}_{m+1} \, + \, 2^{n} \, \sum_{p=1}^{M}\sum_{q=1}^{M} \, c_{-p,-q} \, \ket{2^{m+1}-p}_{m+1}\otimes\ket{2^{m+1}-q}_{m+1} \, . \nonumber
\end{align}
Note that this state is analogous to the state $\ket{\tilde{c}}$ in Eq.~\eqref{new-app-eq-c-state}.

To verify our claim, we traverse through the circuit in Fig.~(\ref{fig:circuit-2d}) and analyze the state after each step: the initial state of the $2n$ qubits is $\ket{0}^{\otimes 2n}$. Then, after applying the $U_{c}$ operator, the state of $2n$ qubits is
\begin{align}
    & \, 2^{n} \, \sum_{p=0}^{M}\sum_{q=0}^{M} \, c_{p,q} \, \ket{0}^{\otimes (n-m-1)}\otimes\ket{p}_{m+1}\otimes\ket{0}^{\otimes (n-m-1)}\otimes\ket{q}_{m+1} \, \\ + \, & \, 2^{n} \, \sum_{p=0}^{M}\sum_{q=1}^{M} \, c_{p,-q} \, \ket{0}^{\otimes (n-m-1)}\otimes\ket{p}_{m+1}\otimes\ket{0}^{\otimes (n-m-1)}\otimes\ket{2^{m+1}-q}_{m+1}
     \nonumber\\ + \, & \, 2^{n} \, \sum_{p=1}^{M}\sum_{q=0}^{M} \, c_{-p,q} \, \ket{0}^{\otimes (n-m-1)}\otimes\ket{2^{m+1}-p}_{m+1}\otimes\ket{0}^{\otimes (n-m-1)}\otimes\ket{q}_{m+1} \, \nonumber\\ + \, & \, 2^{n} \, \sum_{p=1}^{M}\sum_{q=1}^{M} \, c_{-p,-q} \, \ket{0}^{\otimes (n-m-1)}\otimes\ket{2^{m+1}-p}_{m+1}\otimes\ket{0}^{\otimes (n-m-1)}\otimes\ket{2^{m+1}-q}_{m+1} \, . \nonumber
\end{align}
The above state can equivalently be written as
\begin{align}
    & \, 2^{n} \, \sum_{p=0}^{M}\sum_{q=0}^{M} \, c_{p,q} \, \ket{0}^{\otimes (n-m-1)}\otimes\ket{0}\otimes\ket{p}_{m}\otimes\ket{0}^{\otimes (n-m-1)}\otimes\ket{0}\otimes\ket{q}_{m} \, \\ + \, & \, 2^{n} \, \sum_{p=0}^{M}\sum_{q=1}^{M} \, c_{p,-q} \, \ket{0}^{\otimes (n-m-1)}\otimes\ket{0}\otimes\ket{p}_{m}\otimes\ket{0}^{\otimes (n-m-1)}\otimes\ket{1}\otimes\ket{2^{m}-q}_{m}
     \nonumber\\ + \, & \, 2^{n} \, \sum_{p=1}^{M}\sum_{q=0}^{M} \, c_{-p,q} \, \ket{0}^{\otimes (n-m-1)}\otimes\ket{1}\otimes\ket{2^{m}-p}_{m}\otimes\ket{0}^{\otimes (n-m-1)}\otimes\ket{0}\otimes\ket{q}_{m} \, \nonumber\\ + \, & \, 2^{n} \, \sum_{p=1}^{M}\sum_{q=1}^{M} \, c_{-p,-q} \, \ket{0}^{\otimes (n-m-1)}\otimes\ket{1}\otimes\ket{2^{m}-p}_{m}\otimes\ket{0}^{\otimes (n-m-1)}\otimes\ket{1}\otimes\ket{2^{m}-q}_{m} \, . \nonumber
\end{align}
Following the application of the controlled $X^{\otimes (n-m-1)}$ gates in Fig.~(\ref{fig:circuit-2d}), the above state is mapped to
\begin{align}
    & \, 2^{n} \, \sum_{p=0}^{M}\sum_{q=0}^{M} \, c_{p,q} \, \ket{0}^{\otimes (n-m)}\otimes\ket{p}_{m}\otimes\ket{0}^{\otimes (n-m)}\otimes\ket{q}_{m} \, \\ + \, & \, 2^{n} \, \sum_{p=0}^{M}\sum_{q=1}^{M} \, c_{p,-q} \, \ket{0}^{\otimes (n-m)}\otimes\ket{p}_{m}\otimes\ket{1}^{\otimes (n-m)}\otimes\ket{2^{m}-q}_{m}
     \nonumber\\ + \, & \, 2^{n} \, \sum_{p=1}^{M}\sum_{q=0}^{M} \, c_{-p,q} \, \ket{1}^{\otimes (n-m)}\otimes\ket{2^{m}-p}_{m}\otimes\ket{0}^{\otimes (n-m)}\otimes\ket{q}_{m} \, \nonumber\\ + \, & \, 2^{n} \, \sum_{p=1}^{M}\sum_{q=1}^{M} \, c_{-p,-q} \, \ket{1}^{\otimes (n-m)}\otimes\ket{2^{m}-p}_{m}\otimes\ket{1}^{\otimes (n-m)}\otimes\ket{2^{m}-q}_{m} \, . \nonumber
\end{align}
We now denote this state by $\ket{c}_{2n}$ and write it equivalently as
\begin{align}
    \ket{c}_{2n} \, =&\, \, 2^{n} \, \sum_{p=0}^{M}\sum_{q=0}^{M} \, c_{p,q} \, \ket{p}_{n}\otimes\ket{q}_{n} \, + \, 2^{n} \, \sum_{p=0}^{M}\sum_{q=1}^{M} \, c_{p,-q} \, \ket{p}_{n}\otimes\ket{2^{n}-q}_{n} \\
    +&\, \, \, 2^{n} \, \sum_{p=1}^{M}\sum_{q=0}^{M} \, c_{-p,q} \, \ket{2^{n}-p}_{n}\otimes\ket{q}_{n} \, + \, 2^{n} \, \sum_{p=1}^{M}\sum_{q=1}^{M} \, c_{-p,-q} \, \ket{2^{n}-p}_{n}\otimes\ket{2^{n}-q}_{n} \, . \nonumber
\end{align}

Finally, we consider the action of the inverse QFTs. The state of $2n$ qubits following the application of the inverse QFTs is given by
\begin{align}
    \qftd \otimes \qftd \, \ket{c}_{2n} \, =&\, \, \sum_{k=0}^{2^{n}-1}\sum_{j=0}^{2^{n}-1}\sum_{p=0}^{M}\sum_{q=0}^{M} \, c_{p,q} \, e^{-i2\pi p k/2^{n}}e^{-i2\pi q j/2^{n}} \, \ket{k}_{n}\otimes\ket{j}_{n} \, \\ +&\,  \,  \sum_{k=0}^{2^{n}-1}\sum_{j=0}^{2^{n}-1}\sum_{p=0}^{M}\sum_{q=1}^{M} \, c_{p,-q} \, e^{-i2\pi p k/2^{n}}e^{i2\pi q j/2^{n}} \, \ket{k}_{n}\otimes\ket{j}_{n} \nonumber\\
    +&\,  \, \sum_{k=0}^{2^{n}-1}\sum_{j=0}^{2^{n}-1}\sum_{p=1}^{M}\sum_{q=0}^{M} \, c_{-p,q} \, e^{i2\pi p k/2^{n}}e^{-i2\pi q j/2^{n}} \, \ket{k}_{n}\otimes\ket{j}_{n} \, \nonumber\\ +&\, \,  \sum_{k=0}^{2^{n}-1}\sum_{j=0}^{2^{n}-1}\sum_{p=1}^{M}\sum_{q=1}^{M} \, c_{-p,-q} \, e^{i2\pi p k/2^{n}}e^{i2\pi q j/2^{n}} \, \ket{k}_{n}\otimes\ket{j}_{n} \, . \nonumber
\end{align}
By combining terms, we can write this state as
\begin{align}
    \qftd \otimes \qftd \, \ket{c}_{2n} \, =&\, \, \sum_{k=0}^{2^{n}-1}\sum_{j=0}^{2^{n}-1} \left( \, \sum_{p=-M}^{M}\sum_{q=-M}^{M} \, c_{p,q} \, e^{-i2\pi p k/2^{n}}e^{-i2\pi q j/2^{n}} \right)\, \ket{k}_{n}\otimes\ket{j}_{n} \, ,
\end{align}
which is exactly the desired state $\ket{f}_{2n}$ in Eq.~\eqref{app-f2-state}. This verifies our claim that the circuit in Fig.~(\ref{fig:circuit-2d}) can be used for the exact loading of a Fourier series of two or more variables. 

\section{Uniformly controlled rotations} \label{app-ucr}

In the main text, we proposed a cascade of uniformly controlled rotations shown in Fig.~(\hyperlink{fig:zgr_circuit_diagram}{1b}) as one possible method to load the Fourier coefficients to the amplitudes of a quantum state $\ket{\tilde{c}} = U_c \ket{0}^{\otimes m}$. This method has many desirable features. For example, an efficient implementation of this circuit $U_c$ in terms of CNOT gates and rotation gates is known \cite{mottonen}. Moreover, all the angles of rotation can be classically determined given the target state \cite{mottonen}. Since this method plays a significant role in the FSL method, we briefly review the results of \cite{mottonen} in this appendix.

\begin{figure}
    \centering
    \includegraphics[scale=0.675]{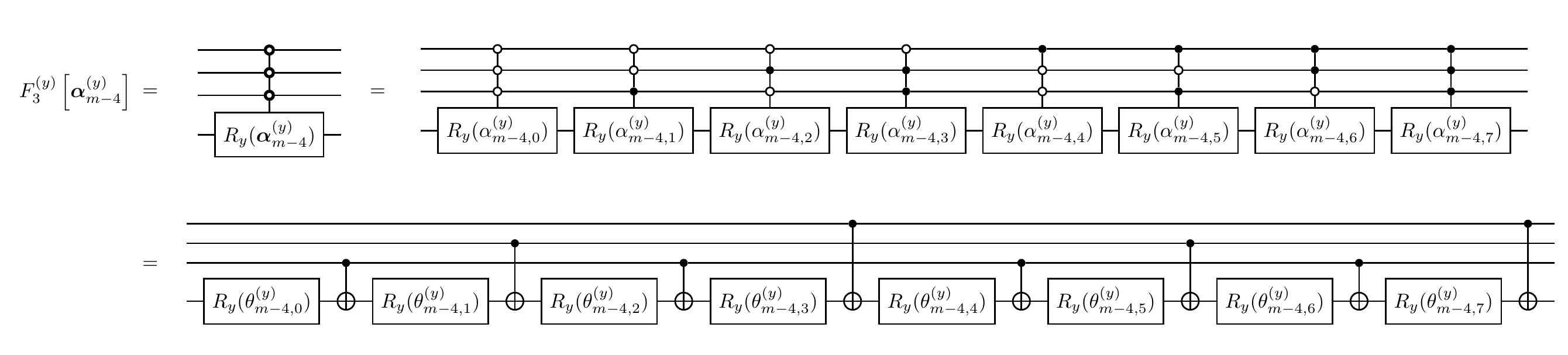}
    \caption{Equivalent implementations of uniformly $3$-controlled rotations. The implementation in terms of rotation gates and CNOT gates was derived in Ref.~\cite{mottonen}.}
    \label{fig-ucr-app}
\end{figure}

As discussed in Sec.~(\ref{sec:method}), the $m$-qubit circuit shown in Fig.~(\hyperlink{fig:zgr_circuit_diagram}{1b}) can be used to prepare an arbitrary $m$-qubit state $\ket{\psi}$. The circuit in Fig.~(\hyperlink{fig:zgr_circuit_diagram}{1b}) has three distinct parts, (i) a $R_z$ gate, (ii) uniformly controlled $R_y$ gates, and (iii) uniformly controlled $R_z$ gates. Hence, the action of this circuit can be succinctly written as
\begin{align}
    \ket{\psi} \, =& \, \, \prod_{j=0}^{m-1} \, F^{(z)}_{j}\left[\boldsymbol{\alpha}_{m-1-j}^{(z)}\right] \, \cdot \, \prod_{j=0}^{m-1} \, F^{(y)}_{j}\left[\boldsymbol{\alpha}_{m-1-j}^{(y)}\right] \, \cdot \, R_{z}(-\phi_{\psi}) \, \ket{0}^{\otimes n} \, . \label{eq-app-ucr}
\end{align}
In this equation, $\boldsymbol{\alpha}^{(y/z)}_{j} = \left(\alpha^{(y/z)}_{j,0},\alpha^{(y/z)}_{j,1}, \dots , \alpha^{(y/z)}_{j,2^{m-1-j}-1} \right) $ is a vector of size $2^{m-1-j}$, and $F^{(y/z)}_{j}\left[\boldsymbol{\alpha}^{(y/z)}_{m-1-j}\right]$ is a uniformly $j$-controlled rotation operator which acts on the $(j+1)^{\text{th}}$ qubit as $R_{y/z}\big(\alpha^{(y/z)}_{m-1-j, k}\big)$ if the state of the first $j$ qubits is $\ket{k}$ for $k = \{0,1,\dots,2^{j}-1\}$ \cite{1998RSPSA.454..313Z,2002quant.ph..8112G,mottonen}. A pictorial representation of $F^{(y)}_{j}$ for $j=3$ is given in Fig.~(\ref{fig-ucr-app}). 

Given the $m$-qubit target state $\ket{\psi} \, = \, \sum_{i=0}^{2^{m}-1} |\psi_{i}| \, e^{i\omega_{i}} \, \ket{i} \, $, the angles $\boldsymbol{\alpha}^{(y)}_{j}$ for $j = \{0,1,\dots,m-1\}$ in Eq.~\eqref{eq-app-ucr} can be determined using \cite{1998RSPSA.454..313Z,2002quant.ph..8112G,mottonen}
\begin{align}
    \alpha_{j,k}^{(y)} \, = \, 2 \arcsin \left[ \frac{\sum_{\ell=0}^{2^{j}-1} \, |\psi_{(2k+1)2^{j}+\ell}|^{2}}{\sum_{\ell=0}^{2^{j+1}-1} \, |\psi_{k 2^{j+1}+\ell}|^{2}} \right] \, . \label{eq-alpha-y}
\end{align}
Moreover, the angles $\boldsymbol{\alpha}^{(z)}_{j}$ for $j = \{0,1,\dots,m-1\}$ and $\phi_{\psi}$ can be determined using \cite{mottonen}
\begin{align}
    \alpha_{j,k}^{(z)} \, = \, 2^{-j} \, \sum_{\ell=0}^{2^{j}} \, \left( \omega_{(2k+1)2^{j}+\ell} - \omega_{k 2^{j+1}+\ell}   \right) \, , \label{eq-alpha-z}
\end{align}
and
\begin{align}
    \phi_{\psi} \, = \, 2^{1-m} \, \sum_{j=0}^{2^{m}-1} \, \omega_{j} \, , \label{eq-phase}
\end{align}
respectively. Hence, given the target state $\ket{\psi}$, all the angles in Eq.~\eqref{eq-app-ucr} can be evaluated on a classical computer.

The next step after finding the angles of uniformly controlled rotations, is to efficiently implement the uniformly controlled gates in terms of simple gates such as rotations and CNOT gates. It was shown in \cite{mottonen} that $F_{j}^{(y/z)}$ for $j>0$ can be implemented using $2^{j}$ $R_{y/z}$ gates acting on the $(j+1)^{\text{th}}$ qubit. The angle of the $k^{\text{th}}$ rotation for $k = \{0,1,\dots,2^{j}-1\}$ is given by \cite{mottonen}
\begin{align}
    \theta_{m-1-j,k}^{(y/z)} \, = \, \sum_{\ell = 0}^{2^{j}-1} \, M_{k\ell} \,  \alpha_{m-1-j,\ell}^{(y/z)}  \, ; \quad\quad\quad M_{k\ell} \, = \, 2^{-j} \, \left(-1\right)^{b_{\ell}\cdot g_{k}} \, ,
\end{align}
where $b_{k}$ and $g_{k}$ denote the binary code and binary Gray code of the integer $k$ respectively. In addition to the $2^{j}$ rotation gates, the efficient implementation of $F_{j}^{(y/z)}$ also has $2^{j}$ CNOT gates acting on the $(j+1)^{\text{th}}$ qubit as shown in Fig.~(\ref{fig-ucr-app}). The last CNOT gate is controlled by the first qubit. Whereas, the control qubit for the $k^{\text{th}}$ CNOT, excluding the last CNOT, has to be determined based on which bit of the binary Gray code of integer $k$ differs from that of $k-1$. For example, the control qubit for the first CNOT in Fig.~(\ref{fig-ucr-app}) is the $3^{\text{rd}}$ qubit because the third bit in the Gray code of $1$ ($001$) and $0$ ($000$) are different. Whereas, the control qubit for the second CNOT is the $2^{\text{nd}}$ qubit because the second bit in the Gray code of $2$ ($011$) and $1$ ($001$) are different.

The Python code used to implement the cascade of uniformly controlled rotations and to perform the pre-processing step discussed above is provided in the code repository accompanying this paper. This finishes our review of Ref.~\cite{mottonen}.

\section{Benchmarking classical pre-processing time} \label{app-classical-time}

\begin{figure}
    \centering
    \includegraphics{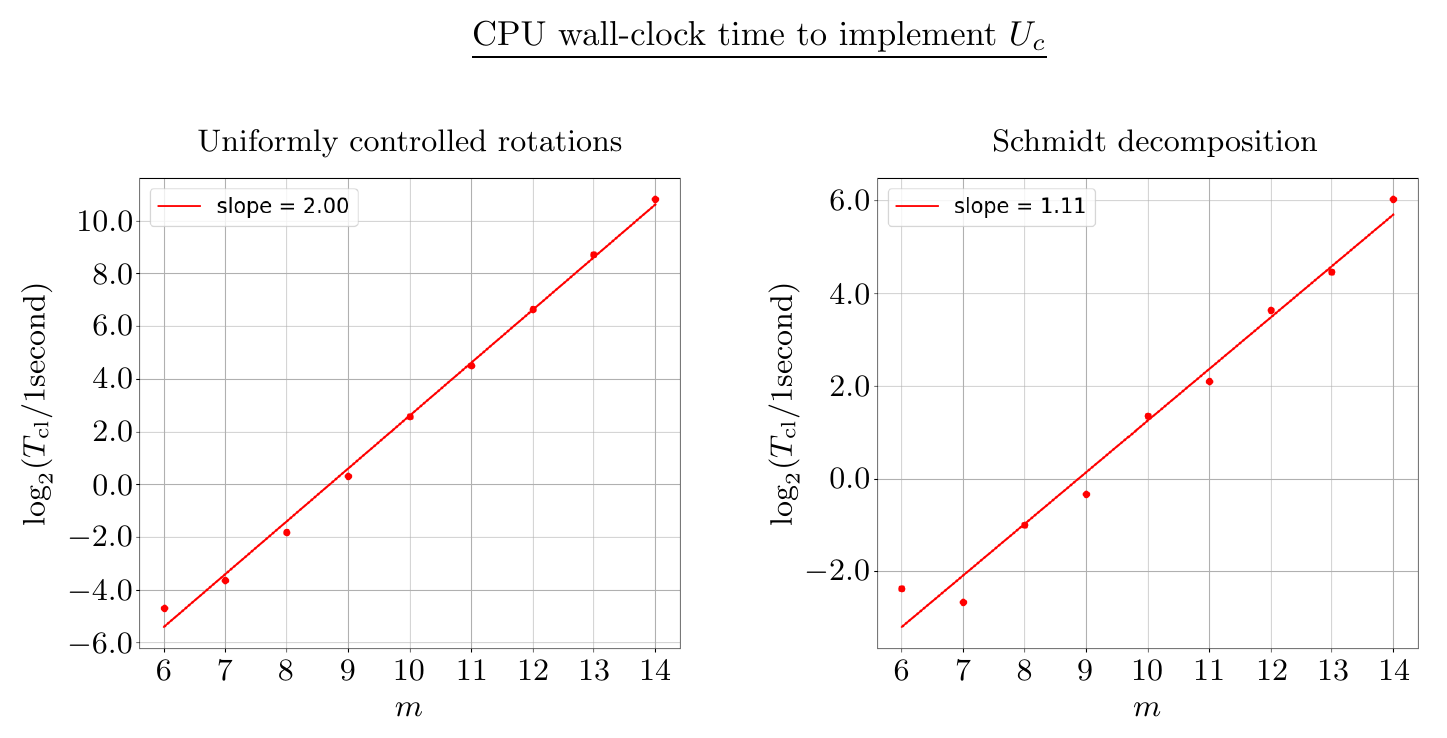}
    \caption{Plots for the CPU wall clock time, $T_{\text{cl}}$, passed during the classical pre-processing step of finding a circuit implementation of the unitary $U_{c}$ using the uniformly controlled rotations (left) and using the Schmidt decomposition method (right). We find that even though $T_{\text{cl}}$ scales polynomially with the number of Fourier coefficients $M = 2^{m+1}$, the values of $T_{\text{cl}}$ are reasonable for large enough values of $M$. For example, $T_{cl} < 6$ seconds for uniformly controlled rotations and $T_{cl} < 3$ seconds for the Schmidt decomposition method for $M \sim 2000$ ($m=10)$.   }
    \label{fig:classical_time}
\end{figure}

An attractive feature of the FSL method is that the classical pre-processing cost needed to implement the FSL circuit is constant with respect to the total number of qubits, $n$. However, an important step during the classical pre-processing involves compiling a circuit implementation of a unitary $U_{c}$ that loads the $M = 2^{(m+1)}$ Fourier coefficients on $m+1$ qubits. This cost scales polynomially with the number of Fourier coefficients. In fact, the asymptotic cost of implementing $U_{c}$ either using the cascade of uniformly controlled rotations as in Fig.~(\hyperlink{fig:zgr_circuit_diagram}{1b}) or using the Schmidt decomposition method as in Fig.~(\hyperlink{fig:svd_circuit}{1c}) scales as $O(M^3)$. If this classical pre-processing cost were too high, then our method would not be suitable for functions for which a large number of Fourier coefficients are needed to approximate the target function with the desired error. For this reason, we benchmark the classical pre-processing time needed to implement the unitary $U_{c}$. More precisely, we recorded the CPU wall clock time, $T_{\text{cl}}$, required to find a circuit implementation of the unitary $U_{c}$ for randomly generated $2^{m+1}$ Fourier coefficients for $6\le m \le 14 $, and the results are presented in Fig.~(\ref{fig:classical_time}). We found that the classical pre-processing time, $T_{\text{cl}}$, for $m=12$ is less than two minutes using the uniformly controlled rotations and is less than a minute using the Schmidt decomposition. The benchmark was performed on a $2020$ $13$-inch Macbook Pro with $3.2$ GHz M1 CPU and $8$ GB main memory. This benchmark demonstrates that the classical pre-processing time needed to implement the FSL method is computationally tractable even for sufficiently large values of $m$. 

\begin{figure}
    \centering
    \includegraphics{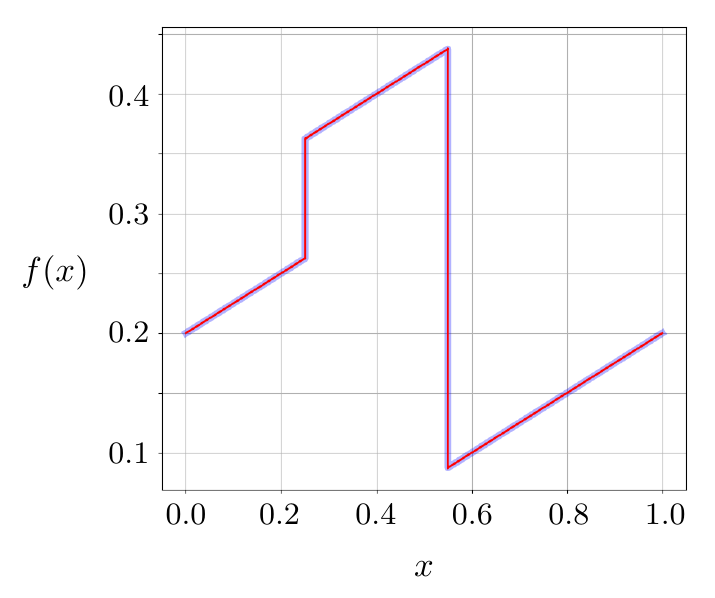}
    \caption{Quantum circuit simulation of loading a piece-wise discontinuous function into a quantum state of $n=20$ qubits using the FSL method with a quarter million Fourier coefficients. The total CPU wall clock time for this simulation was around $15$ minutes which included the classical pre-processing time needed to find a circuit implementation of loading Fourier coefficients and the time needed to simulate the execution of the circuit. }
    \label{fig:m17}
\end{figure}

Furthermore, we recorded the total compute time needed to perform a classical simulation of the FSL method's quantum circuit loading the discontinuous function we presented in Fig.~(\hyperlink{fig:discontinuous}{2f})  with $m=17$ (i.e., approximately a quarter million Fourier coefficients) and $n=20$. We found that it took a total of around $15$ minutes ($13$ minutes for finding $U_{c}$ using the Schmidt decomposition method and $2$ minutes for executing the circuit in Qiskit). The results of the simulation are presented in Fig.~(\ref{fig:m17}).

\section{Error analysis} \label{app-fid-analysis}

\begin{figure}
    \centering
    \includegraphics[scale=1.1]{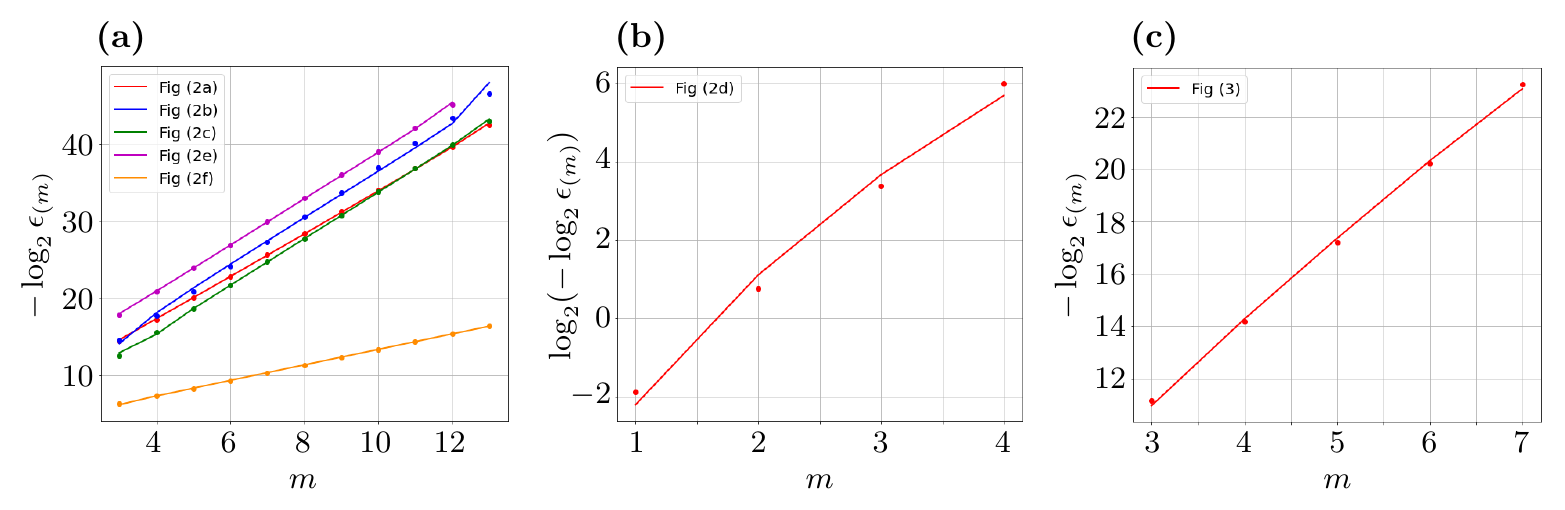}
    \caption{Plots for infidelity, $\epsilon_{(m)}$, as a function of $m$ for all of the functions considered in Sec.~(\ref{sec:method}). The colored lines are used to plot the values of the infidelities, whereas the colored dots denote the linear function determined by linear fitting the values of the infidelities. The slope of the linear fit function describing the infidelity for the function in Fig.~(\protect\hyperlink{fig:dg}{2f}) is  approximately $1$, whereas the slope is approximately $3$ for the functions in Fig.~(\protect\hyperlink{fig:dg}{2a}), Fig.~(\protect\hyperlink{fig:dg}{2b}), Fig.~(\protect\hyperlink{fig:dg}{2c}), Fig.~(\protect\hyperlink{fig:dg}{2e}), and Fig.~(\ref{fig:2d-statevector}). Plot \textbf{(b)} illustrates the doubly exponential decay of infidelity for the function in Fig.~(\protect\hyperlink{fig:dg}{2d}).}
    \label{fig:error_plots}
\end{figure}

In the main text, we stated that the infidelity of the FSL method decays exponentially with $m$ at a rate that depends on the smoothness of the target function. In Fig.~(\ref{fig:error_plots}), we explicitly verify this exponential decay of infidelity for all of the functions considered in Sec.~(\ref{sec:method}). In particular, we show that the infidelity decays as $O(2^{-m})$ for the discontinuous function in Fig.~(\hyperlink{fig:dg}{2f}), whereas it decays as $O(2^{-3m})$ for continuous functions such as $x^{x}$ in Fig.~(\hyperlink{fig:dg}{2a}), sinc function in Fig.~(\hyperlink{fig:dg}{2b}), reflected put option function in Fig.~(\hyperlink{fig:dg}{2c}), hyperbolic tangent function in Fig.~(\hyperlink{fig:dg}{2e}), and the two-dimensional sinc function considered in Fig.~(\ref{fig:2d-statevector}). Moreover, for an infinitely differentiable function such as the wavefunction of the first excited state of the quantum harmonic oscillator considered in Fig.~(\hyperlink{fig:qho}{2d}), the infidelity decays doubly exponential in $m$.

In the rest of this appendix, we argue that the exponential decay of infidelity is true in general. We do this by deriving a bound on infidelity as a function of $m$. For simplicity, we only consider functions of a single variable in our analysis, but the generalization to multivariate functions is straightforward. Our derivation is similar to the analysis in Ref.~\cite{truncation_error} in which the bounds on the truncation error, as a function of the number of terms in the Fourier series, were obtained. 

Note that any function encoding state $\ket{f} \, = \, \sum_{k=0}^{2^{n}-1} \, f_{k} \ket{k}$ can exactly be written as $\ket{f} \, = \, \qftd \ket{c} \, $, where
\begin{align}
    \ket{c} \, = \, \sum_{k=0}^{2^{n-1}} \, c_{k} \ket{k} \, + \, \sum_{k=1}^{2^{n-1}-1} \, c_{-k} \ket{2^{n}-k} \, ,
\end{align}
and $c_{k}$ are the \textit{discrete} Fourier coefficients of $f(x)$:
\begin{align}
    c_{k} \, = \, \frac{1}{2^{n/2}} \, \sum_{\ell=0}^{2^{n}-1} \, f_{\ell} \, e^{i2\pi k \ell / 2^{n}} \, . \label{eq-dis-c}
\end{align}
The normalization of the state $\ket{f}$ implies that the state $\ket{c}$ is normalized. That is,
\begin{align}
    \sum_{k = -2^{n-1}+1}^{2^{n-1}} \, |c_{k}|^{2} \, = \, 1 \, . \label{eq-norm-1}
\end{align}

The FSL method approximates the state $\ket{f}$ by the state $\ket{f_{(m)}} \, = \, \qftd \ket{c_{(m)}} \, $, where
\begin{align}
    \ket{c_{(m)}} \, = \, \mathcal{N}^{-1/2} \, \sum_{k=0}^{2^{m}-1} \, c_{k} \ket{k} \, + \, \mathcal{N}^{-1/2} \, \sum_{k=1}^{2^{m}-1} \, c_{-k} \ket{2^{n}-k} \, ,
\end{align}
and $\mathcal{N}$ is normalization constant:
\begin{align}
    \mathcal{N} \, = \, \sum_{k = -2^{m}+1}^{2^{m}-1} \, |c_{k}|^{2} \, . \label{eq-norm-2}
\end{align}
The infidelity between the state $\ket{f}$ and $\ket{f_{(m)}}$ can be calculated in terms of the Fourier coefficients as follows:
\begin{align}
    \epsilon_{(m)} \, =&\, \, 1 - \left| \braket{f | f_{(m)}} \right|^{2} \, , \nonumber\\
    =&\, \, 1 - \left| \braket{c | c_{(m)}} \right|^{2} \, , \nonumber\\
    =&\, \,    1 -  \mathcal{N}^{-1} \, \left( \sum_{k = -2^{m}+1}^{2^{m}-1} \, |c_{k}|^{2} \right)^{2} \, , \nonumber\\
    =&\, \,    1 -    \sum_{k = -2^{m}+1}^{2^{m}-1} \, |c_{k}|^{2} \, , \nonumber\\
    =&\, \,  \sum_{k = 2^{m}}^{2^{n-1}} \, |c_{k}|^{2} \, + \, \sum_{k = 2^{m}}^{2^{n-1}-1} \, |c_{-k}|^{2} \, , \label{eq-em-form}
\end{align}
where we have used Eq.~\eqref{eq-norm-1} in the last step and Eq.~\eqref{eq-norm-2} in the step before that.

In order to derive a bound on the infidelity, we first need to derive a bound on the Fourier coefficients.
To do this, we define a \textit{difference} operator $\Delta$ such that
\begin{align}
    \Delta f_{\ell} \, =: \, f_{\ell+1} - f_{\ell} \, .
\end{align}
From this definition, we deduce that the action of this difference operator on a product of functions is given by
\begin{align}
    \Delta( f_{\ell} g_{\ell} ) \, = \, (\Delta f_{\ell}) \cdot g_{\ell+1} \, + \, f_{\ell} \cdot (\Delta g_{\ell} ) \, .
\end{align}
This is a discrete analog of the action of the derivative operator on a product of functions. Now, if we take $g_{\ell} \, = \, e^{i 2\pi k \ell/2^{n}}$ in the above expression and rearrange the terms, we get
\begin{align}
    f_{\ell} e^{i 2\pi k \ell/2^{n}} \, = \, \frac{1}{e^{i 2\pi k/2^{n}} - 1} \, \Delta( f_{\ell} e^{i 2\pi k \ell/2^{n}} ) \, - \, \frac{1}{1 - e^{-i 2\pi k/2^{n}}} \, (\Delta f_{\ell}) \cdot e^{i 2\pi k \ell/2^{n}} \, .
\end{align}
Note we have applied the identity $\Delta e^{i 2\pi k \ell/2^{n}} \, = \, \left( e^{i 2\pi k/2^{n}} - 1  \right) \, \cdot \, e^{i 2\pi k \ell/2^{n}} \, $ to obtain the above expression. Now using Eq.~\eqref{eq-dis-c}, we find that the Fourier coefficient $c_{k}$ can be written as
\begin{align}
    c_{k} \, =&\, \, \frac{1}{2^{n/2}} \, \frac{1}{e^{i 2\pi k/2^{n}} - 1} \, \sum_{\ell=0}^{2^{n}-1} \Delta( f_{\ell} e^{i 2\pi k \ell/2^{n}} ) \, - \frac{1}{2^{n/2}} \, \frac{1}{1 - e^{-i 2\pi k/2^{n}}} \, \sum_{\ell=0}^{2^{n}-1} \, (\Delta f_{\ell}) \cdot e^{i 2\pi k \ell/2^{n}} \, ,\nonumber\\
    =&\, \, \frac{1}{2^{n/2}} \, \frac{1}{e^{i 2\pi k/2^{n}} - 1} \, \Big(f_{2^{n}} - f_{0} \Big) \, - \frac{1}{2^{n/2}} \, \frac{1}{1 - e^{-i 2\pi k/2^{n}}} \, \sum_{\ell=0}^{2^{n}-1} \, (\Delta f_{\ell}) \cdot e^{i 2\pi k \ell/2^{n}} \, , \nonumber\\
    =&\, \, - \frac{1}{2^{n/2}} \, \frac{1}{1 - e^{-i 2\pi k/2^{n}}} \, \sum_{\ell=0}^{2^{n}-1} \, (\Delta f_{\ell}) \cdot e^{i 2\pi k \ell/2^{n}} \, , \label{eq-ck-1}
\end{align}
where we have used the periodicity of $f$ to cancel the first term. Now, using the triangle inequality, we get
\begin{align}
    |c_{k}| \, =&\, \, \frac{1}{2^{n/2}} \, \frac{1}{2\sin(\pi k/2^{n})} \, \left| \sum_{\ell=0}^{2^{n}-1} \, (\Delta f_{\ell}) \cdot e^{i 2\pi k \ell/2^{n}}\right| \, , \nonumber\\
    \le&\, \, \frac{1}{2^{n/2}} \, \frac{1}{2\sin\big(\pi k/2^{n}\big)} \,  \sum_{\ell=0}^{2^{n}-1} \, \big| (\Delta f_{\ell}) \big| \, , \nonumber\\
    =&\, \, \frac{1}{2^{n/2}} \, \frac{1}{2\sin\big(\pi k/2^{n}\big)} \,   \Big\Vert \ket{\Delta f} \Big\Vert_{1} \, , \label{eq-ck-bound}
\end{align}
where $\Big\Vert \ket{\Delta f} \Big\Vert_{1}$ is the $1$-norm of the state $\ket{\Delta f} \, = \, \sum_{\ell=0}^{2^{n}-1} \, \Delta f_{\ell} \ket{\ell} \, $.

Due to this bound on the Fourier coefficients, the  infidelity in Eq.~\eqref{eq-em-form} is bounded by
\begin{align}
    \epsilon_{(m)} \, \le \, \frac{\, \Big\Vert \ket{\Delta f} \Big\Vert_{1}^{2}}{2^{n+1}} \, \sum_{k=2^{m}}^{2^{n-1}} \frac{1}{\Big(\sin\big(\pi k/2^{n}\big)\Big)^{2}} \, .
\end{align}
Bounding the sum by an integral provides the bound
\begin{align}
    \epsilon_{(m)} \, \le&\, \, \frac{ \Big\Vert \ket{\Delta f} \Big\Vert_{1}^{2}}{2^{n+1}} \, \int_{k=2^{m}}^{2^{n-1}} dk \, \frac{1}{\Big(\sin\big(\pi k/2^{n}\big)\Big)^{2}} \, ,\nonumber\\
    =&\, \, \frac{ 1 }{2\pi} \, \Big\Vert \ket{\Delta f} \Big\Vert_{1}^{2} \,  \cot\left( \pi 2^{m}/2^{n} \right) \, . \label{eq-em-bound-int}
\end{align}

The bound derived above holds for an arbitrary number of qubits, $n$. To find the asymptotic limit of this bound, we first need to determine how the $1$-norm of the state $\ket{\Delta f} $ scales with large values of $n$. Note that the $1$-norm of $\ket{\Delta f} $ is determined by the values of $f_{\ell}$ at local maximas, local minimas, and the end-points of the domain. That is,
\begin{align}
    \Big\Vert \ket{\Delta f} \Big\Vert_{1} \, =&\,  \, \sum_{\ell=0}^{2^{n}-1} \, \big| (\Delta f_{\ell}) \big| \, ,\nonumber\\
    =&\,  \, - \big(\text{sgn}(\Delta f_{0}) - \text{sgn}(\Delta f_{2^{n}-1})\big) \, f_{0} \, + \, 2  \sum_{\ell\in\ell_{\text{max}}}  f_{\ell} \, - \, 2  \sum_{\ell\in\ell_{\text{min}}}  f_{\ell} \,  , \label{eq-1norm-form}
\end{align}
where $\ell_{\text{max}}$ is the set of points when $f_{\ell}$ is locally maximum (i.e., $f_{\ell}>f_{\ell \pm 1}$) and $\ell_{\text{min}}$ is the set of points when $f_{\ell}$ is locally minimum (i.e., $f_{\ell}<f_{\ell \pm 1} \, $). The normalization condition for the state $\ket{f}$ implies that $f_{\ell} \, \sim 1/\sqrt{2^{n}}$ for large values of $n$. This means that the $1$-norm also scales as
\begin{align}
    \Big\Vert \ket{\Delta f} \Big\Vert_{1} \sim \frac{C}{\sqrt{2^{n}}} \, ,
\end{align}
where $C > 0$ is a constant. With this scaling result, we find that the bound on $\epsilon_{(m)}$ approaches 
\begin{align}
    \epsilon_{(m)} \, \le \,  \, \frac{C^{2}}{\pi^{2} 2^{m+1}} \label{eq-bound-p1}
\end{align}
in the large limit as $n \to \infty$. 

The bound in Eq.~\eqref{eq-bound-p1} is valid for any periodic target function, $f$. We can obtain stronger bounds if we make certain assumptions about the \textit{continuity} of $\Delta^{p}f_{k}$, where
\begin{align}
    \Delta^{p} f_{\ell} \, =: \, \Delta^{p-1} f_{\ell+1} - \Delta^{p-1} f_{\ell} \, ,
\end{align}
where $p$ is a positive integer, and $f_{\ell}$ is periodically extended beyond $\ell \in \{ 0,1,\cdots, 2^{n}-1\}$ according to $f_{2^{n}+\ell} = f_{\ell}$. Let us assume that the $\Delta^{q}f_{\ell}$ for all $q \le p$ is \textit{continuous} by which we mean that $\Delta^{q}f_{\ell} \sim (1/2^{n})^{q+1/2}$ for $q \le p$ and that $\Delta^{q}f_{\ell} \sim (1/2^{n})^{p+1/2}$ for $q \ge p$. This implies that 
\begin{align}
    \Big\Vert \ket{\Delta^{p+1} f} \Big\Vert_{1} \sim \frac{C_{p}}{(2^{n})^{p+1/2}} \, . \label{eq-bound-1-norm-del-p}
\end{align}
To see why this leads to a stronger bound, we repeat the analysis in Eq.~\eqref{eq-ck-1} $p$ times to get
\begin{align}
    c_{k} \, = \,  \frac{1}{2^{n/2}} \, \left(\frac{-1}{1 - e^{-i 2\pi k/2^{n}}}\right)^{p+1} \,\, \sum_{\ell=0}^{2^{n}-1} \, (\Delta^{p+1} f_{\ell}) \cdot e^{i 2\pi k \ell/2^{n}} \, .
\end{align}
This leads to a stronger bound of the Fourier coefficients,
\begin{align}
    |c_{k}| \, \le \, \frac{1}{2^{n/2}} \, \frac{1}{\left(2\sin\big(\pi k/2^{n}\big)\right)^{p+1}} \,   \Big\Vert \ket{\Delta^{p+1} f} \Big\Vert_{1} \, ,
\end{align}
which leads to a stronger bound on infidelity,
\begin{align}
    \epsilon_{(m)} \, \le&\, \, \frac{\Big\Vert \ket{\Delta^{p+1} f} \Big\Vert_{1}^{2} \,}{2^{n+1+2p}} \, \sum_{k=2^{m}}^{2^{n-1}} \frac{1}{\Big(\sin\big(\pi k/2^{n}\big)\Big)^{2p+2}} \, , \nonumber\\
    \le&\, \, \frac{\Big\Vert \ket{\Delta^{p+1} f} \Big\Vert_{1}^{2} \,}{2^{n+1+2p}} \, \int_{k=2^{m}}^{2^{n-1}} dk \, \frac{1}{\Big(\sin\big(\pi k/2^{n}\big)\Big)^{2p+2}} \, , \nonumber\\
    =&\, \, \frac{\Big\Vert \ket{\Delta^{p+1} f} \Big\Vert_{1}^{2} \,}{2^{2p+2} \pi} \, \left[B\Big(1;-\frac{p+1}{2},\frac{1}{2}\Big) \, - \, B\Big(\sin^{2}\left(\frac{\pi 2^{m}}{2^{n}}\right) ;-\frac{p+1}{2},\frac{1}{2}\Big) \right] \, ,
\end{align}
where $B(x;a,b)$ is the incomplete Beta function. Asymptotically, this bound approaches 
\begin{align}
    \epsilon_{(m)} \, \le \,  \, \frac{C_{p}^{2}}{(2p+1) \pi^{2p+2}  2^{(m+1)(2p+1)}} \, .
\end{align}
This concludes our discussion of the upper bound on the infidelity.

\section{Details of experiments on the Quantiuum System Model H1 Quantum Computers} \label{app-experiments}

In Sec.~(\ref{sec:exp_results}), we presented the results of the experiments we performed on the Quantinuum H$1$-$1$ and H$1$-$2$ quantum computers. Here, we provide further details about those experiments and their implementation.  

\begin{figure}
    \centering
    \includegraphics{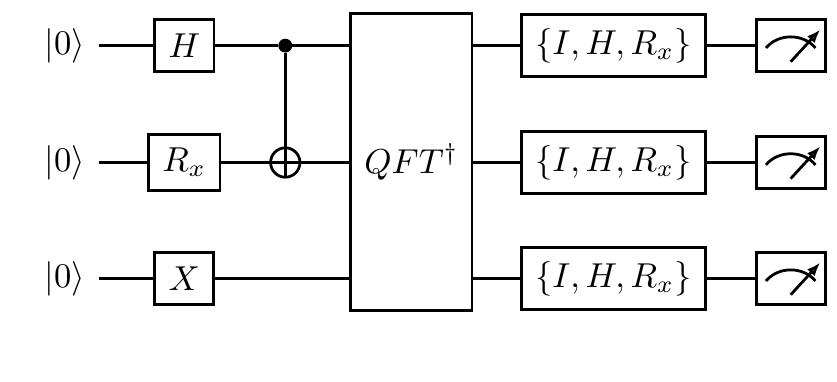}
    \caption{The FSL circuit used to load the complex-valued function $f(x) = \left(\cos(2\pi x) - 1.5 i \cos(6\pi x) \right)/\sqrt{13}$ on $3$-qubit state. In order to verify the prepared state, we apply single qubit gates from $\{I, H, R_{x}(\pi/2)\}$ to each qubit before measuring it in the computational basis. We repeat this process for all operations in the set $\{I, H, R_{x}(\pi/2)\}^{\otimes 3}$ as depicted in the above figure. The results of these $27$ experiments are processed to reconstruct the density matrix through maximum likelihood estimation.}
    \label{fig:tomo_circuit}
\end{figure}

The first experiment we performed involved loading a complex-valued function $f(x) = \left( \cos(2\pi x) - 1.5 i \cos(6\pi x) \right)/\sqrt{13} \, $. Since this function is already in the form of a Fourier series, the pre-processing step of approximating a function by a truncated Fourier series is not needed. Moreover, since there is no truncation error in this case, the FSL method can load this function exactly. The FSL circuit used to load this function into a $3$-qubit state is shown in Fig.~(\ref{fig:tomo_circuit}). It is worthwhile to point out that this circuit is a special case of the circuit shown in Fig.~(\hyperlink{fig:state_prep}{1a}) as the unitary $U_{c}$ is acting on all of the $n=3$ qubits. This, of course, is only possible for functions that are sufficiently sparse in Fourier space so that all the non-zero Fourier coefficients can be loaded with a shallow circuit. 

To verify the state prepared by the circuit in Fig.~(\ref{fig:tomo_circuit}), we performed quantum state tomography where we applied single-qubit gates from the set $\{I, H, R_{x}(\pi/2)\}$ on each of the qubits before measuring them. The reason for using $H$ rather than $R_{y}(\pi/2)$ for rotating the measurement basis was the observation that the inverse QFT, used in the FSL circuit, ends with a $H$ gate. Therefore, in some of the $27$ experiments, we were able to ignore two successive $H$ gates due to their cancellation. 

As we discussed in Sec.~(\ref{sec:exp_results}), we performed maximum likelihood estimation to reconstruct the density matrix from the measurement results. Here, we discuss the results of an alternate reconstruction of the state. Recall that a $3$-qubit density matrix can be written as: $\rho \, = \, \frac{1}{8} \, \sum_{\mu_{1}=0}^{3}\sum_{\mu_{2}=0}^{3}\sum_{\mu_{3}=0}^{3} \, s_{\mu_{1}\mu_{2}\mu_{3}} \, \sigma_{\mu_{1}}\otimes\sigma_{\mu_{2}}\otimes\sigma_{\mu_{3}} \, $, where $\{\sigma_{0}, \sigma_{1}, \sigma_{2}, \sigma_{3}\} = \{I, \sigma_{x}, \sigma_{y}, \sigma_{z}\}$ are the standard Pauli matrices. Thus, the reconstruction of the state is reduced to the reconstruction of the coefficients $s_{\mu_{1}\mu_{2}\mu_{3}}$. Since $s_{\mu_{1}\mu_{2}\mu_{3}} \, = \, \text{tr} \left( \rho \cdot \sigma_{\mu_{1}}\otimes\sigma_{\mu_{2}}\otimes\sigma_{\mu_{3}} \right) \, $ are just the expectation values of tensor products of Pauli operators, they can be estimated from the results of the measurements in $27$ different bases. We found that the matrix reconstructed in this way has a fidelity of $94.0\%$ which is almost the same as the fidelity we obtained through maximum likelihood estimation. However, this matrix is not a valid density matrix as it possesses some negative eigenvalues. Nevertheless, we find it encouraging that the fidelities obtained using two different methods are almost in agreement with each other.   

In the rest of the experiments we performed, we used the FSL method to load $\sqrt{f(x)}$ into a quantum state for a given function $f(x)$ and then compared the amplitudes of the prepared state to the value of the target function.
The precise expressions for the functions we considered are:  
\begin{align}
    \text{Bi-modal Gaussian:} \quad\quad f(x) \, \sim&\, \, \left(1-\lambda\right) e^{-(x-1/4)^{2}/2\sigma^{2}} \, + \, \lambda e^{-(x-3/4)^{2}/2\sigma^{2}} \, \quad\quad\quad \text{with} \, (\lambda,\sigma) = (0.3,0.1) \, , \label{eq-1d-bimodal}\\
    \text{Log-normal:} \,\,\,\,\,\,\, \quad\quad\quad\quad f(x) \, \sim&\, \, \frac{1}{x} e^{-\left(\log\left(x/q\right)\right)^{2}/2\sigma^{2}} \, \quad\quad\quad\quad\quad\quad\quad\quad\quad\quad\quad\quad\quad\,  \text{with} \, (q,\sigma) = (0.2,0.5) \,  \, , \label{eq-lognormal}\\
    \quad\text{Lorentzian:} \quad\,\,\, \quad\quad\quad\quad f(x) \, \sim&\, \, \frac{1}{1+(x-0.5)^{2}/\sigma^{2}} \, \quad\quad\quad\quad\quad\quad\quad\quad\quad\quad\quad\quad\quad  \text{with} \,\,  \sigma = 0.1 \,  \,  , \label{eq-cauchy}\\
     \text{`Spiky':} \,\,\,\,\,\,\,\,\quad\quad\quad\quad\quad\quad f(x) \, \sim&\, \, \left( \cos(4\pi x) \, + \lambda \, \cos(20\pi x) \right)^{2} \quad\quad\quad\quad\quad\quad\quad\quad\quad  \text{with} \,\,  \lambda = 2.5 \,  \,  ,\label{eq-spiky-app}
\end{align}
where $\sim$ denotes equality up to normalization. In addition to these functions of a single variable, we also considered the following function of two variables:
\begin{align}
    f(x,y) \, \sim \, \exp\left(-(x-\mu_{1})^{2}/\sigma_{11}^{2} \, - \, (y-\mu_{1})^{2}/\sigma_{12}^{2} \right) \, + \, \lambda \, \exp\left(-(x-\mu_{2})^{2}/\sigma_{21}^{2} \, - \, (y-\mu_{2})^{2}/\sigma_{22}^{2} \right) \, ,
\end{align}
with $\mu_{1} = 0.65$, $\mu_{2} = 0.35$, $\sigma_{11} = \sigma_{22} = \sqrt{1/50}$, $\sigma_{12} = \sqrt{1/40}$, $\sigma_{21} = \sqrt{1/30}$, and $\lambda = 0.5 \, $.

To load these functions on a quantum computer using the FSL method, we need to find the first few dominant Fourier coefficients of the square root of these functions. For bi-modal Gaussian function, log-normal distribution, and Lorentzian function, we used the first $8$, $16$, and $16$ Fourier coefficients, respectively. For the two-variable function, we used the first $64$ Fourier coefficients. Once the Fourier coefficients are determined, we need to find a suitable unitary $U_{c}$ which can load these Fourier coefficients. For all the functions except the `spiky' function in Eq.~\eqref{eq-spiky-app}, we implemented $U_{c}$ using the Schmidt-decomposition circuit shown in Fig.~(\hyperlink{fig:svd_circuit}{1c}). We first determined the unitary operators $U$ and $V$ in Fig.~(\hyperlink{fig:svd_circuit}{1c}) by performing the Schmidt decomposition of the state $\ket{\tilde{c}}$ from Eq.~\eqref{eq-c-state}. Then, we decomposed the unitary operators $U$ and $V$ into single and two-qubit gates using the built-in transpiler available in Qiskit \cite{Qiskit}. 

\begin{figure}
    \centering
    \includegraphics{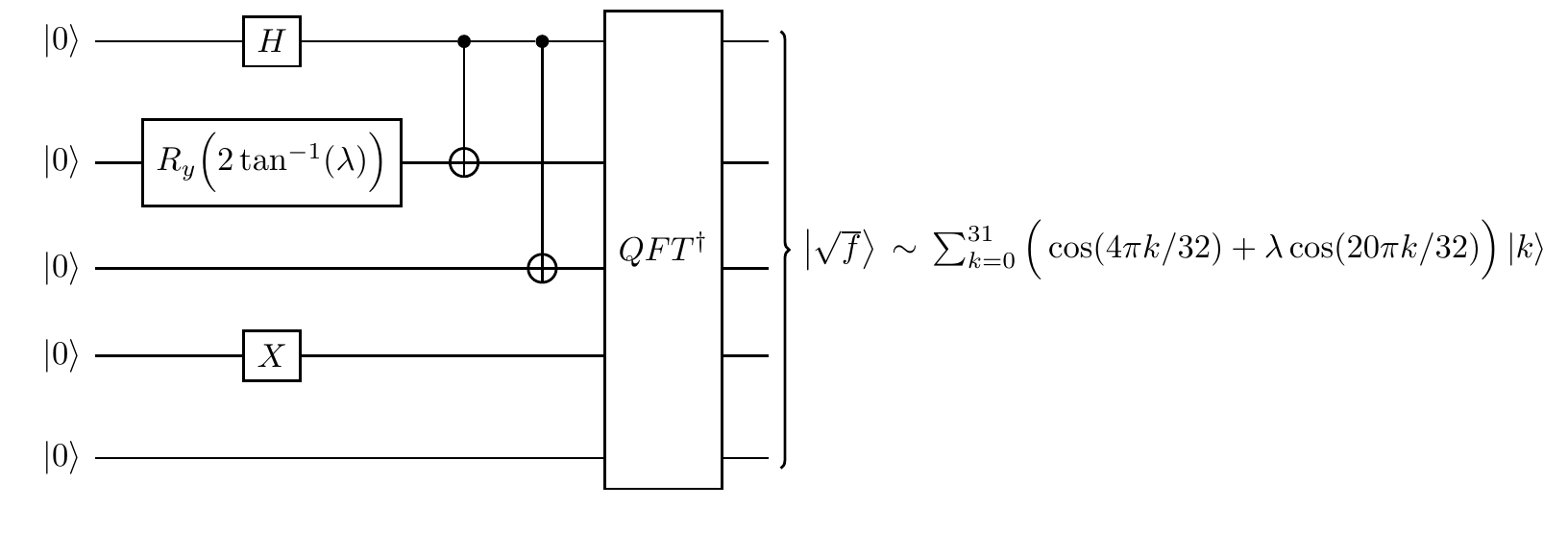}
    \caption{The quantum circuit to load the `spiky' function in Eq.~\eqref{eq-spiky-app} into a quantum state of $5$-qubit. Since there were only two Fourier modes present in this case, we were able to construct an efficient to load the Fourier coefficients. }
    \label{fig:spiky-circuit}
\end{figure}

On the other hand, when loading the `spiky' function, we do not need to perform any such classical pre-processing given that only two Fourier modes are active. This is another instance in which the target state is suitably sparse in Fourier space such that we are able to construct a shallow quantum circuit to load all the non-zero Fourier coefficients. The full FSL circuit that we used to load the `spiky' function is shown in Fig.~(\ref{fig:spiky-circuit}). 

Before we executed the FSL circuit on the Quantinuum H$1$-$1$ and H$1$-$2$ quantum computers, we reduced the gate count of the circuits as much as possible in order to reduce the effect of hardware noise.  One obvious but crucial simplification that we exploited was to ignore all the SWAP gates that appear in the implementation of the inverse QFT. Instead of applying these SWAP gates, we manually changed the ordering of the qubits prior to applying the inverse QFT without SWAPs.

The experiments we performed on the Quantinuum H$1$-$1$ and H$1$-$2$ quantum computers were subject to various sources of noise. A list of sources of error present in System Model H$1$ quantum computers is available in the product data sheet \cite{h1-data}. For instance, the typical state preparation and measurement (SPAM) error is around $3\times 10^{-3}$, whereas the typical infidelities for single qubit and two qubit gates are $4 \times 10^{-5}$ and $3 \times 10^{-3}$, respectively.

Since we are making use of multiple mid-circuit measurements in our experiments, another source of error is the decoherence of the unmeasured qubits. This measurement crosstalk error is around $2\times 10^{-5}$ on average. Additionally, the error due to dephasing is $4\times 10^{-4}$ on average per depth-$1$ circuit execution time (the time required to apply a single round of single-qubit and two-qubit gates on all pairs of the qubits) \cite{h1-data}. 

The hardware performance specifications for the Quantinuum System Model H$1$ quantum computers provided in \cite{h1-data} are correct as of October $3$rd, $2022$. However, we performed our experiments over the course of several months, and the hardware specifications fluctuated during that time. Therefore, we provide in Table.~(\ref{table-record}) the gate infidelities, measurement crosstalk errors, and SPAM errors as measured by Quantinuum for each of the dates we have performed the experiments \footnote{These specifications were provided by Quantinuum upon request.}. 

\begin{table}
    \centering
    \begin{tabular}{|m{8.05em}||m{7.0em} m{8.5em} m{8.5em} m{8.5em} || m{8.5em}|} 
    \hline
     &  SPAM & {Measurement} & Single-qubit & Two-qubit & {Experiments} \\  
        {Dates} & error  & crosstalk error & gate infidelity  & gate infidelity &  performed   \\
         & (Average)  & \,\,\,\,(Worst) & (Average) & (Average) &  \\ \hline\hline
        $12/04/2021$ & $3.7 \times 10^{-3}$ & $5.0 \times 10^{-4}$ & $4.4 \times 10^{-5}$ & $3.0 \times 10^{-3}$ & Fig.~(\hyperlink{fig:lognormal_exp}{5b})   \\ 
        \hline
        $12/07/2021$ & $3.4 \times 10^{-3}$ & $4.0 \times 10^{-4}$ & $3.9 \times 10^{-5}$  & $2.9 \times 10^{-3}$  & Fig.~(\hyperlink{fig:cauchy_exp}{5c}) \\
        \hline 
        $12/15/2021$ & $4.0 \times 10^{-3}$ & $2.0 \times 10^{-4}$ & $7.1 \times 10^{-5}$  & $2.1 \times 10^{-3}$  & Figs.~(\hyperlink{fig:lognormal_exp}{5b}) and ~(\hyperlink{fig:cauchy_exp}{5c})\\ 
        \hline 
        $01/12/2022$ & $3.3 \times 10^{-3}$ & $3.0 \times 10^{-4}$ & $5.9 \times 10^{-5}$ & $3.2 \times 10^{-3}$  & Fig.~(\hyperlink{fig:spiky_exp}{5d}) \\ 
        \hline 
        $02/17/2022$ & $4.0 \times 10^{-3}$ & $6.0 \times 10^{-4}$ & $5.7 \times 10^{-5}$ & $2.9 \times 10^{-3}$  & Fig.~(\ref{fig:2d_state_expr}) \\ 
        \hline 
        $04/14/2022$ & $3.0 \times 10^{-3}$ & $4.0 \times 10^{-4}$ & $5.3 \times 10^{-5}$ & $2.9 \times 10^{-3}$ & Fig.~(\ref{fig:2d_state_expr}) \\ 
        \hline 
        $04/29/2022$ & $3.0 \times 10^{-3}$ & $3.0 \times 10^{-4}$ &  $3.5 \times 10^{-5}$ & $2.5 \times 10^{-3}$ & Fig.~(\hyperlink{fig:bimodal_exp}{5a}) \\ 
        \hline 
        $05/17/2022$ & $2.9 \times 10^{-3}$ & $2.0 \times 10^{-4}$ & $5.3 \times 10^{-5}$  & $2.8 \times 10^{-3}$ & Fig.~(\hyperlink{fig:bimodal_exp}{5a}) \\ 
        \hline 
        $05/18/2022$ & $3.3 \times 10^{-3}$ & $2.0 \times 10^{-4}$ & $6.0 \times 10^{-5}$  & $2.5 \times 10^{-3}$ & Fig.~(\ref{fig:2d_state_expr}) \\ 
        \hline 
        $06/21/2022$ & $2.7 \times 10^{-3}$ & $2.0 \times 10^{-4}$ & $4.6 \times 10^{-5}$  & $2.3 \times 10^{-3}$ & Fig.~(\ref{fig:3qubit_tomography}) \\ 
        \hline 
        $06/22/2022$ & $2.4 \times 10^{-3}$ & $4.0 \times 10^{-4}$ & $3.4 \times 10^{-5}$  & $2.4 \times 10^{-3}$ & Fig.~(\ref{fig:3qubit_tomography}) \\ 
        \hline 
        $06/23/2022$ & $2.6 \times 10^{-3}$ & $1.0 \times 10^{-4}$ & $2.9 \times 10^{-5}$  & $2.1 \times 10^{-3}$ & Fig.~(\ref{fig:3qubit_tomography}) \\ 
        \hline 
        $07/12/2022$ & $2.6 \times 10^{-3}$ & $2.0 \times 10^{-4}$ & $5.6 \times 10^{-5}$  & $2.4 \times 10^{-3}$ & Fig.~(\ref{fig:2d_state_expr}) \\ 
        \hline 
        $07/22/2022$ & $2.7 \times 10^{-3}$ & $2.0 \times 10^{-4}$ & $4.1 \times 10^{-5}$ & $2.1 \times 10^{-3}$ & Fig.~(\ref{fig:3qubit_tomography}) \\ 
        \hline 
        $08/09/2022$ & $3.4 \times 10^{-3}$ & $2.0 \times 10^{-4}$ & $4.8 \times 10^{-5}$  & $2.0 \times 10^{-3}$ & Fig.~(\ref{fig:3qubit_tomography}) \\ 
        \hline 
        $08/23/2022$ & $2.8 \times 10^{-3}$ & $1.0 \times 10^{-4}$ & $5.5 \times 10^{-5}$  & $2.0 \times 10^{-3}$ & Fig.~(\ref{fig:2d_state_expr}) \\ 
        \hline 
        $09/23/2022$ & $3.2 \times 10^{-3}$ & $1.0 \times 10^{-4}$ & $7.1 \times 10^{-5}$  & $3.4 \times 10^{-3}$ & Fig.~(\ref{fig:2d_state_expr}) \\ 
        \hline 
    \end{tabular}
    \caption{A record of hardware performance specifications for the Quantinuum System Model H$1$ quantum computers used on the days the experiments were performed. These specifications were measured by Quantinuum.}
    \label{table-record}
\end{table}

\section{Image loading using FSL method} \label{app-image}

Digital image processing is useful in many fields such as artificial intelligence, computer vision, and medical imagining to name a few. Quantum computers are expected to speed up various image processing tasks \cite{Zhang2014QSobelAN,Yao2017QuantumIP} provided there exists an efficient way to load images on a quantum computer. In this appendix, we argue that the FSL method can be modified to load images on a quantum computer. This is not surprising as commonly used classical image compression algorithms, such as JHEP, are based on spectral transforms. 

We can view a grayscale image $\mI$ as a two-dimensional matrix whose elements $\mI_{jk}$ store the brightness of the corresponding pixel. The matrix elements can be normalized such that $\mI_{jk} = 0$ for black-colored pixels and $\mI_{jk} = 1$ for white-colored pixels. For simplicity, we will assume that the image $\mI$ is composed of $2^{n}\times 2^{n}$ pixels. One way to encode the image $\mI$ on a quantum computer is the Flexible Representation of Quantum Images (FRQI). FRQI stores image data as the following quantum state of $(2n+1)$-qubits \cite{Le2011AFR}:
\begin{align}
    \ket{\mI} \, = \, \frac{1}{2^{n} } \, \sum_{j=0}^{2^{n}-1}\sum_{k=0}^{2^{n}-1} \, \ket{\mI_{jk}} \, \otimes \ket{j} \otimes \ket{k} \, , \label{eq-image-state}
\end{align}
where
\begin{align}
    \ket{\mI_{jk}} \, = \, \cos\left( \frac{\pi \mI_{jk}}{2} \right) \, \ket{0} \, + \, \sin\left( \frac{\pi \mI_{jk}}{2} \right) \, \ket{1} \, .
\end{align}
It follows that, $\ket{\mI_{jk}} = \ket{0}$ represents a purely black pixel, whereas $\ket{\mI_{jk}} = \ket{1}$ represents a purely white pixel. 

\begin{figure}
    \centering
    \hypertarget{fig:fsl_image}{}
    \hypertarget{fig:moon}{}
    \includegraphics{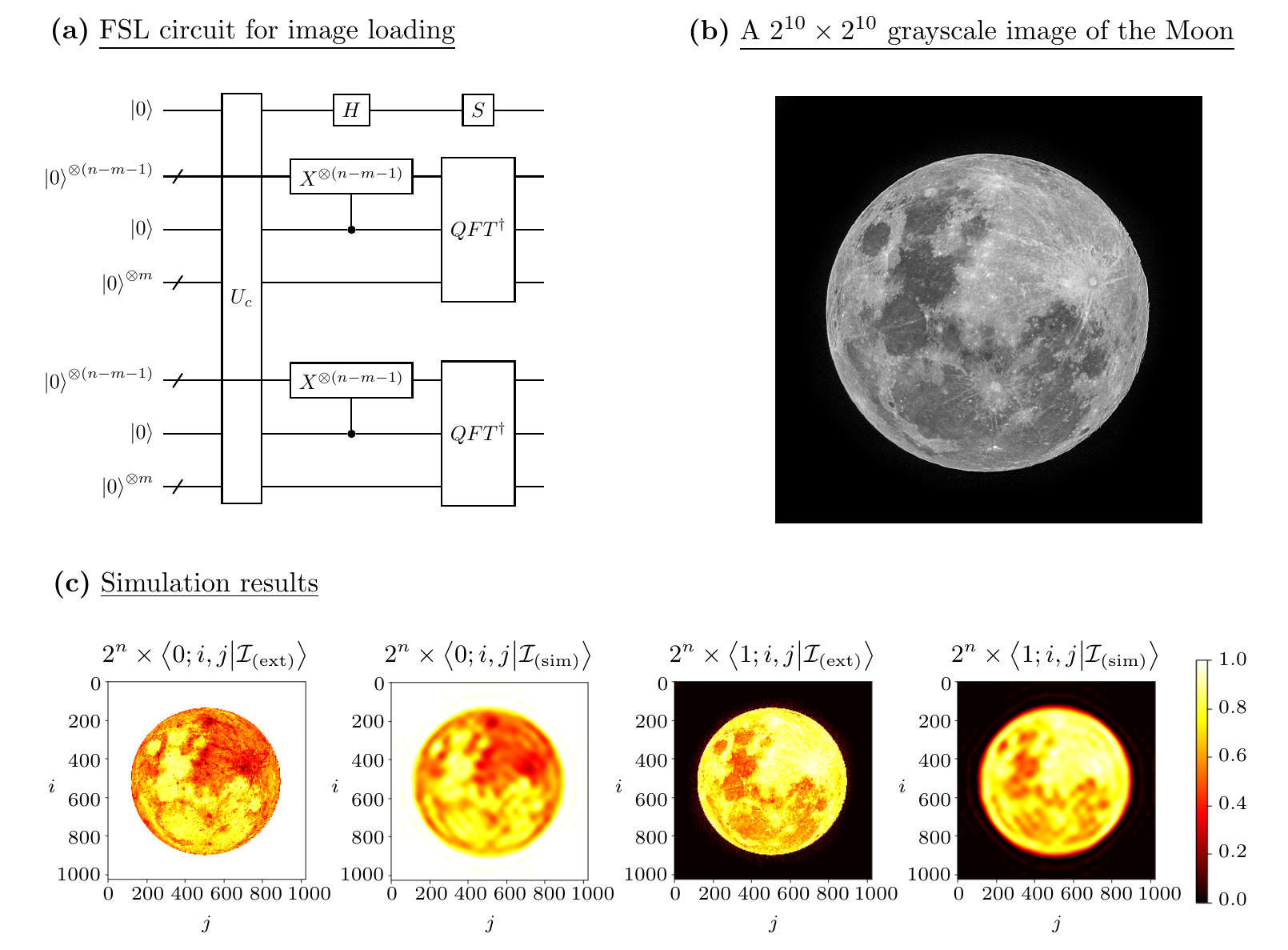}
    \caption{(a) Quantum circuit to prepare the FRQI representation of an image using the FSL method. The first qubit stores the brightness of a pixel, the next $n$ qubits store the horizontal coordinates of a pixel, and the last $n$ qubits store the vertical coordinates of a pixel. (b) A grayscale image of the Moon \cite{cite-unsplash} that we used as a test case for the FSL method. (c) Simulation results of approximately preparing the FRQI representation of the image of the Moon using the FSL method with $m=4$. As can be seen from these results, the state prepared by the FSL method $\ket{\mI_{(\text{sim})}}$ is a good approximation of the exact state $\ket{\mI_{(\text{ext})}}$ which can be further improved by taking higher values of $m$.}
    \label{fig:fsl_image}
\end{figure}

While the state $\ket{\mI}$ could be prepared using uniformly controlled rotations, that would require $O(2^{n})$ quantum gates \cite{Le2011AFR}. We can do better by observing that the problem of preparing $\ket{\mI}$ can be reduced to the problem of loading functions on a quantum computer. This becomes apparent if we rewrite $\ket{\mI}$ in Eq.~\eqref{eq-image-state} as
\begin{align}
    \ket{\mI} \, = \, \frac{1}{2^{n} } \, \ket{0} \otimes \, \left[ \sum_{j=0}^{2^{n}-1}\sum_{k=0}^{2^{n}-1} \, \cos\left( \frac{\pi \mI_{jk}}{2} \right) \ket{j} \otimes \ket{k} \right] \, + \, \frac{1}{2^{n} } \, \ket{1} \otimes \, \left[ \sum_{j=0}^{2^{n}-1}\sum_{k=0}^{2^{n}-1} \, \sin\left( \frac{\pi \mI_{jk}}{2} \right) \, \ket{j} \otimes \ket{k} \right] \, . \label{eq-image-state-e3}
\end{align}
It follows that we may prepare $\ket{\mI}$ by initializing the first qubit in a (non-uniform) superposition, and then loading $\cos\left( \frac{\pi \mI}{2} \right)$ and $\sin\left( \frac{\pi \mI}{2} \right)$ on the other $2n$ qubits when the first qubit is in $\ket{0}$ and $\ket{1}$ state, respectively. While controlled versions of any of the methods discussed in Sec.~(\ref{sec:discussion}) can be used to prepare the state $\ket{\mI}$ as written in Eq.~\eqref{eq-image-state-e3}, the FSL method is more suitable and efficient. This is because only a shallow depth portion of the quantum circuit used by the FSL method depends on the target function i.e., $U_c$ in Fig.~(\hyperlink{fig:state_prep}{1a}) and Fig.~(\ref{fig:circuit-2d}). The rest of the FSL circuit is independent of the target function. Therefore, unlike other methods, we need only vary $U_c$ instead of the entire circuit in order to prepare $\ket{\mI}$. This renders the FSL method very efficient for loading images on quantum computers.

In order to construct the modified quantum circuit implementing the FSL method for image loading, notice that we can further rewrite the state $\ket{\mI}$ as
\begin{align}
    \ket{\mI} \, = \, \frac{1}{\sqrt{2}} \ket{+i} \otimes \ket{g^{+}} \, + \, \frac{1}{\sqrt{2}} \ket{-i} \otimes \ket{g^{-}} \, , 
\end{align}
where $\ket{\pm i} \, = \, \frac{1}{\sqrt{2}} \ket{0} \pm \frac{i}{\sqrt{2}} \ket{1} \, $, and 
\begin{align}
    \ket{g^{\pm}} \, = \, \frac{1}{2^{n}} \, \sum_{j=0}^{2^{n}-1}\sum_{k=0}^{2^{n}-1} \, \exp\left( \mp \frac{i\pi \mI_{jk}}{2} \right) \ket{j} \otimes \ket{k} \, .
\end{align}
The advantage of writing the state $\ket{\mI}$ in this way is that the Fourier coefficients of the state $\ket{g^{+}}$ and $\ket{g^{-}}$ are no longer independent. Hence, we only need to find the discrete Fourier coefficients of $\ket{g^{+}}$ which reduces the classical pre-processing time. If we now assume that $\ket{g^{\pm}}$ can be well-approximated by $4^{(m+1)}$ Fourier modes, then the state $\ket{\mI}$ can be approximately prepared using the circuit shown in Fig.~(\hyperlink{fig:fsl_image}{12a}). In this circuit, the unitary $U_{c}$ acts on $2(m+1)+1$ qubits as
\begin{align}
    U_{c} \left(\ket{0}\otimes \ket{0}^{\otimes 2(m+1)} \right) \, = \, \ket{0} \otimes \ket{\tilde{c}^{+}} \, + \, \ket{1} \otimes \ket{\tilde{c}^{-}} \,  ,
\end{align}
where $\ket{\tilde{c}^{\pm}}$ are $2(m+1)$-qubit states encoding the Fourier coefficients of $\ket{g^{\pm}} \, $ and are analogous to the state $\ket{\tilde{c}}$ in Eq.~\eqref{eq-2d-fsl-tc}.

In order to assess the accuracy of the FSL method, we 
chose a $2^{10}\times 2^{10}$ pixel grayscale image of the Moon provided in Fig.~(\hyperlink{fig:moon}{12b}). We performed a classical simulation of the version of the quantum circuit shown in Fig.~(\hyperlink{fig:fsl_image}{12a}) for loading the image of the Moon into a state of $21$ qubits using the `statevector\textunderscore simulator' provided by Qiskit \cite{Qiskit}. The resulting image-encoding state $\ket{\mI}$ was loaded with the infidelity of around $10^{-2}$ for $m=4$, and the results of the simulation are shown in  Fig.~(\hyperlink{fig:fsl_image}{12c}).

\end{document}